# STOCHASTIC DEFLATOR FOR AN ECONOMIC SCENARIO GENERATOR WITH FIVE FACTORS

**Po-Keng Cheng*** **Frédéric Planchet***

Univ Lyon - Université Claude Bernard Lyon 1,

ISFA, Laboratoire SAF EA2429, F-69366, Lyon, France

Prim'Act, 42 avenue de la Grande Armée, 75017 Paris, France

**Abstract**

In this paper, we implement a stochastic deflator with five economic and financial risk factors: interest rates, market price of risk, stock prices, default intensities, and convenience yields. We examine the deflator with different financial assets, such as stocks, zero-coupon bonds, vanilla options, and corporate coupon bonds. We find required regularity conditions to implement our stochastic deflator. Our numerical results show the reliability of the deflator approach in pricing financial derivatives.

## Summary

* Po-Keng Cheng and Frédéric Planchet are researcher at SAF laboratory (EA n°2429). Frédéric Planchet is also consulting actuary at Prim'Act. Contact: ansd39@gmail.com / frederic@planchet.net.

## 1. INTRODUCTION

The Arrow-Debreu model of general equilibrium introduced the existence of an equilibrium in which the allocation of consumption and production is Pareto optimal with a system of prices for contingent commodities.[1] Their works have inspired tremendous research in fields of macroeconomics, financial economics, and asset pricing theory. Based on the concept of Arrow-Debreu securities, researchers had developed the fundamental theorems of asset pricing, which the second theorem tells us that an arbitrage-free market is complete if and only if the equivalent martingale measure is unique.[2]

In the case of Brownian diffusion, the Girsanov's Theorem enables us to change probability measure from a physical world to a risk-neutral world. Under risk-neutral measure, we have a closed-form solution for Black-Scholes options pricing model. However, we wouldn't always have analytical solutions for various classes of stochastic processes, which motivates us to study numerical methods for approximating solutions. In this paper, we investigate stochastic deflator approach for pricing of life insurance contracts.

Due to the complicatedness of life insurance contracts and interactions among economic and financial risk factors, a reliable tool for asset/liability management (ALM) and calculations of reserves would be demanded. In practice, "economic scenario generators" assist insurers in pricing insurance contracts and managing long-term risk[3].

The usual pricing scheme is as follows.

Fig. 1 - **Calculating the best estimate reserve for a life insurance contract**

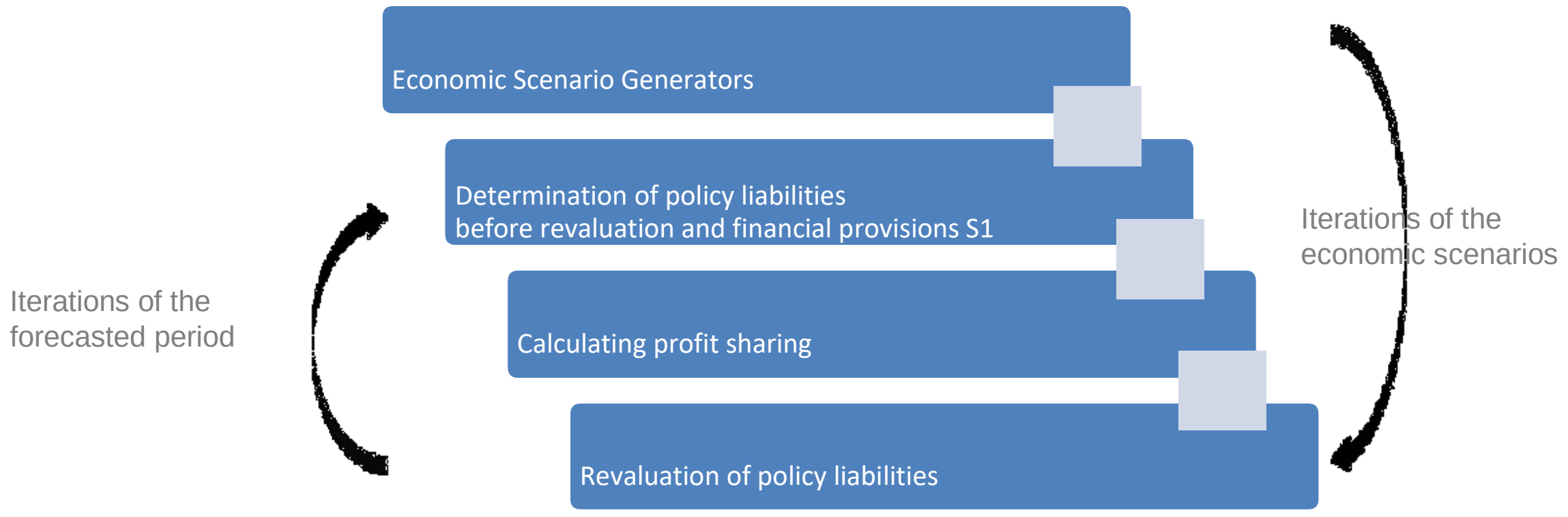

Usually, economic scenarios are computed under a risk-neutral measure; the actualization process involving risk-free rate is quite simple, numerically speaking. However, we like to point out that many "unusual" scenarios occur (e.g. 10-year rate $\geq$ 50%) under risk-neutral measure, which increases the difficulty to justify the calibration of "reaction functions" embedded in the ALM-projection model used to compute cash flows.

For example, the lapse rate is often a function of the difference between the revalorization rate of the contract and a reference rate; the parameters are calibrated observing “usual” values of economic parameters but may become difficult later to justify for atypical values of economic risk factors. We could use a stochastic deflator to address this problem, using only scenarios under physical measure[4]. The numerical calculations become tedious due to the complexity of the deflator, which involves a risk-free rate process and a change of measure between physical and risk-neutral measure. But the benefit is that we could calculate the deflator separately and multiply the deflator with projected cash flows for pricing insurance contracts.

[1] See, for example, Arrow and Debreu (1954), Geanakoplos (1989), and Mas-Colell et al. (1995) Chapter 19.
[2] See, for example, Harrison and Kreps (1979), Harrison and Pliska (1981), Delbaen and Schachermayer (1994), and Shreve (2004) Chapter 5.4.
[3] See, for example, Varnell (2011), Laurent et al. (2016) Chapters 3,4, and 5, and Pedersen et al. (2016).

In this paper, we adopt the deflator approach initiated by Dastarac and Sauveplane (2010) and include the processes of default and convenience yield from Longstaff et al. (2005) to calculate prices for financial derivatives. We compare the values calculated from the deflator approach with the values suggested by analytical formulas in simple cases. We find required regularity conditions to implement our stochastic deflator. Also, our numerical results show the reliability in statistics of the deflator approach for quite simple financial derivatives. Our goal is then to use this deflator to compute best estimates for a life insurance contract.

The remainder of the paper is organized as follows. Section 2 shows the deflator approach. Section 3 discusses the implementation of time discretization. Section 4 presents the numerical results. Section 5 concludes.

## 2. DEFLATOR APPROACH

Before discussing and deriving the general form of deflator, we need to generate correlated Brownian motions for the stochastic processes in our model. In our model, we consider the processes of interest rates, market prices of risk, stock prices, default intensities and convenience yields. Sections 2 and 3 discuss the technical details of implementations of the deflator and time discretization. Readers who are familiar with stochastic deflator and time discretization could directly skip to numerical results in Section 4.

### 2.1. GENERATE CORRELATED BROWNIAN MOTIONS

Let the Brownian motion part of each process $\mathbf{W}_{ESG}$ and the correlations matrix $\mathbf{C}_{ESG}$ among interest rates, stock prices, default intensities, and convenience yields be as follows.[5]

$$\mathbf{W}_{ESG} = \begin{bmatrix} W_r \\ W_S \\ W_\chi \\ W_\gamma \end{bmatrix}, \quad \mathbf{C}_{ESG} = \begin{array}{c} \\ W_r \\ W_S \\ W_\chi \\ W_\gamma \end{array} \begin{array}{c} \begin{array}{cccc} W_r & W_S & W_\chi & W_\gamma \end{array} \\ \begin{bmatrix} 1 & \rho_{rS} & \rho_{r\chi} & \rho_{r\gamma} \\ \rho_{rS} & 1 & \rho_{S\chi} & \rho_{S\gamma} \\ \rho_{r\chi} & \rho_{S\chi} & 1 & \rho_{\chi\gamma} \\ \rho_{r\gamma} & \rho_{S\gamma} & \rho_{\chi\gamma} & 1 \end{bmatrix} \end{array} \tag{1}$$

---

[4] See, for example, Bonnin et al. (2014), Borel-Mathurin et al. (2015), and Vedani et al. (2017).

[5] Here the correlations matrix $\mathbf{C}_{ESG}$ describes the linear correlation between each two processes of their Brownian motion parts. For discussions of dependence structure among random variables, see, for example, McNeil et al. (2005) Chapter 5 and Rachev et al. (2011) Chapter 2.6.4 .

Denote $r$ interest rate; $S$ stock price; $\chi$ default intensity; $\gamma$ convenience yield; $W_i,\ i=r,S,\chi,\gamma$ Brownian motion part of each process; and $\rho_{jk},\ j\neq k$ correlation between each two processes. To generate correlated $W_i,\ i=r,S,\chi,\gamma$, we require four independent Brownian motions $W_i,\ i=0,1,2,3$. Following is the construction of $\mathbf{W}_{ESG}$, technical details are provided in Appendix 1 of Supplementary materials.

$$\mathbf{W}_{ESG}=\begin{bmatrix} W_r \\ W_S \\ W_\chi \\ W_\gamma \end{bmatrix}=\begin{bmatrix} 1 & 0 & 0 & 0 \\ \rho_{rS} & \sqrt{1-\rho_{rS}^2} & 0 & 0 \\ \rho_{r\chi} & \rho'_{S\chi} & \rho'_{\chi\chi} & 0 \\ \rho_{r\gamma} & \rho''_{S\gamma} & \rho''_{\chi\gamma} & \rho''_{\gamma\gamma} \end{bmatrix}\begin{bmatrix} W_0 \\ W_1 \\ W_2 \\ W_3 \end{bmatrix} \quad (2)$$

where

$$\rho'_{S\chi}=\frac{\rho_{S\chi}-\rho_{rS}\rho_{r\chi}}{\sqrt{1-\rho_{rS}^2}},\ \rho'_{\chi\chi}=\sqrt{\frac{1-\rho_{rS}^2-\rho_{r\chi}^2-\rho_{S\chi}^2+2\rho_{rS}\rho_{r\chi}\rho_{S\chi}}{1-\rho_{rS}^2}},\ \rho''_{S\gamma}=\frac{\rho_{S\gamma}-\rho_{rS}\rho_{r\gamma}}{\sqrt{1-\rho_{rS}^2}},$$

$$\rho''_{\chi\gamma}=\frac{\rho_{\chi\gamma}-\rho_{r\chi}\rho_{r\gamma}-\rho_{S\chi}\rho_{S\gamma}-\rho_{rS}^2\rho_{\chi\gamma}+\rho_{rS}\rho_{r\chi}\rho_{S\gamma}+\rho_{rS}\rho_{r\gamma}\rho_{S\chi}}{\sqrt{1+\rho_{rS}^4-2\rho_{rS}^3\rho_{r\chi}\rho_{S\chi}-2\rho_{rS}^2+\rho_{rS}^2\rho_{r\chi}^2+\rho_{rS}^2\rho_{S\chi}^2-\rho_{r\chi}^2-\rho_{S\chi}^2+2\rho_{rS}\rho_{r\chi}\rho_{S\chi}}},$$

$$\rho''_{\gamma\gamma}=\sqrt{1-\rho_{r\gamma}^2-\rho''^2_{S\gamma}-\rho''^2_{\chi\gamma}}.$$

### 2.2. GENERAL FORM OF DEFLATOR WITH FIVE FACTORS[6]

Let $r(t)$, $B(t)$, $P(t,T,r(t))$, $S(t)$, $\chi(t)$, $\gamma(t)$, $D(t)$ be processes of interest rate, short-term saving, zero coupon bond of no risk with maturity $T$, stock price, default density, convenience yield, and deflator respectively.[7] Denote $E(\cdot)$ expectation under physical measure and $E^{\mathrm{Q}}(\cdot)$ expectation under risk-neutral measure. Let a discount process $\delta(t)$ equal $e^{-\int_0^t r(s)ds}$. For a nonnegative random variable $X$, we would like to have $E^{\mathrm{Q}}(\delta(t)X)=E[D(t)X]$ (i.e. $D(t)=\delta(t)\frac{d\mathbf{Q}}{d\mathbf{P}}$, where $\frac{d\mathbf{Q}}{d\mathbf{P}}$ is a Radon-Nikodym derivative). We describe the dynamics of each process in the following paragraphs, in a quite general Markovian framework.

---

[6] For discussions of stochastic deflator in insurance, see, for example, Dastarac and Sauveplane (2010) and Caja and Planchet (2011). For a reference of stochastic calculus related to Itô's lemma and Girsanov's Theorem, see Shreve (2004) Chapters 4 and 5.

[7] Note that research studies of pricing, default and liquidity are fundamental no matter which currency we use.

### 2.2.1. Dynamics of each process

For simplicity, we present the dynamics of each process in matrix as follows.[8]

$$\begin{bmatrix} dr(t) \\ dB(t) \\ \dfrac{dP(t,T,r(t))}{P(t,T,r(t))} \\ \dfrac{dS(t)}{S(t)} \\ d\chi(t) \\ d\gamma(t) \end{bmatrix} = \begin{bmatrix} \alpha(t,r(t)) & \beta(t,r(t)) & 0 & 0 & 0 \\ B(t)r(t) & 0 & 0 & 0 & 0 \\ \tilde{\mu}(t,r(t)) & \tilde{\sigma}(t,r(t)) & 0 & 0 & 0 \\ \mu_S(t) & 0 & \sigma_S(t) & 0 & 0 \\ e-f\chi(t) & 0 & 0 & \sigma_\chi\sqrt{\chi(t)} & 0 \\ 0 & 0 & 0 & 0 & \eta \end{bmatrix} \begin{bmatrix} dt \\ dW_r(t) \\ dW_S(t) \\ dW_\chi(t) \\ dW_\gamma(t) \end{bmatrix}$$

$\alpha(t,r(t))$ and $\beta(t,r(t))$ are the drift term and the diffusion term of interest rate process $r(t)$ respectively. $B(t)r(t)$ is the drift term of short-term saving process $B(t)$. $\dfrac{dP(t,T,r(t))}{P(t,T,r(t))}$ is the process of zero coupon bond of no risk with maturity $T$.

We would like to derive the drift term $\tilde{\mu}(t,r(t))$ and the diffusion term $\tilde{\sigma}(t,r(t))$ for $P(t,T,r(t))$, in which technical details are provided in Appendix 2 of Supplementary materials. From Appendix 2 of Supplementary materials, $\tilde{\mu}(t,r(t))$ equals $r(t)+\tilde{\sigma}(t,r(t))\theta(t)$ and $\tilde{\sigma}(t,r(t))$ equals $\dfrac{P_r\beta(t,r(t))}{P(t,r(t))}$. Here, $\theta(t)$ is the process of market price of risk under probability measure $\mathbf{Q}'$ and $P_r$ is the first partial derivative of $P(t,T,r(t))$ with respect to $r(t)$, $\dfrac{\partial P}{\partial r}$.

$\mu_S(t)$ and $\sigma_S(t)$ are the drift term and the diffusion term respectively of stock price $S(t)$. $\chi(t)$ and $\gamma(t)$ are the processes of default density and convenience yield respectively, following the model settings in Longstaff et al. (2005).[9] $e-f\chi(t)$ and $\sigma_\chi\sqrt{\chi(t)}$ are the drift term and the diffusion term of $\chi(t)$ respectively. $B(t)r(t)$ is the diffusion term of $\gamma(t)$.

---

[8] One more benefit for matrix is that we could do some analyses on the coefficient matrix, e.g. eigenvalues and eigenvectors of the coefficient matrix.

[9] Different from Longstaff et al. (2005), $\eta$ in our model could be negative. The regularity condition introduced later require $\eta$ equalling $-\dfrac{\gamma(t)r(t)}{\rho_{r\gamma}\theta(t)}$. If we impose $\eta$ to be positive, then $\rho_{r\gamma}$ has to be negative (positive) when $\gamma(t)$ is positive (negative) given $r(t)$ and $\theta(t)$ are positive in our numerical examples later in Section 4. The switch of sign of $\rho_{r\gamma}$ could be a further research question.

Based on equation (2), we could rewrite the dynamics of each process as follows.

$$\begin{bmatrix} dr(t) \\ dB(t) \\ dP(t,T,r(t)) \\ dS(t) \\ d\chi(t) \\ d\gamma(t) \end{bmatrix} = \begin{bmatrix} \alpha(t,r(t)) & \beta(t,r(t)) & 0 & 0 & 0 \\ B(t)r(t) & 0 & 0 & 0 & 0 \\ P(t,T,r(t))\tilde{\mu}(t,r(t)) & P(t,T,r(t))\tilde{\sigma}(t,r(t)) & 0 & 0 & 0 \\ S(t)\mu_S(t) & S(t)\sigma_S(t)\rho_{rS} & S(t)\sigma_S(t)\sqrt{1-\rho_{rS}^2} & 0 & 0 \\ e-f\chi(t) & \sigma_\chi\rho_{r\chi}\sqrt{\chi(t)} & \sigma_\chi\rho'_{S\chi}\sqrt{\chi(t)} & \sigma_\chi\rho'_{\chi\chi}\sqrt{\chi(t)} & 0 \\ 0 & \eta\rho_{r\gamma} & \eta\rho''_{S\gamma} & \eta\rho''_{\chi\gamma} & \eta\rho''_{\gamma\gamma} \end{bmatrix} \begin{bmatrix} dt \\ dW_r(t) \\ dW_1(t) \\ dW_2(t) \\ dW_3(t) \end{bmatrix} \quad (3)$$

### 2.2.2. General form of deflator

We are now able to derive the general form of deflator. First, let

$$dD(t) = \Omega(t)dt + \Phi(t)dW_r(t) + \Psi(t)dW_1(t) + \Gamma(t)dW_2(t) + \mathrm{I}(t)dW_3(t). \quad (4)$$

We would like to have $D(t)B(t)$, $D(t)P(t,T,r(t))$, $D(t)S(t)$, $D(t)\chi(t)$, and $D(t)\gamma(t)$ be $\mathbf{P}$-martingales.

By Itô product rule, we have $d\left[D(t)X(t)\right] = X(t)dD(t) + D(t)dX(t) + dX(t)dD(t)$ for a stochastic process $X(t)$. We derive $\Omega(t)$, $\Phi(t)$, $\Psi(t)$, $\Gamma(t)$, and $\mathrm{I}(t)$ step by step in Appendix 3 of Supplementary materials.

Let
$$\begin{cases} K_\Psi(t) = \dfrac{\Psi(t)}{D(t)} = \dfrac{r(t)+\theta(t)\sigma_S(t)\rho_{rS}-\mu_S(t)}{\sigma_S(t)\sqrt{1-\rho_{rS}^2}} \\[2ex] K_\Gamma(t) = \dfrac{\Gamma(t)}{D(t)} = \dfrac{\theta(t)\rho_{r\chi}}{\rho'_{\chi\chi}} + \dfrac{r(t)\chi(t)-e+f\chi(t)}{\sigma_\chi\rho'_{\chi\chi}\sqrt{\chi(t)}} + \dfrac{\rho'_{S\chi}\left[\mu_S(t)-r(t)-\theta(t)\sigma_S(t)\rho_{rS}\right]}{\rho'_{\chi\chi}\sigma_S(t)\sqrt{1-\rho_{rS}^2}} \\[2ex] K_\mathrm{I}(t) = \dfrac{\mathrm{I}(t)}{D(t)} = \dfrac{\rho_{r\gamma}\theta(t)}{\rho''_{\gamma\gamma}} + \dfrac{r(t)\gamma(t)}{\eta\rho''_{\gamma\gamma}} - \dfrac{\rho''_{\chi\gamma}\rho_{r\chi}\theta(t)}{\rho''_{\gamma\gamma}\rho'_{\chi\chi}} + \dfrac{\rho''_{\chi\gamma}\left[e-r(t)\chi(t)-f\chi(t)\right]}{\rho''_{\gamma\gamma}\rho'_{\chi\chi}\sigma_\chi\sqrt{\chi(t)}} \\[2ex] \qquad + \dfrac{\left(\rho''_{S\gamma}\rho'_{\chi\chi}-\rho''_{\chi\gamma}\rho'_{S\chi}\right)\left[\mu_S(t)-r(t)-\rho_{rS}\theta(t)\sigma_S(t)\right]}{\rho''_{\gamma\gamma}\rho'_{\chi\chi}\sigma_S(t)\sqrt{1-\rho_{rS}^2}} \end{cases}$$
, then

$$dD(t) = -D(t)r(t)dt - D(t)\theta(t)dW_r(t) + D(t)K_\Psi(t)dW_1(t) + D(t)K_\Gamma(t)dW_2(t) + D(t)K_\mathrm{I}(t)dW_3(t).$$

We have the general form of deflator $D(t)$ as follows.

$$\begin{aligned} D(t) = D(0)\,\mathbf{exp}&\left\{-\int_0^t r(s)ds - \int_0^t \frac{1}{2}\left[\theta^2(s)+K_\Psi^2(s)+K_\Gamma^2(s)+K_\mathrm{I}^2(s)\right]ds\right\} \\ \times\,\mathbf{exp}&\left[-\int_0^t \theta(s)dW_r(s) + \int_0^t K_\Psi(s)dW_1(s) + \int_0^t K_\Gamma(s)dW_2(s) + \int_0^t K_\mathrm{I}(s)dW_3(s)\right] \end{aligned}$$

In addition, we require $\begin{cases} \mu_S(t) = r(t) + \theta(t)\sigma_S(t)\rho_{rS} \\ e = r(t)\chi(t) + f\chi(t) + \sigma_\chi \rho_{r\chi}\theta(t)\sqrt{\chi(t)} \\ \eta = -\dfrac{\gamma(t)r(t)}{\rho_{r\gamma}\theta(t)} \end{cases}$ as regularity conditions, such that $\delta(t)S(t)$, $\delta(t)\chi(t)$, and $\delta(t)\gamma(t)$ are $\mathbf{Q}$-martingales (technical details are provided in Appendix 4 of Supplementary materials). As a result, $K_\Psi(t) = 0$, $K_\Gamma(t) = 0$, and $K_\mathrm{I}(t) = 0$. Then, $dD(t) = -D(t)r(t)dt - D(t)\theta(t)dW_r(t)$.[10]

We could rewrite the general form of deflator $D(t)$ as follows.[11]

$$dD(t) = -D(t)r(t)dt - D(t)\theta(t)dW_r(t) \quad (5)$$

$$D(t) = D(0)\mathbf{exp}\left[-\int_0^t r(s)ds - \frac{1}{2}\int_0^t \theta^2(s)ds - \int_0^t \theta(s)dW_r(s)\right] \quad (6)$$

---

[10] We will show later that one more regularity condition is required for the diffusion term of stock price in our model, but the deflator remains the same as shown in equation (5).

[11] The disappearance of $K_\Psi(t)$ term also tells us that stocks are financial derivatives. Given a rate of time value (growth) $r(t)$ and a rate of market risk $\theta(t)$, the proper expected return of a stock in our model is equal to $r(t) + \theta(t)\sigma_S(t)\rho_{rS}$. Also, the regularity conditions tell us more about the relations among interest rates, market prices of risk, stock prices, default intensities, and convenience yields in our model based on how we derive the deflator. For example, if we choose $\mathbf{exp}\left[\chi(t)\right]$ instead of $\chi(t)$ to derive the deflator, the regularity conditions will be different.

## 3. IMPLEMENTATION OF TIME DISCRETIZATION

From equations (3) and (5) with regularity conditions required in Section 2.2.2,

$$\begin{bmatrix} dr(t) \\ dB(t) \\ dP(t,T,r(t)) \\ dS(t) \\ d\chi(t) \\ d\gamma(t) \\ dD(t) \end{bmatrix} = \begin{bmatrix} \alpha(t,r(t)) & \beta(t,r(t)) & 0 & 0 & 0 \\ B(t)r(t) & 0 & 0 & 0 & 0 \\ P(t,T,r(t))\left[r(t)+\tilde{\sigma}(t,r(t))\theta(t)\right] & P(t,T,r(t))\tilde{\sigma}(t,r(t)) & 0 & 0 & 0 \\ S(t)\left[r(t)+\theta(t)\sigma_S(t)\rho_{rS}\right] & S(t)\sigma_S(t)\rho_{rS} & S(t)\sigma_S(t)\sqrt{1-\rho_{rS}^2} & 0 & 0 \\ r(t)\chi(t)+\sigma_\chi\rho_{r\chi}\theta(t)\sqrt{\chi(t)} & \sigma_\chi\rho_{r\chi}\sqrt{\chi(t)} & \sigma_\chi\rho'_{S\chi}\sqrt{\chi(t)} & \sigma_\chi\rho'_{\chi\chi}\sqrt{\chi(t)} & 0 \\ 0 & -\frac{\gamma(t)r(t)}{\theta(t)} & -\frac{\gamma(t)r(t)}{\rho_{r\gamma}\theta(t)}\rho''_{S\gamma} & -\frac{\gamma(t)r(t)}{\rho_{r\gamma}\theta(t)}\rho''_{\chi\gamma} & -\frac{\gamma(t)r(t)}{\rho_{r\gamma}\theta(t)}\rho''_{\gamma\gamma} \\ -D(t)r(t) & -D(t)\theta(t) & 0 & 0 & 0 \end{bmatrix} \begin{bmatrix} dt \\ dW_r(t) \\ dW_1(t) \\ dW_2(t) \\ dW_3(t) \end{bmatrix}.$$

To implement the deflator approach, we need to discretize time steps for each process. We discuss the time discretization here. We adopt the Euler method, the Milstein method, and the simplified Second Milstein method for time discretization in our model.[12]

Denote a stochastic process $X(t)$ with its dynamics $dX(t) = b_X(t, X(t))dt + \sigma_X(t, X(t))dW_X(t)$ where $W_X(t)$ is the Brownian part of $X(t)$. We partition the time $[0,T]$ into $N$ segments with each length equalling $(T-0)/N$, then we have a time discretization $\Pi_N = \Pi_N([0,T])$ with $0 = t_0 < t_1 < \cdots < t_N = T$.

### 3.1. EULER METHOD

In Euler method, we approximate $X(t)$ by $Y_t$ discretely, in which $Y_{i+1} = Y_i + b_X(t_i, Y_i)(t_{i+1} - t_i) + \sigma_X(t_i, Y_i)(W_{i+1} - W_i)$, $i = 0,1,\ldots,N-1$, $W_i$ is the value of a Brownian motion at time period $i$, and $Y_0$ is equal to $X(0)$. Denote $\Delta t_i = t_{i+1} - t_i = (T-0)/N$ and $\Delta W_{k,i} = W_{k,t_{i+1}} - W_{k,t_i}$, $k = r,1,2,3$.

We present the approximations of $r(t)$, $B(t)$, $P(t,T,r(t))$, $S(t)$, $\chi(t)$, $\gamma(t)$, $D(t)$ by the Euler method in matrix as follows.

$$\begin{bmatrix} r_{i+1} \\ B_{i+1} \\ P_{i+1} \\ S_{i+1} \\ \chi_{i+1} \\ \gamma_{i+1} \\ D_{i+1} \end{bmatrix} = \begin{bmatrix} r_i \\ B_i \\ P_i \\ S_i \\ \chi_i \\ \gamma_i \\ D_i \end{bmatrix} + \begin{bmatrix} \alpha_{t_i,r_i} & \beta_{t_i,r_i} & 0 & 0 & 0 \\ B_i r_i & 0 & 0 & 0 & 0 \\ P_i\left(r_i + \tilde{\sigma}_{t_i,r_i}\theta_i\right) & P_i\tilde{\sigma}_{t_i,r_i} & 0 & 0 & 0 \\ S_i\left(r_i + \theta_i\sigma_{S,t_i}\rho_{rS}\right) & S_i\sigma_{S,t_i}\rho_{rS} & S_i\sigma_{S,t_i}\sqrt{1-\rho_{rS}^2} & 0 & 0 \\ r_i\chi_i + \sigma_\chi\rho_{r\chi}\theta_i\sqrt{\chi_i} & \sigma_\chi\rho_{r\chi}\sqrt{\chi_i} & \sigma_\chi\rho'_{S\chi}\sqrt{\chi_i} & \sigma_\chi\rho'_{\chi\chi}\sqrt{\chi_i} & 0 \\ 0 & -\frac{\gamma_i r_i}{\theta_i} & -\frac{\gamma_i r_i}{\rho_{r\gamma}\theta_i}\rho''_{S\gamma} & -\frac{\gamma_i r_i}{\rho_{r\gamma}\theta_i}\rho''_{\chi\gamma} & -\frac{\gamma_i r_i}{\rho_{r\gamma}\theta_i}\rho''_{\gamma\gamma} \\ -D_i r_i & -D_i\theta_i & 0 & 0 & 0 \end{bmatrix} \begin{bmatrix} \Delta t_i \\ \Delta W_{r,i} \\ \Delta W_{1,i} \\ \Delta W_{2,i} \\ \Delta W_{3,i} \end{bmatrix} \quad (7)$$

[12] For references of time discretization, see, for example, Kloeden and Platen (1992), Iacus (2009), and Glasserman (2013).

### 3.2. MILSTEIN METHOD

Denote $\sigma_{X'} = \dfrac{\partial \sigma_X(t,x)}{\partial x}$, we approximate $X(t)$ by $Y_t$ discretely as

$$Y_{i+1} = Y_i + b_X(t_i, Y_i)(t_{i+1} - t_i) + \sigma_X(t_i, Y_i)(W_{i+1} - W_i) + \frac{1}{2}\sigma_X(t_i, Y_i)\sigma_{X'}(t_i, Y_i)\left[(W_{i+1} - W_i)^2 - (t_{i+1} - t_i)\right].$$

We present the approximations of $r(t)$, $B(t)$, $P(t,T,r(t))$, $S(t)$, $\chi(t)$, $\gamma(t)$, $D(t)$ by the Milstein method in matrix as follows.

$$\begin{bmatrix} r_{i+1} \\ B_{i+1} \\ P_{i+1} \\ S_{i+1} \\ \chi_{i+1} \\ \gamma_{i+1} \\ D_{i+1} \end{bmatrix} = \begin{bmatrix} r_i \\ B_i \\ P_i \\ S_i \\ \chi_i \\ \gamma_i \\ D_i \end{bmatrix} + \begin{bmatrix} \alpha_{t_i,r_i} & \beta_{t_i,r_i} & 0 & 0 & 0 \\ B_i r_i & 0 & 0 & 0 & 0 \\ P_i\left(r_i + \tilde{\sigma}_{t_i,r_i}\theta_i\right) & P_i\tilde{\sigma}_{t_i,r_i} & 0 & 0 & 0 \\ S_i\left(r_i + \theta_i\sigma_{S,t_i}\rho_{rS}\right) & S_i\sigma_{S,t_i}\rho_{rS} & S_i\sigma_{S,t_i}\sqrt{1-\rho_{rS}^2} & 0 & 0 \\ r_i\chi_i + \sigma_\chi\rho_{r\chi}\theta_i\sqrt{\chi_i} & \sigma_\chi\rho_{r\chi}\sqrt{\chi_i} & \sigma_\chi\rho'_{S\chi}\sqrt{\chi_i} & \sigma_\chi\rho'_{\chi\chi}\sqrt{\chi_i} & 0 \\ 0 & -\dfrac{\gamma_i r_i}{\theta_i} & -\dfrac{\gamma_i r_i}{\rho_{r\gamma}\theta_i}\rho''_{S\gamma} & -\dfrac{\gamma_i r_i}{\rho_{r\gamma}\theta_i}\rho''_{\chi\gamma} & -\dfrac{\gamma_i r_i}{\rho_{r\gamma}\theta_i}\rho''_{\gamma\gamma} \\ -D_i r_i & -D_i\theta_i & 0 & 0 & 0 \end{bmatrix} \begin{bmatrix} \Delta t_i \\ \Delta W_{r,i} \\ \Delta W_{1,i} \\ \Delta W_{2,i} \\ \Delta W_{3,i} \end{bmatrix}$$

$$+\frac{1}{4}\begin{bmatrix} 2\beta_{t_i,r_i}\beta_{r,t_i,r_i} & 0 & 0 & 0 \\ 0 & 0 & 0 & 0 \\ 2\tilde{\sigma}_{t_i,r_i}^2 & 0 & 0 & 0 \\ 2\sigma_{S,t_i}^2\rho_{rS}^2 & 2\sigma_{S,t_i}^2\left(1-\rho_{rS}^2\right) & 0 & 0 \\ \sigma_\chi^2\rho_{r\chi}^2 & \sigma_\chi^2\rho_{S\chi}'^2 & \sigma_\chi^2\rho_{\chi\chi}'^2 & 0 \\ 2\gamma_i\left(\dfrac{r_i}{\theta_i}\right)^2 & 2\gamma_i\left(\dfrac{\rho''_{S\gamma}r_i}{\rho_{r\gamma}\theta_i}\right)^2 & 2\gamma_i\left(\dfrac{\rho''_{\chi\gamma}r_i}{\rho_{r\gamma}\theta_i}\right)^2 & 2\gamma_i\left(\dfrac{\rho''_{\gamma\gamma}r_i}{\rho_{r\gamma}\theta_i}\right)^2 \\ 2D_i\theta_i^2 & 0 & 0 & 0 \end{bmatrix} \begin{bmatrix} \left(\Delta W_{r,i}\right)^2 - \Delta t_i \\ \left(\Delta W_{1,i}\right)^2 - \Delta t_i \\ \left(\Delta W_{2,i}\right)^2 - \Delta t_i \\ \left(\Delta W_{3,i}\right)^2 - \Delta t_i \end{bmatrix} \quad (8)$$

### 3.3. SIMPLIFIED SECOND MILSTEIN METHOD

We advance to multi-dimensional case in this sub-section. Let $X_t$ be multi-dimensional stochastic processes with the dynamics $dX_t = a(t, X_t)dt + b(t, X_t)dW_t$, where $X_t$ is a $d\times 1$ vector, $a(t, X_t)$ is a $d\times 1$ vector, $b(t, X_t)$ is a $d\times m$ matrix, and $W_t$ is a $m\times 1$ vector. $d$ is the number of different stochastic processes in $X_t$, and $m$ is the number of independent Brownian motions involved in $X_t$.

For a continuously twice differentiable function $f(t, x_{d\times 1})$, we could write $df(t, X_t)$ by Itô formula for multi-dimensional case as follows.

$$df\left(t,X_t\right)=\left[\frac{\partial f\left(t,X_t\right)}{\partial t}+\sum_{i=1}^{d}\frac{\partial f\left(t,X_t\right)}{\partial x_i}a_i\left(t,X_t\right)+\frac{1}{2}\sum_{i,j=1}^{d}\frac{\partial^2 f\left(t,X_t\right)}{\partial x_i\partial x_j}\Sigma_{t,ij}\right]dt + \sum_{i=1}^{d}\sum_{k=1}^{m}b_{ik}\left(t,X_t\right)\frac{\partial f\left(t,X_t\right)}{\partial x_i}dW_{t,k},\ \Sigma_t=b\left(t,X_t\right)b^T\left(t,X_t\right) \tag{9}$$

In equation 11, $a_i\left(t,X_t\right)$ is the element of $i^{th}$ row of $a\left(t,X_t\right)$, $b_{ik}\left(t,X_t\right)$ is the element of $b\left(t,X_t\right)$ at its $i^{th}$ row and $k^{th}$ column, $b^T\left(t,X_t\right)$ is the transpose of $b\left(t,X_t\right)$, $\Sigma_{t,ij}$ is the element of $\Sigma_t$ at its $i^{th}$ row and $j^{th}$ column, and $W_{t,k}$ is the element of $k^{th}$ row of $W_t$. Next, we introduce operators $L^0$ and $L^k$ and rewrite $df\left(t,X_t\right)$ for multi-dimensional case.

$$L^0=\frac{\partial}{\partial t}+\sum_{i=1}^{d}a_i\left(t,X_t\right)\frac{\partial}{\partial x_i}+\frac{1}{2}\sum_{i,j=1}^{d}\Sigma_{t,ij}\frac{\partial^2}{\partial x_i\partial x_j} \tag{10}$$

$$L^k=\sum_{i=1}^{d}b_{ik}\left(t,X_t\right)\frac{\partial}{\partial x_i},\ \forall k=1,\ldots,m \tag{11}$$

$$df\left(t,X_t\right)=L^0 f\left(t,X_t\right)dt+\sum_{k=1}^{m}L^k f\left(t,X_t\right)dW_{t,k} \tag{12}$$

We approximate $X_t$ by $Y_t$ discretely by simplified Second Milstein method, where $Y_t$ is a $d\times 1$ vector. For each $i=1,\ldots,d$,

$$Y_{n+1,i}=Y_{n,i}+a_i\left(n,Y_n\right)\Delta t+\sum_{k=1}^{m}b_{ik}\left(n,Y_n\right)\Delta W_{n,k}+\frac{1}{2}L^0 a_i\left(n,Y_n\right)\left(\Delta t\right)^2 +\frac{1}{2}\sum_{k=1}^{m}\left[L^k a_i\left(n,Y_n\right)+L^0 b_{ik}\left(n,Y_n\right)\right]\Delta W_{n,k}\Delta t+\frac{1}{2}\sum_{k=1}^{m}\sum_{j=1}^{m}L^j b_{ik}\left(n,Y_n\right)\left(\Delta W_{n,j}\Delta W_{n,k}-V_{jk}\right) \tag{13}$$

$Y_{n+1,i}$ is the element of $i^{th}$ row of $Y_t$ in the time step $n+1$. $V_{jk}$ is an independent random variable with probabilities $\Pr\left(V_{jk}=\Delta t\right)=\Pr\left(V_{jk}=-\Delta t\right)=\frac{1}{2}$ for $j<k$, $V_{kj}=-V_{jk}$ for $j>k$, and $V_{jk}=\Delta t$ for $j=k$. The following are the $X_t$, $a\left(t,X_t\right)$, $W_t$, and $b\left(t,X_t\right)$ in our model.

$$X_t=\begin{bmatrix} r(t)\\ \theta(t)\\ B(t)\\ P\left(t,T,r(t)\right)\\ S(t)\\ \chi(t)\\ \gamma(t)\\ D(t)\end{bmatrix},\ a\left(t,X_t\right)=\begin{bmatrix} \alpha\left(t,r(t)\right)\\ a_\theta-b_\theta\theta(t)\\ B(t)r(t)\\ P\left(t,T,r(t)\right)\left[r(t)+\tilde{\sigma}\left(t,r(t)\right)\theta(t)\right]\\ S(t)\left[r(t)+\theta(t)\sigma_S(t)\rho_{rS}\right]\\ r(t)\chi(t)+\sigma_\chi\rho_{r\chi}\theta(t)\sqrt{\chi(t)}\\ 0\\ -r(t)D(t)\end{bmatrix},W_t=\begin{bmatrix} W_r(t)\\ W_1(t)\\ W_2(t)\\ W_3(t)\\ W_\theta(t)\end{bmatrix},$$

$$b(t,X_t)=\begin{bmatrix} \beta(t,r(t)) & 0 & 0 & 0 & 0 \\ 0 & 0 & 0 & 0 & \sigma_\theta\sqrt{\theta(t)} \\ 0 & 0 & 0 & 0 & 0 \\ P(t,T,r(t))\tilde{\sigma}(t,r(t)) & 0 & 0 & 0 & 0 \\ S(t)\sigma_S(t)\rho_{rS} & S(t)\sigma_S(t)\sqrt{1-\rho_{rS}^2} & 0 & 0 & 0 \\ \sigma_\chi\rho_{r\chi}\sqrt{\chi(t)} & \sigma_\chi\rho'_{S\chi}\sqrt{\chi(t)} & \sigma_\chi\rho'_{\chi\chi}\sqrt{\chi(t)} & 0 & 0 \\ -\frac{\gamma(t)r(t)}{\theta(t)} & -\frac{\rho''_{S\gamma}\gamma(t)r(t)}{\rho_{r\gamma}\theta(t)} & -\frac{\rho''_{\chi\gamma}\gamma(t)r(t)}{\rho_{r\gamma}\theta(t)} & -\frac{\rho''_{\gamma\gamma}\gamma(t)r(t)}{\rho_{r\gamma}\theta(t)} & 0 \\ -\theta(t)D(t) & 0 & 0 & 0 & 0 \end{bmatrix}$$

## 4. NUMERICAL RESULTS

We implement the deflator approach with three methods for time discretization and adopt CIR interest rate model for short-term saving.[13] In addition, we also incorporate parallel computing with a variance technique, antithetic sampling in our algorithm.[14] In CIR interest rate model, $dr(t)=\left[a_r-b_r r(t)\right]dt+\sigma_r\sqrt{r(t)}d\tilde{W}_r(t);\ a_r,b_r,\sigma_r>0$ . The process of interest rate is defined under probability measure $\mathbf{Q}'$. To convert the process into physical measure $\mathbf{P}$, we have to consider the process of market price of risk $\theta(t)$. From Section 2.2.1, we let $d\tilde{W}_r(t)=\theta(t)dt+dW_r(t)$. Thus, we could rewrite $dr(t)$ in $\mathbf{P}$-measure as $dr(t)=\left[a_r-b_r r(t)+\theta(t)\sigma_r\sqrt{r(t)}\right]dt+\sigma_r\sqrt{r(t)}dW_r(t)$.

Let $\theta(t)$ also be CIR process here and $W_\theta$ is an independent Brownian motion of $W_i,\ i=r,1,2,3$.[15] The dynamics of $\theta(t)$ is

$$d\theta(t)=\left[a_\theta-b_\theta\theta(t)\right]dt+\sigma_\theta\sqrt{\theta(t)}dW_\theta(t);a_\theta,b_\theta,\sigma_\theta>0.$$

In CIR interest rate model, the price of zero coupon bond of no risk with maturity $T$, $P(t,T,r(t))$, is equal to $e^{-r(t)C_P(t,T)-A_P(t,T)}$ where

$$C_P(t,T)=\frac{\sinh(\gamma_{CIR}(T-t))}{\gamma_{CIR}\cosh(\gamma_{CIR}(T-t))+\frac{1}{2}b_r\sinh(\gamma_{CIR}(T-t))},\ \gamma_{CIR}=\frac{1}{2}\sqrt{b_r^2+2\sigma_r^2},$$

---

[13] Here we choose CIR interest rate model because the model has a closed-form formula for prices of zero-coupon bonds of no risk.

[14] The R codes are available from the authors by inquiry. For examples of computing time, user CPU time is 3.363 s, system CPU time is 0.286 s, and elapsed time is 15.747 s in 2500 simulations; user CPU time is 2828.413 s, system CPU time is 161.860 s, and elapsed time is 5127.524 s in 1000000 simulations.

[15] Here we choose $\theta(t)$ to be CIR process, so that $\theta(t)$ would be positive in any time period $t$.

and $A_P(t,T) = -\frac{2a_r}{\sigma_r^2}\ln\left[\frac{\gamma_{CIR}e^{\frac{1}{2}b_r(T-t)}}{\gamma_{CIR}\cosh\left(\gamma_{CIR}(T-t)\right)+\frac{1}{2}b_r\sinh\left(\gamma_{CIR}(T-t)\right)}\right]$.

Note that $\sinh u = \frac{e^u - e^{-u}}{2}$, $\cosh u = \frac{e^u + e^{-u}}{2}$, and $P(0,T,r(0)) = e^{-r(0)C_P(0,T)-A_P(0,T)}$.[16]

To calculate the option price for stock under CIR interest rate process analytically, we use the formula proposed by Kim (2002), which we leave the technical detail in Appendix 5 of Supplementary materials.[17] In addition, notice that the drift term of stock prices in Kim (2002) is a constant, which is different from our model in which $\mu_S(t) = r(t) + \theta(t)\sigma_S(t)\rho_{rS}$. Thus, our numerical results could be different from the value suggested by the formula in Kim (2002).

We present the approximations of $r(t)$, $\theta(t)$, $P(t,T,r(t))$ in matrix by the Euler method and the Milstein method as follows.

First, we rewrite the dynamics of $r(t)$, $\theta(t)$, $P(t,T,r(t))$ in matrix.

$$\begin{bmatrix} dr(t) \\ d\theta(t) \\ dP(t,T,r(t)) \end{bmatrix} = \begin{bmatrix} a_r - b_r r(t) + \theta(t)\sigma_r\sqrt{r(t)} & \sigma_r\sqrt{r(t)} & 0 \\ a_\theta - b_\theta\theta(t) & 0 & \sigma_\theta\sqrt{\theta(t)} \\ P(t,T,r(t))r(t) + P_r\sigma_r\sqrt{r(t)}\theta(t) & P_r\sigma_r\sqrt{r(t)} & 0 \end{bmatrix} \begin{bmatrix} dt \\ dW_r(t) \\ dW_\theta(t) \end{bmatrix} \quad (14)^{18}$$

Then, $\begin{bmatrix} r_{i+1} \\ \theta_{i+1} \\ P_{i+1} \end{bmatrix} = \begin{bmatrix} r_i \\ \theta_i \\ P_i \end{bmatrix} + \begin{bmatrix} a_r - b_r r_i + \theta_i\sigma_r\sqrt{r_i} & \sigma_r\sqrt{r_i} & 0 \\ a_\theta - b_\theta\theta_i & 0 & \sigma_\theta\sqrt{\theta_i} \\ P_i r_i + P_{r,t_i}\sigma_r\sqrt{r_i}\theta_i & P_{r,t_i}\sigma_r\sqrt{r_i} & 0 \end{bmatrix} \begin{bmatrix} \Delta t_i \\ \Delta W_{r,i} \\ \Delta W_{\theta,i} \end{bmatrix}$ by the Euler method;

and $\begin{bmatrix} r_{i+1} \\ \theta_{i+1} \\ P_{i+1} \end{bmatrix} = \begin{bmatrix} r_i \\ \theta_i \\ P_i \end{bmatrix} + \begin{bmatrix} a_r - b_r r_i + \theta_i\sigma_r\sqrt{r_i} & \sigma_r\sqrt{r_i} & 0 \\ a_\theta - b_\theta\theta_i & 0 & \sigma_\theta\sqrt{\theta_i} \\ P_i r_i + P_{r,t_i}\sigma_r\sqrt{r_i}\theta_i & P_{r,t_i}\sigma_r\sqrt{r_i} & 0 \end{bmatrix} \begin{bmatrix} \Delta t_i \\ \Delta W_{r,i} \\ \Delta W_{\theta,i} \end{bmatrix} + \frac{1}{4}\begin{bmatrix} \sigma_r^2 & 0 \\ 0 & \sigma_\theta^2 \\ 0 & 0 \end{bmatrix}\begin{bmatrix} (\Delta W_{r,i})^2 - \Delta t_i \\ (\Delta W_{\theta,i})^2 - \Delta t_i \end{bmatrix}$

by the Milstein method.

The details of implementation of simplified Second Milstein method is provided in the Appendix 6 of Supplementary materials.

[16] See, for example, Shreve (2004) Chapter 6.

[17] For references of option pricing under stochastic interest rates, see, for example, Shreve (2004) Chapter 9, and Brigo and Mercurio (2006) Chapter 3 and Appendix B.

[18] For process of $P(t,T,r(t))$, plug $\tilde{\sigma}(t,r(t)) = \frac{P_r\beta(t,r(t))}{P(t,r(t))}$ and $\beta(t,r(t)) = \sigma_r\sqrt{r(t)}$ in equation (3). Note that $P_r = -C_P(t,T)e^{-r(t)C_P(t,T)-A_P(t,T)}$ in CIR interest rate model.

### 4.1. AN NUMERICAL EXAMPLE WITH CIR MODEL

The following is the settings describing the dynamics of each process in our example.[19]

$$\begin{cases} d\theta(t) = \left[0.05 - 0.01\theta(t)\right]dt + 0.01\sqrt{\theta(t)}dW_\theta(t),\ \theta(0) = 0.3 \\ dr(t) = \left[0.02 - 0.04r(t)\right]dt + 0.01\sqrt{r(t)}d\tilde{W}_r(t),\ r(0) = 0.02 \\ dS(t) = S(t)\left[r(t) + 0.2\rho_{rS}\theta(t)\right]dt + 0.2S(t)dW_S(t),\ S(0) = 1 \\ d\chi(t) = \left[r(t)\chi(t) + 0.01\rho_{r\chi}\theta(t)\sqrt{\chi(t)}\right]dt + 0.01\sqrt{\chi(t)}dW_\chi(t),\ \chi(t) = 0.05 \\ d\gamma(t) = -\dfrac{\gamma(t)r(t)}{\rho_{r\gamma}\theta(t)}dW_\gamma(t),\ \gamma(0) = 0.01 \end{cases}$$

$$T = 1,\ \Delta t = 0.01,\ \rho_{rS} = 0.6,\ \rho_{r\chi} = 0.7,\ \rho_{r\gamma} = 0.5,\ \rho_{S\chi} = 0.1,\ \rho_{S\gamma} = 0.3,\ \rho_{\chi\gamma} = 0.1$$

Let $D(0)$ equal 1 in equation (10). The deflator approach tells us that for a nonnegative random variable $X(t)$, we would have $E^{\mathbf{Q}}\left[\delta(t)X(t)\right] = E\left[D(t)X(t)\right]$.

#### 4.1.1. Zero-coupon bond of no risk with maturity $T$

The price of a zero-coupon bond of no risk with maturity $T$ at time period $T$ is equal to 1.

$$D(0)P(0,T,r(0)) = P(0,T,r(0)) = E^{\mathbf{Q}}\left[\delta(T)P(T,T,r(T))\right] = E\left[D(T)P(T,T,r(T))\right] = E\left[D(T)\right]$$

(15)

Tables 1 shows the numerical results. Figures 2, 3, 4, 5, 6, 7, 8, and 9 show the convergence of approximations to expected values, i.e. $P(0,T,r(0))$, and the differences between approximations and expected values. In general, we could see that the simplified Second Milstein method provides better approximations and converges faster than the Euler method and the Milstein method do. This could be explained by convergence order in which the simplified Second Milstein method has larger weak order of convergence.[20]

#### 4.1.2. Corporate coupon bond

Longstaff et al. (2005) assumed the independence among interest rate, default intensity, and convenience yield. Thus, we let $\rho_{r\chi} = 0$, $\rho_{r\gamma} = 0$, $\rho_{S\chi} = 0$, $\rho_{S\gamma} = 0$, and $\rho_{\chi\gamma} = 0$.[21] To accommodate the three risk factors (interest rate, default intensity, and convenience yield) with deflator, we let $dB(t)=B(t)\left[r(t)+\chi(t)+\gamma(t)\right]dt$.[22] In addition, notice again that the formula provided in Longstaff et al. (2005) is not directly applicable after we require regularity conditions in our model, which we leave technical detail of the formula in Longstaff et al. (2005) in Appendix 7 of Supplementary materials.

[19] Here we provide a numerical example for the model, in which the chosen values for model settings could be different. In our example, there are strong positive correlations between interest rates and other factors (i.e. stock prices, default densities, and convenience yields), but weak positive correlations between each two of stock prices, default densities, and convenience yields. Note that the Feller condition holds in our numerical examples, e.g. $2 \times 0.05 > 0.01^2$.

[20] The Euler method and the Milstein method have weak order of convergence 1, and the simplified Second Milstein method has weak order of convergence 2, see, for example, Glasserman (2013) Chapter 6.

[21] We let $dW_i dW_j = 0$ here, i.e. pairwise independence.

To implement the deflator, we look at the original definition of $CB(c,\omega,T)$.

$$\begin{aligned}CB(c,\omega,T)=&E\left\{c\int_0^T \exp\left[-\int_0^t \left(r(s)+\chi(s)+\gamma(s)\right)ds\right]dt\right\}+E\left\{\exp\left[-\int_0^T \left(r(s)+\chi(s)+\gamma(s)\right)ds\right]\right\}\\&+E\left\{(1-\omega)\int_0^T \chi_t \exp\left[-\int_0^t \left(r(s)+\chi(s)+\gamma(s)\right)ds\right]dt\right\}\end{aligned} \quad (16)$$

For the time period $t$ when a bond holder receives a coupon or a fraction of the par value of the bond (because of default), the payoff at that time period $t$ is equal to $c$ or $(1-\omega)$ multiply the par value of the bond respectively. Thus, we could implement the deflator as follows, the details of implementation of time discretization is provided in Appendix 8 of Supplementary materials.

$$D(0)CB(c,\omega,T)=E\left[D(T)\right]+cE\left[\int_0^T D(t)dt\right]+(1-\omega)E\left[\int_0^T \chi(t)D(t)dt\right] \quad (17)$$[23]

Tables 2 shows the numerical results. Figures 10, 11, 12, and 13 show the convergence of approximations and the differences between approximations and expected values.

### 4.2. ONE MORE REGULARITY CONDITION REQUIRED FOR THE DIFFUSION TERM IN STOCK PRICE

Up to Section 4.1, we successfully implement the deflator approach for zero-coupon bond of no risk with maturity $T$ and corporate coupon bond. However, we notice that one more regularity condition is required for the diffusion term in stock price. As illustrative examples, Figures 14 and 15 show the instability of the deflator approach corresponding to stock price in Second Milstein method with 10000 simulations after projecting longer than 15 years.

Recall that we would have $E^{\mathbf{Q}}\left[\delta(t)X(t)\right]=E\left[D(t)X(t)\right]$ for a nonnegative random variable $X(t)$. We calculate the price of Put option of $S(T)$ with strike $K$ equalling to 2, and expect the following equations to hold.

$$D(0)S(0)=S(0)=E^{\mathbf{Q}}\left[\delta(T)S(T)\right]=E\left[D(T)S(T)\right]=1 \quad (18)$$

$$D(0)Put(0,S(0),T,K)=Put(0,S(0),T,K)=E^{\mathbf{Q}}\left[\delta(T)(K-S(T))^+\right]=E\left[D(T)(K-S(T))^+\right] \quad (19)$$

---

[22] Recall that $D(t)=Discount\ factor\cdot\frac{d\mathbf{Q}}{d\mathbf{P}}$, $D(t)$ could not be the same given different discount factors with the same Radon-Nikodym derivative, i.e. different discount factors imply different risk-neutral worlds.

[23] We approximate $\int_0^T D(t)dt$ by $\sum_{t_i=0}^{T}\frac{\left(D_{t_i}+D_{t_{i+1}}\right)}{2}\Delta t_i$ ; similarly, we approximate $\int_0^T \chi(t)D(t)dt$ by $\sum_{t_i=0}^{T}\frac{\left(\chi_{t_i}D_{t_i}+\chi_{t_{i+1}}D_{t_{i+1}}\right)}{2}\Delta t_i$.

From equations (6) and (18), we derive one more regularity condition $\sigma_S\left(t\right)=\rho_{rS}\theta\left(t\right)\pm\theta\left(t\right)\sqrt{\rho_{rS}^2-1}$ , technical details are provided in Appendix 9 of Supplementary materials.

We could see that $\sigma_S\left(t\right)$ is a complex number if $\rho_{rS}\neq 1$ (so that $\left|\rho_{rS}\right|<1$). We choose $\rho_{rS}$ equalling 1 here as an example, then $\sigma_S\left(t\right)$ is equal to $\theta\left(t\right)$. Then, we could reduce the matrix form of equation (2) as follows.

$$\mathbf{W}_{ESG}=\begin{bmatrix} W_r \\ W_S \\ W_\chi \\ W_\gamma \end{bmatrix}=\begin{bmatrix} 1 & 0 & 0 \\ 1 & 0 & 0 \\ \rho_{r\chi} & \sqrt{1-\rho_{r\chi}^2} & 0 \\ \rho_{r\gamma} & \rho'_{\chi\gamma} & \rho'_{\gamma\gamma} \end{bmatrix}\begin{bmatrix} W_0 \\ W_2 \\ W_3 \end{bmatrix}, \tag{20}$$

$$\text{where } \rho'_{\chi\gamma}=\frac{\rho_{\chi\gamma}-\rho_{r\chi}\rho_{r\gamma}}{\sqrt{1-\rho_{r\chi}^2}},\ \rho'_{\gamma\gamma}=\sqrt{\frac{1-\rho_{r\chi}^2-\rho_{r\gamma}^2-\rho_{\chi\gamma}^2+2\rho_{r\chi}\rho_{r\gamma}\rho_{\chi\gamma}}{1-\rho_{r\chi}^2}}$$

From equations (3) and (5) with regularity conditions required in Section 2.2.2 and here (i.e. $\sigma_S\left(t\right)=\theta\left(t\right)$),

$$\begin{bmatrix} dr(t) \\ d\theta(t) \\ dB(t) \\ dP(t,T,r(t)) \\ dS(t) \\ d\chi(t) \\ d\gamma(t) \\ dD(t) \end{bmatrix}=\begin{bmatrix} \alpha(t,r(t)) & \beta(t,r(t)) & 0 & 0 & 0 \\ a_\theta-b_\theta\theta(t) & 0 & 0 & 0 & \sigma_\theta\sqrt{\theta(t)} \\ B(t)r(t) & 0 & 0 & 0 & 0 \\ P(t,T,r(t))\left[r(t)+\tilde{\sigma}(t,r(t))\theta(t)\right] & P(t,T,r(t))\tilde{\sigma}(t,r(t)) & 0 & 0 & 0 \\ S(t)\left[r(t)+\theta^2(t)\right] & S(t)\theta(t) & 0 & 0 & 0 \\ r(t)\chi(t)+\sigma_\chi\rho_{r\chi}\theta(t)\sqrt{\chi(t)} & \sigma_\chi\rho_{r\chi}\sqrt{\chi(t)} & \sigma_\chi\sqrt{\left(1-\rho_{r\chi}^2\right)\chi(t)} & 0 & 0 \\ 0 & -\frac{\gamma(t)r(t)}{\theta(t)} & -\frac{\gamma(t)r(t)}{\rho_{r\gamma}\theta(t)}\rho'_{\chi\gamma} & -\frac{\gamma(t)r(t)}{\rho_{r\gamma}\theta(t)}\rho'_{\gamma\gamma} & 0 \\ -r(t)D(t) & -D(t)\theta(t) & 0 & 0 & 0 \end{bmatrix}\begin{bmatrix} dt \\ dW_r(t) \\ dW_2(t) \\ dW_3(t) \\ dW_\theta(t) \end{bmatrix}.$$

Figures 16, 17, 18, and 19 show the numerical results of Second Milstein method with 10000 simulations after projecting 100 years.[24]

### 4.3. Discussions

Given the variance of a random variable $X$, $Var\left(X\right)$, the variance of $\frac{1}{n}X$ is equal to $\frac{1}{n^2}Var\left(X\right)$. Suppose the risk factors and parameters involved are constant at time period $t$, $D\left(t\right)$ is lognormal distributed. With the sample size being equal to $n$, the mean of

[24] Note that the formulas in Longstaff et al. (2005) and Kim (2002) are not applicable for longer periods, i.e. $C_{CB}\left(t\right)=\exp\left(\frac{\eta^2t^3}{6}\right)$ in Longstaff et al. (2005) and the square root term $\sqrt{r_0-\theta_{Kim}\left(1-e^{-\kappa_{Kim}T}\right)}$ of $C_{11}$ are not computable when $t$ is larger.

$D(t)$, $E\left[D(t)\right]$, equals $\frac{1}{n}\sum_{n\ trials} D(t)$; and its variance $Var\left(E\left[D(t)\right]\right)$ is equal to $\frac{1}{n}Var\left(D(t)\right)$. We could calculate its 95% confidence interval as follows.[25]

$$CI_{95\%} = E\left[D(t)\right] + \frac{Var\left(E\left[D(t)\right]\right)}{2} \pm t_{d.f.=n-1}\sqrt{\frac{Var\left(E\left[D(t)\right]\right)}{n} + \frac{\left[Var\left(E\left[D(t)\right]\right)\right]^2}{2(n-1)}} \quad (21)$$

Here $t_{d.f.=n-1}$ is the $t$ statistics with degree of freedom equalling $n-1$. For example, in our numerical results of Second Milstein method with antithetic sampling and sample size equalling 2500, the 95% confidence interval of $E\left[D(T)\right]$ is equal to $\left[0.9714838, 0.9718523\right]$.

Suppose the weights of investment in a portfolio on stock, zero coupon bond of no risk with maturity $T$, and corporate coupon bond equal $w_S$, $w_P$, and $w_{CB}$ respectively. Theoretically, the variance of the portfolio is equal to $\sum_{i=S,P,CB} w_i^2 Var(i) + 2\sum_{j,k=S,P,CB;\, j\neq k} w_j w_k Cov(j,k)$, where $Cov(j,k)$ is the covariance between $j$ and $k$. Given stochastic differential equations of two normalized stochastic processes $dX$ and $dY$, we could calculate $Cov(X,Y)$ by $dXdY$. The multiplication of lognormal random variables is again lognormal distributed, and the sum of lognormal random variables most likely behaves as either normal or lognormal distributions (so that we could still calculate the confidence interval).[26] As a numerical example, we let $w_S$, $w_P$, and $w_{CB}$ be 0.15, 0.65, and 0.2 respectively. Figure 20 shows the comparison of histograms with/without antithetic sampling of the portfolio.

However, the risk factors and parameters involved are not constant. For example, $r(t)$, $\theta(t)$ and $\chi(t)$ in our numerical examples are not constant. In addition, $r(t)$ rises sharply over long time periods as we could see in Figures 21. By switching the coefficients in the drift term of $\theta(t)$, $0.01-0.05\theta(t)$, we could alleviate this situation observed in Figure 22. In Appendix 10 of Supplementary materials, we show that the mean and variance of the interest rate process $r(t)$ behave like the mean and variance of a CIR process asymptotically. In addition, we observe negative values of $D(t)$ while implementing time discretization over long time periods, which could result from discretization bias and no differentiability of Brownian motion.[27]

We provide one more example related to insurance contract. From Bonnin et al. (2014), the flow of benefits for a saving contract $\Lambda(\tau)$ at time $\tau$ is equal to $VR(\tau\wedge T)\delta(\tau\wedge T)$, in which $\tau\wedge T$ is the minimum of $\tau$ and $T$, $VR(t)$ is the value of a saving contract with its instantaneous accumulation rate $r_s(t)$ at time $t$, and $\delta(t)$ is the discount factor. Figure 23 shows the expected value of $\Lambda(\tau)$ at each year $\tau$, denote $E\left[\Lambda(\tau)|\tau\right]$. Suppose $\tau$ is uniform distributed at time interval $(0,T)$, then we could calculate the best estimated of a saving contract $BEL(0,T)$ as average of the expected value of $\Lambda(\tau)$ at each year $\tau$, $BEL(0,T) \approx \frac{1}{T}\sum_{\tau=1}^{T} E\left[\Lambda(\tau)|\tau\right]$. In our example here, $BEL(0,T) \approx 0.1490473$.

[25] See, for example, Olsson (2005).

[26] See, for example, Dufresne (2004), Lo (2012), and Gulisashvili and Tankov (2016).

[27] See, for example, Glasserman (2013) Chapter 6.3.3, and Mörters and Peres (2010) Chapter 1.3.

Further study would be to investigate the situation when the diffusion term in stock price is a complex number and when the observed estimated processes are deviated from the processes with required regularity conditions.

## 5. Conclusion

In this paper, we derive the general form of deflator for four risk factors: interest rates, stock prices, default intensities, and convenience yields and then we find the regularity conditions for the deflator. We examine the deflator with different financial derivatives, comparing the numerical results with values calculated from closed-form formulas. We find required regularity conditions to implement our stochastic deflator. Our results indicate the reliability in statistics of the deflator for financial asset pricing.

Except the benefit that we could compute best estimate value by simply averaging the multiplication of deflator and projected cash flows, the fact that we observe data only in physical world would provide the motivation for us to use deflator for the convenience to estimate parameters of "reaction functions" in an ALM projection model as in Chapter 4 of Laurent et al. (2016).

Further work would be to compare the best estimate values of a life insurance contract by the deflator approach under physical measure and risk-neutral measure. More importantly, how to handle the situation when the diffusion term in stock price is a complex number and when the observed estimated processes are deviated from the processes with required regularity conditions would be further research directions.

**Acknowledgement**

Travaux réalisés dans le cadre de la Chaire « Data Analytics & Models for Insurance », placée sous l'égide de la Fondation du Risque en partenariat avec l'UCBL et BNP Paribas Cardif.

Authors also thanks for comments from anonymous referees.

## 7. TABLES AND FIGURES

**Tab. 1. Zero coupon bond of no risk with maturity $T$**

| # of Simulations | $E[D(T)]$ | $E[D(T)P(T,T,r(T))]$ | $P(0,T,r(0))$ |
|---|---|---|---|
| **Euler method** | | | |
| 2500 | 0.967646653771768 | 0.967552873529761 | |
| 5000 | 0.968545566976094 | 0.968451758793272 | |
| 10000 | 0.971055574970586 | 0.970961666405972 | |
| 100000 | 0.969964099445665 | 0.969870204097554 | 0.970957220487724 |
| 250000 | 0.970861713011257 | 0.970767802364827 | |
| 500000 | 0.970697882733399 | 0.970603984186568 | |
| 1000000 | 0.971001216710056 | 0.970907309007362 | |
| **Milstein method** | | | |
| 2500 | 0.969203041882130 | 0.969109232549717 | |
| 5000 | 0.968364554353440 | 0.968270761261484 | |
| 10000 | 0.972015353313569 | 0.971921436693164 | |
| 100000 | 0.969796855984494 | 0.969702984763468 | 0.970957220487724 |
| 250000 | 0.970693248836965 | 0.970599353426790 | |
| 500000 | 0.970367620896040 | 0.970273749033845 | |
| 1000000 | 0.970880147642130 | 0.970786256114490 | |
| **Second Milstein method** | | | |
| 2500 | 0.971985461477127 | 0.971985482351078 | |
| 5000 | 0.973386426354929 | 0.973386494993151 | |
| 10000 | 0.970979266245196 | 0.970979241795446 | |
| 100000 | 0.972928673921800 | 0.972928710489062 | 0.970957220487724 |
| 250000 | 0.970544559568911 | 0.970544535585851 | |
| 500000 | 0.970579400873414 | 0.970579381127470 | |
| 1000000 | 0.970630578417248 | 0.970630560272500 | |

**Tab. 2. Corporate coupon bond**

| # of Simulations | Deflator | Longstaff et al. (2005) |
|---|---|---|
| **Euler method** | | |
| 2500 | 1.03001393884536 | |
| 5000 | 1.02907100520235 | |
| 10000 | 1.03312948158384 | |
| 100000 | 1.03162033904634 | 1.03313616115971 |
| 250000 | 1.03255241573140 | |
| 500000 | 1.03237954508927 | |
| 1000000 | 1.03271899062813 | |
| **Milstein method** | | |
| 2500 | 1.02572713722014 | |
| 5000 | 1.02914390668280 | |
| 10000 | 1.03349796386179 | |
| 100000 | 1.03135227993609 | 1.03313616115971 |
| 250000 | 1.03247145621524 | |
| 500000 | 1.03214895930589 | |
| 1000000 | 1.03261933018597 | |
| **Second Milstein method** | | |
| 2500 | 1.03584708542322 | |
| 5000 | 1.03243391095283 | |
| 10000 | 1.03330315641903 | |
| 100000 | 1.03468566679080 | 1.03313616115971 |
| 250000 | 1.03243619762781 | |
| 500000 | 1.03234018373030 | |
| 1000000 | 1.03229088096527 | |

Fig. 2 - **Zero coupon bond,** $E\left[D(T)\right]$

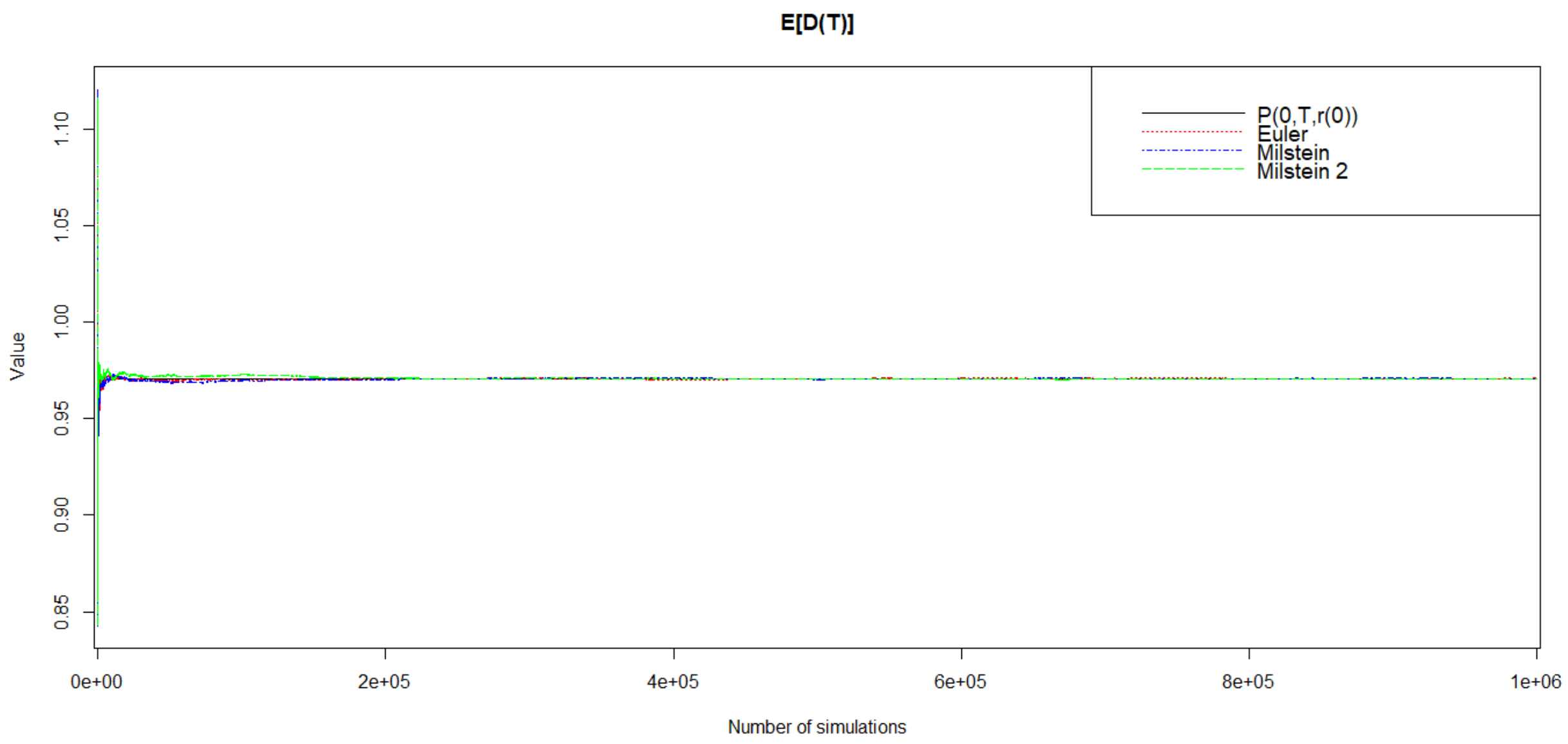

Fig. 3 - **Differences between approximation and expected value of Zero coupon bond,** $E\left[D(T)\right]$

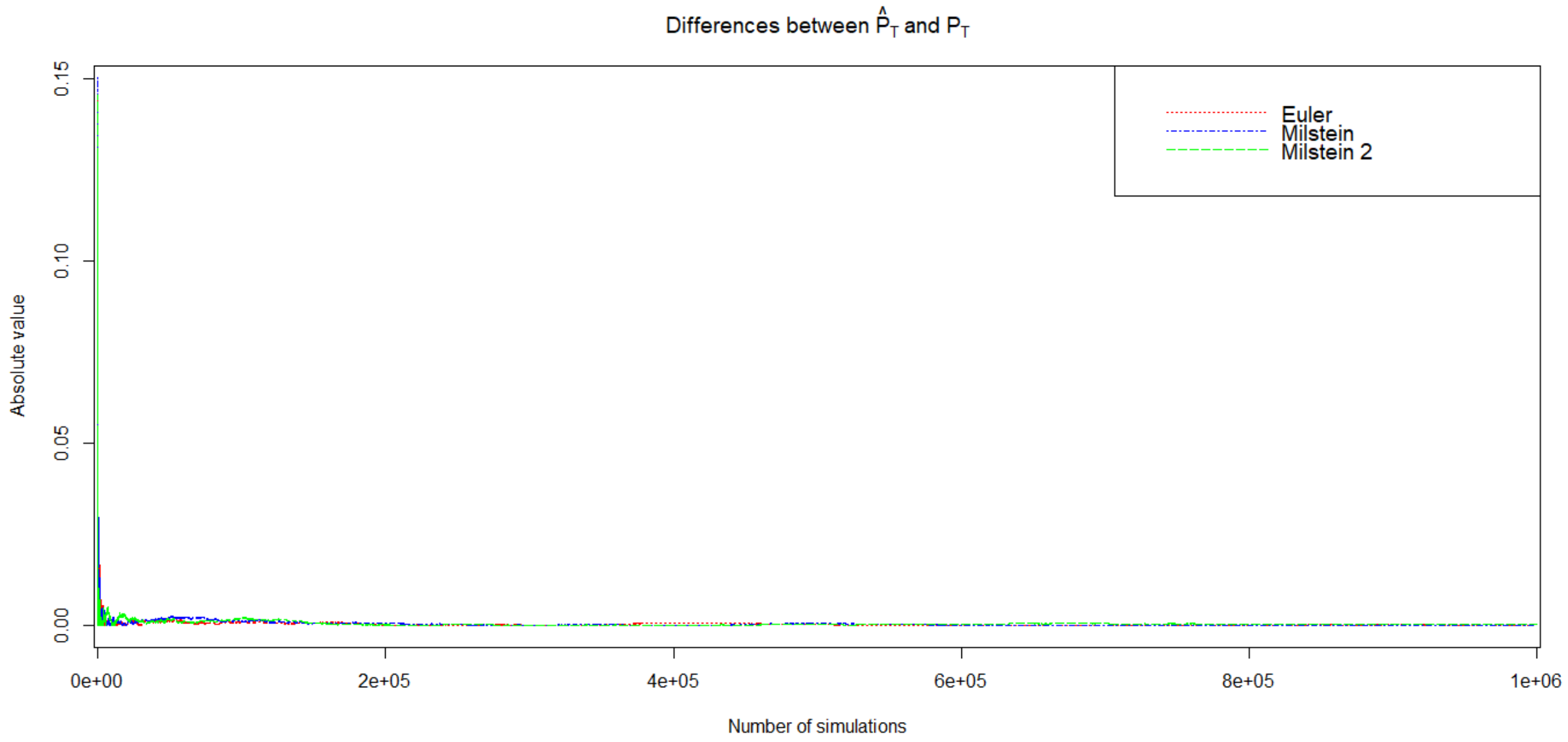

Fig. 4 - **Zero coupon bond,** $E\left[D(T)\right]$**, number of simulations less than 10000**

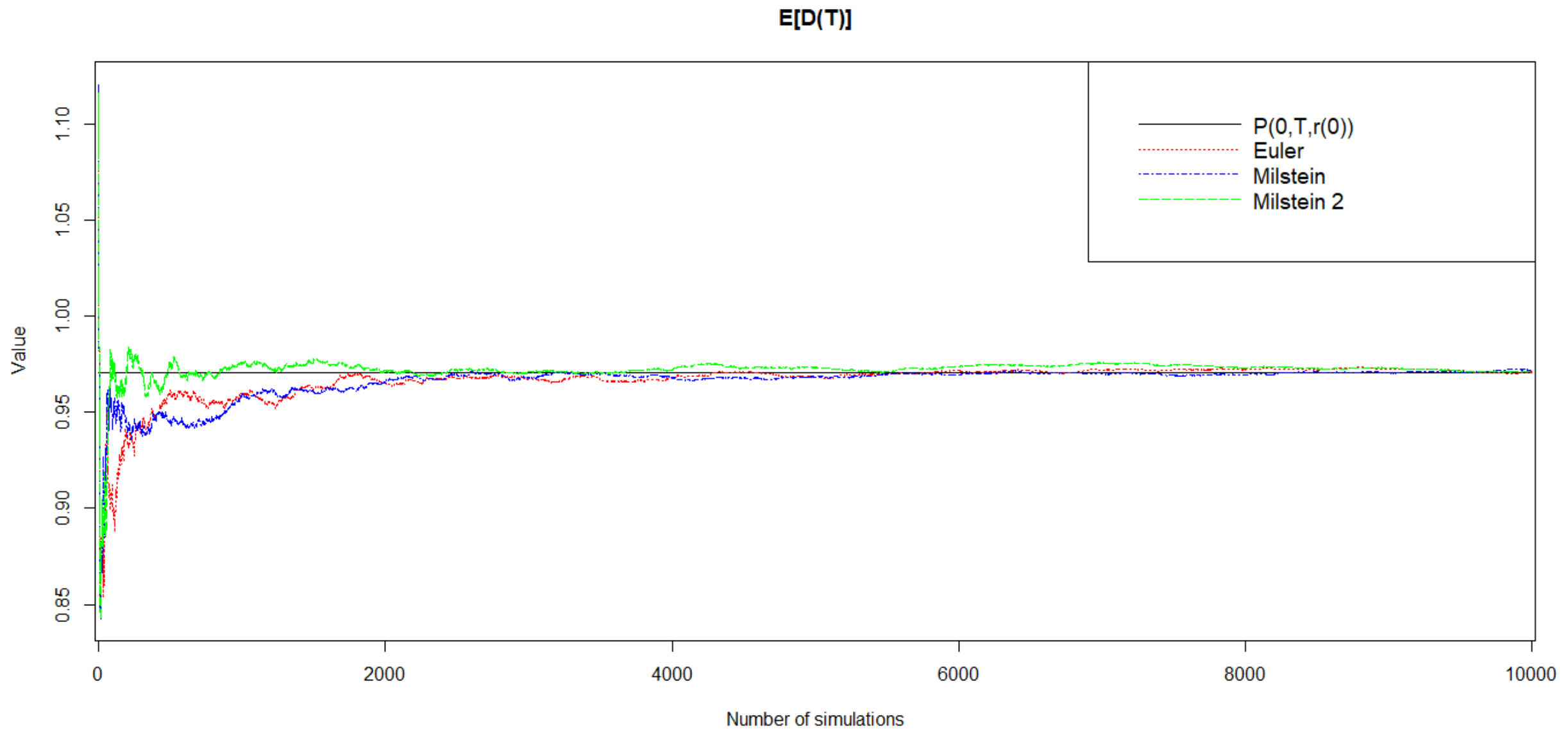

Fig. 5 - **Differences between approximation and expected value of Zero coupon bond,** $E\left[D(T)\right]$**, number of simulations less than 10000**

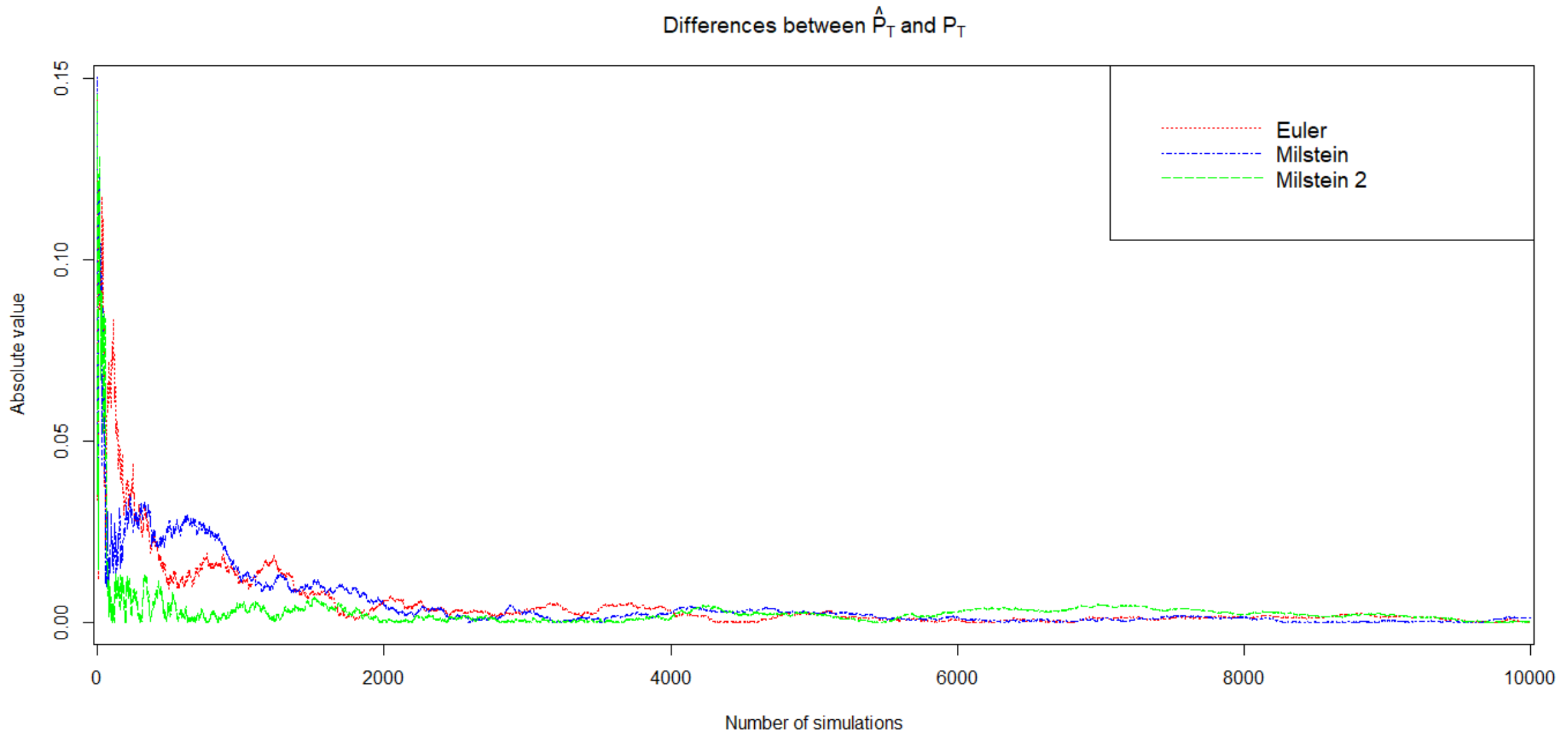

Fig. 6 - **Zero coupon bond,** $E\left[D(T)P\left(T,T,r(T)\right)\right]$

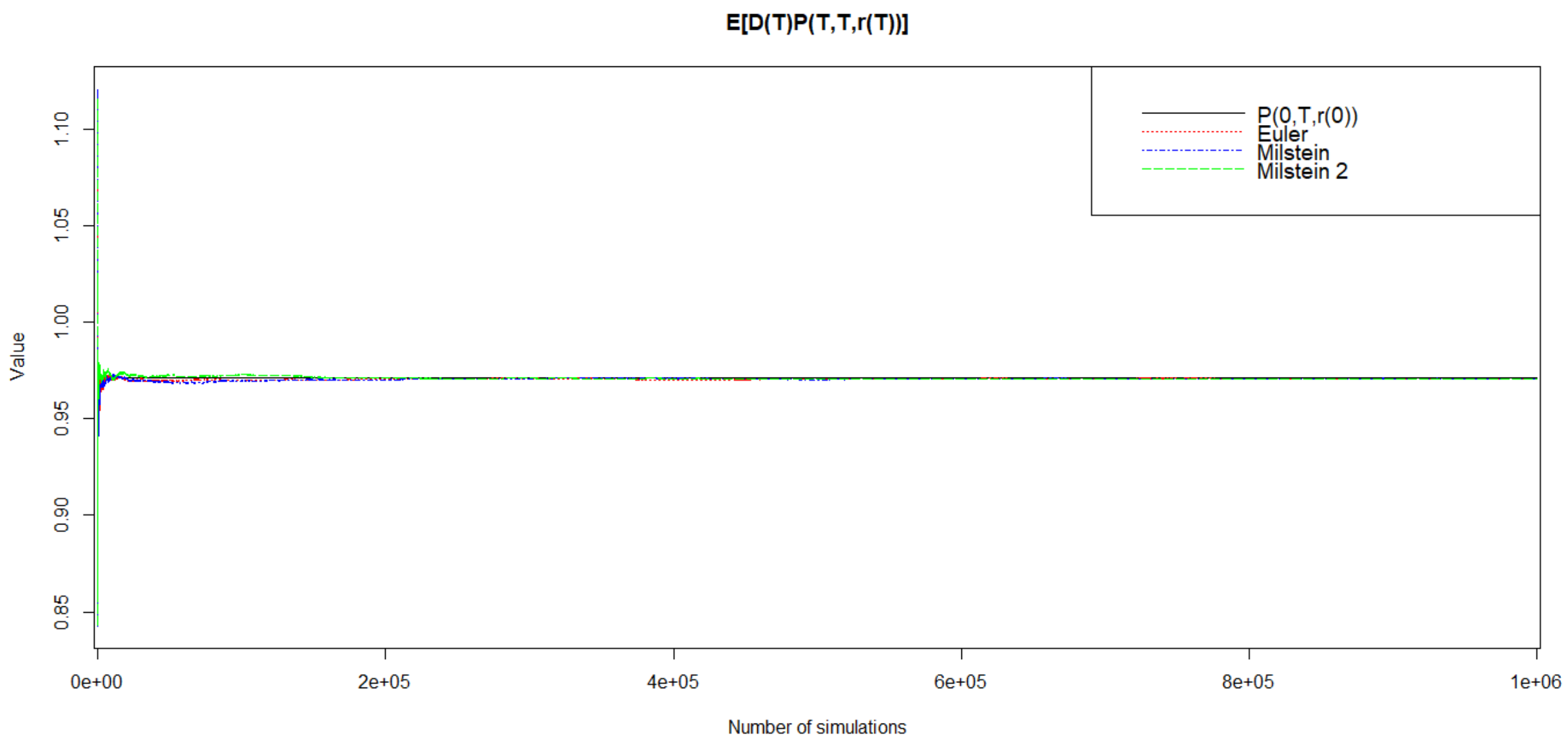

Fig. 7 - **Differences between approximation and expected value of Zero coupon bond,** $E\left[D(T)P\left(T,T,r(T)\right)\right]$

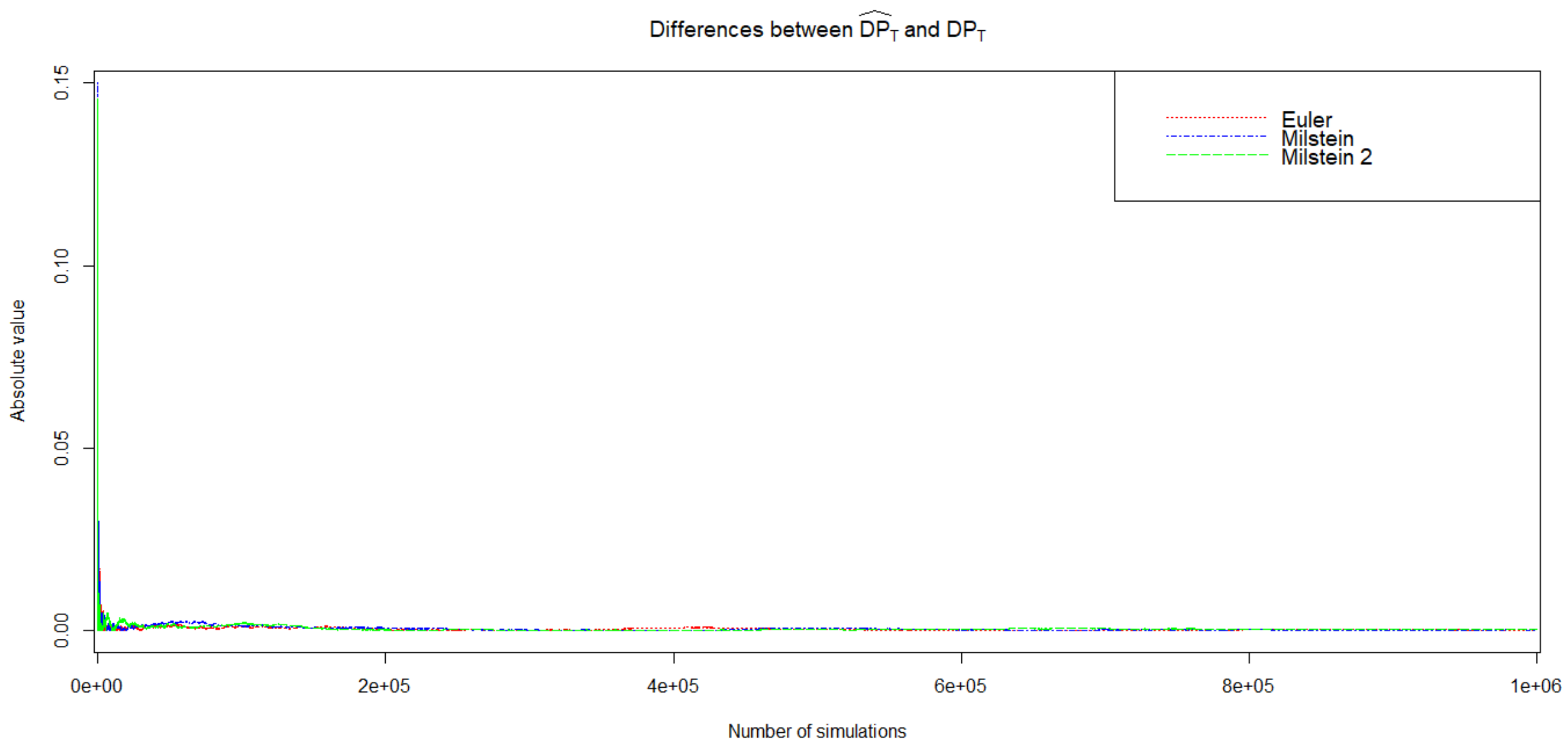

Fig. 8 - **Zero coupon bond,** $E\left[D(T)P(T,T,r(T))\right]$**, number of simulations less than 10000**

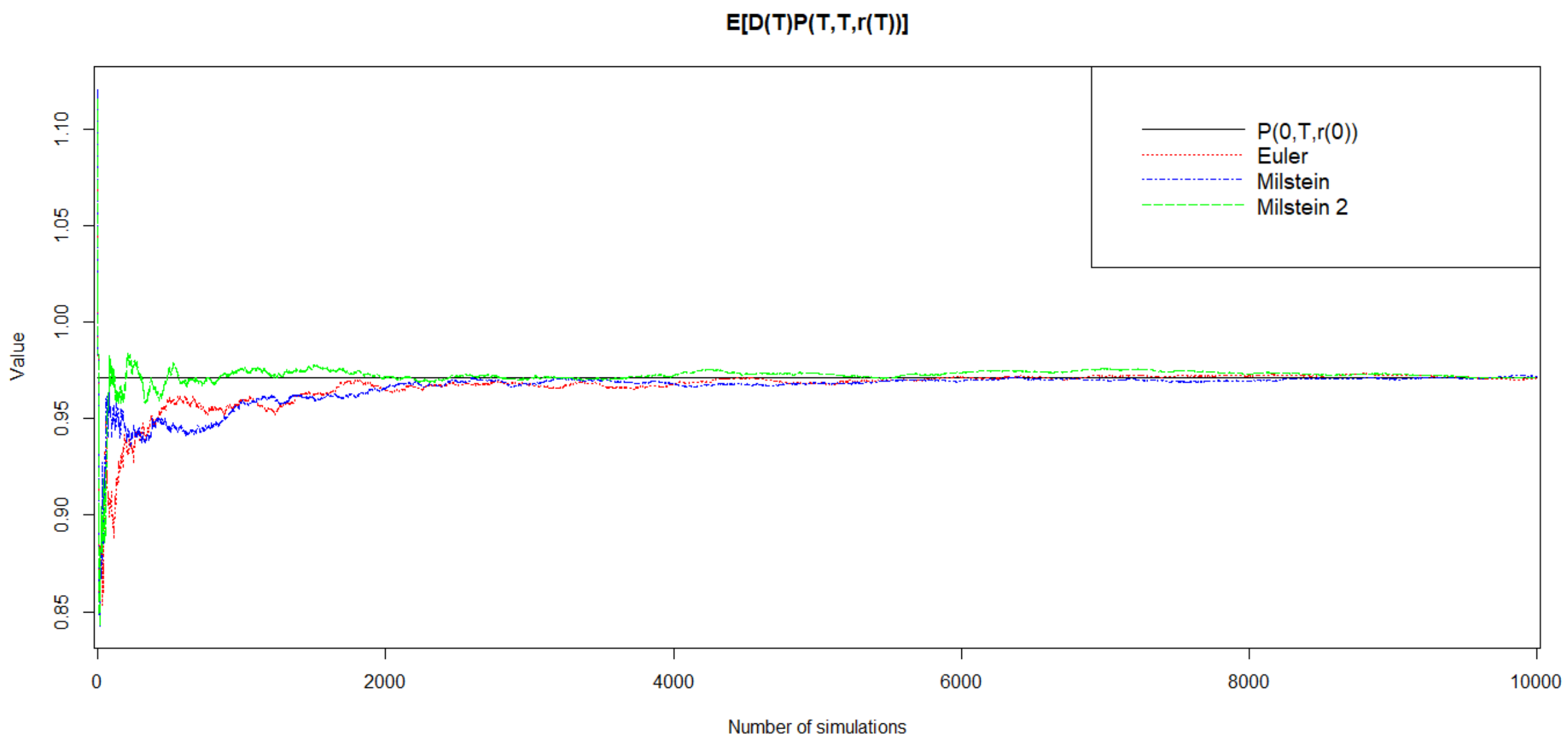

Fig. 9 - **Differences between approximation and expected value of Zero coupon bond,** $E\left[D(T)P(T,T,r(T))\right]$**, number of simulations less than 10000**

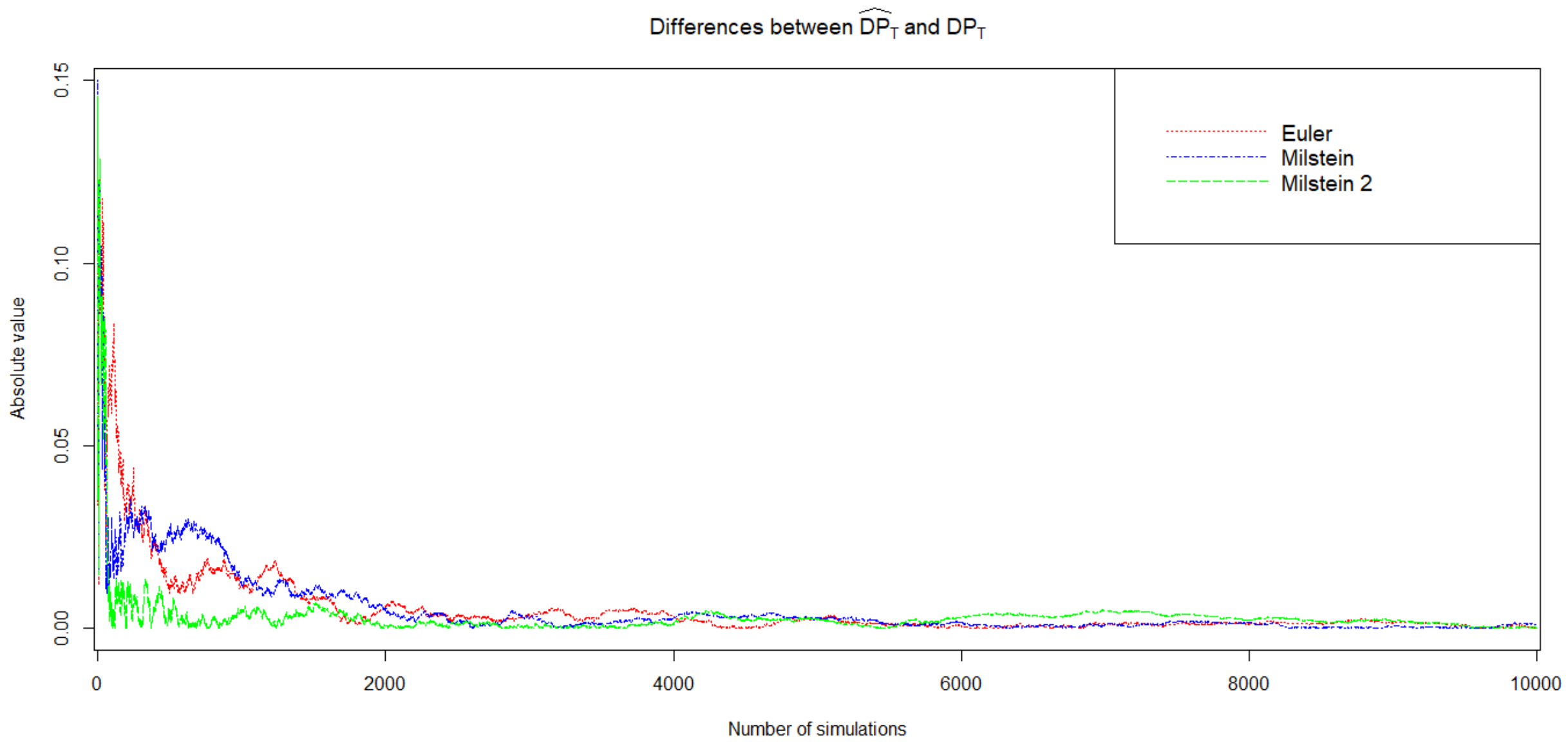

Fig. 10 - **Corporate coupon bond**

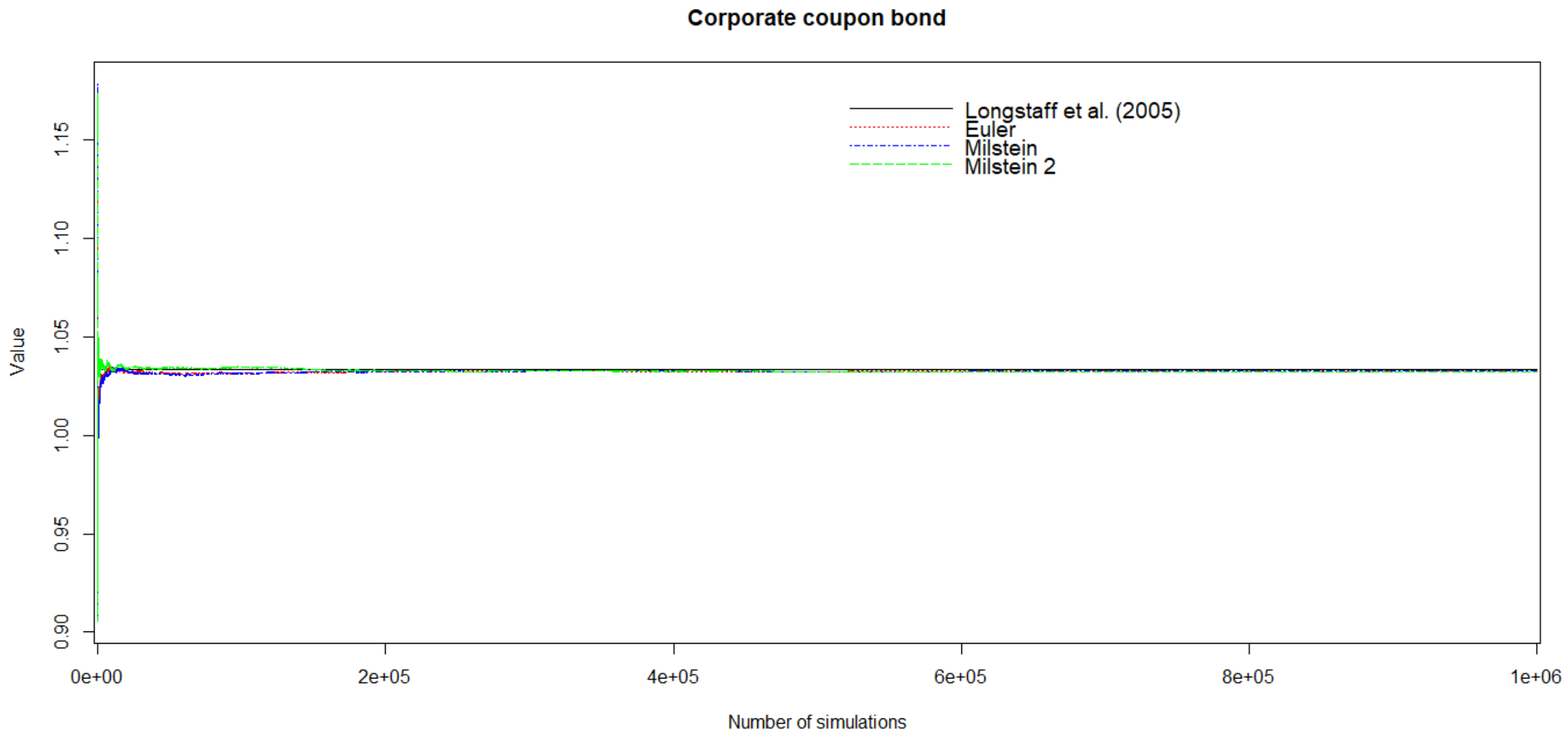

Fig. 11 - **Differences between approximation and expected value of Corporate coupon bond in Longstaff et al. (2005)**

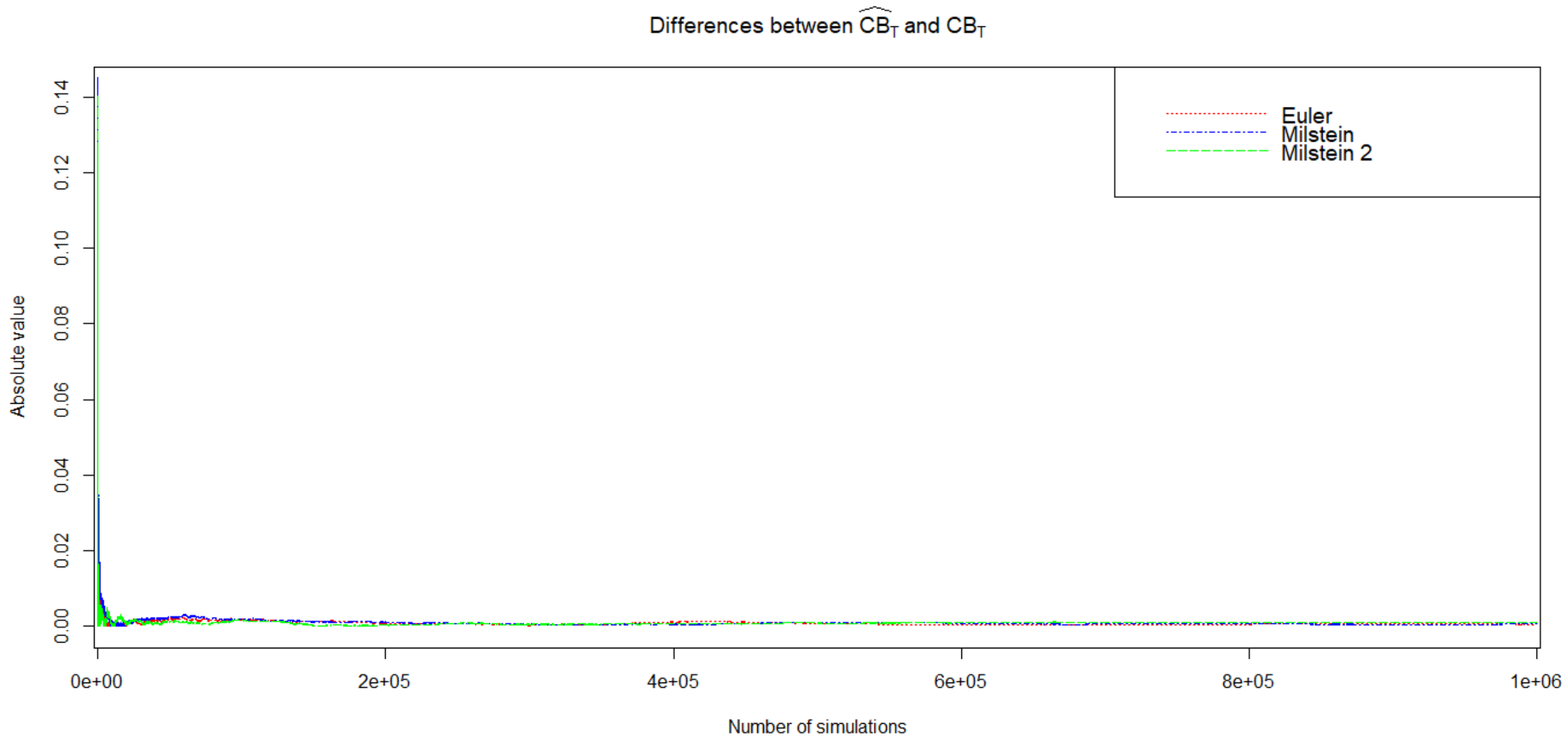

Fig. 12 - **Corporate coupon bond, number of simulations less than 10000**

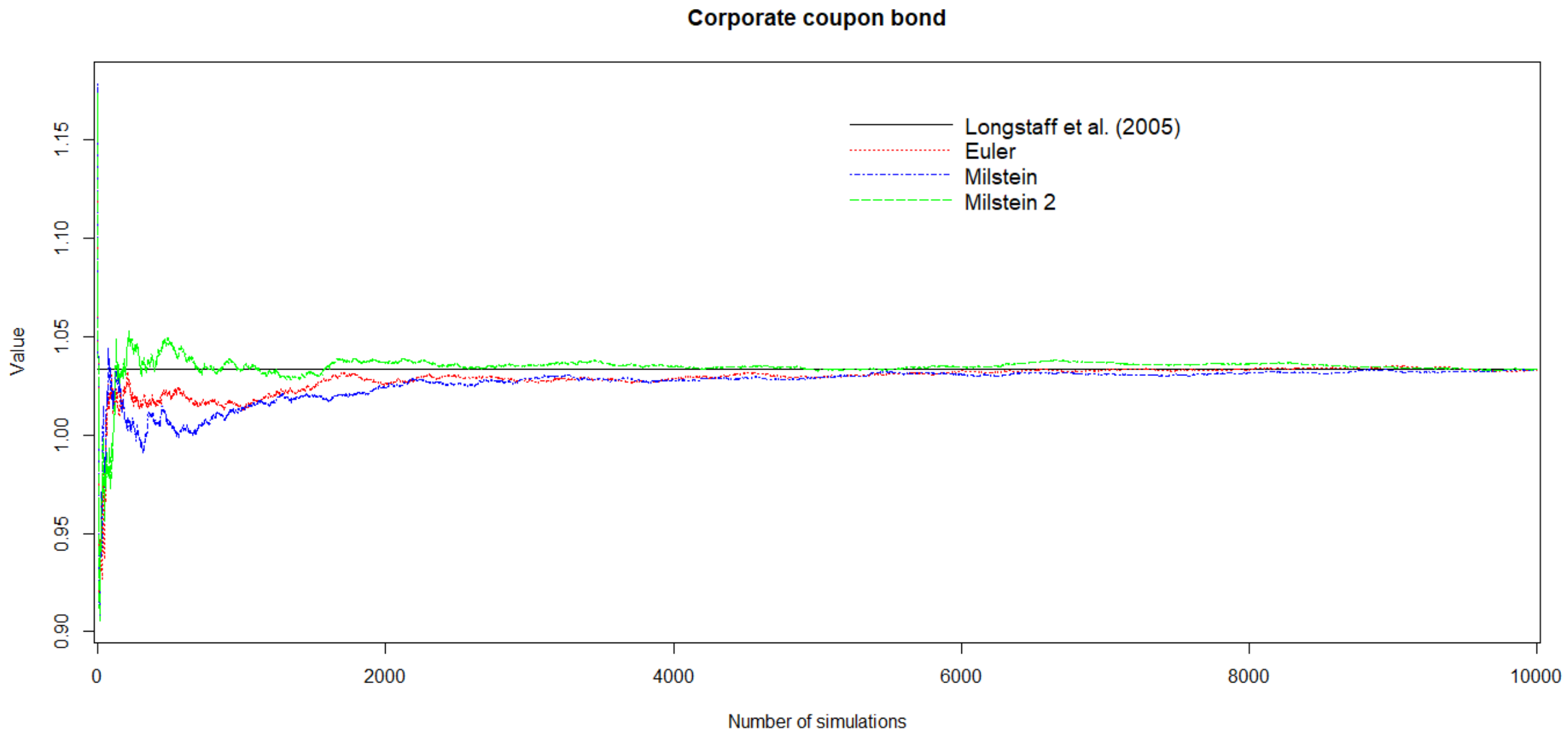

Fig. 13 - **Differences between approximation and expected value of Corporate coupon bond in Longstaff et al. (2005), number of simulations less than 10000**

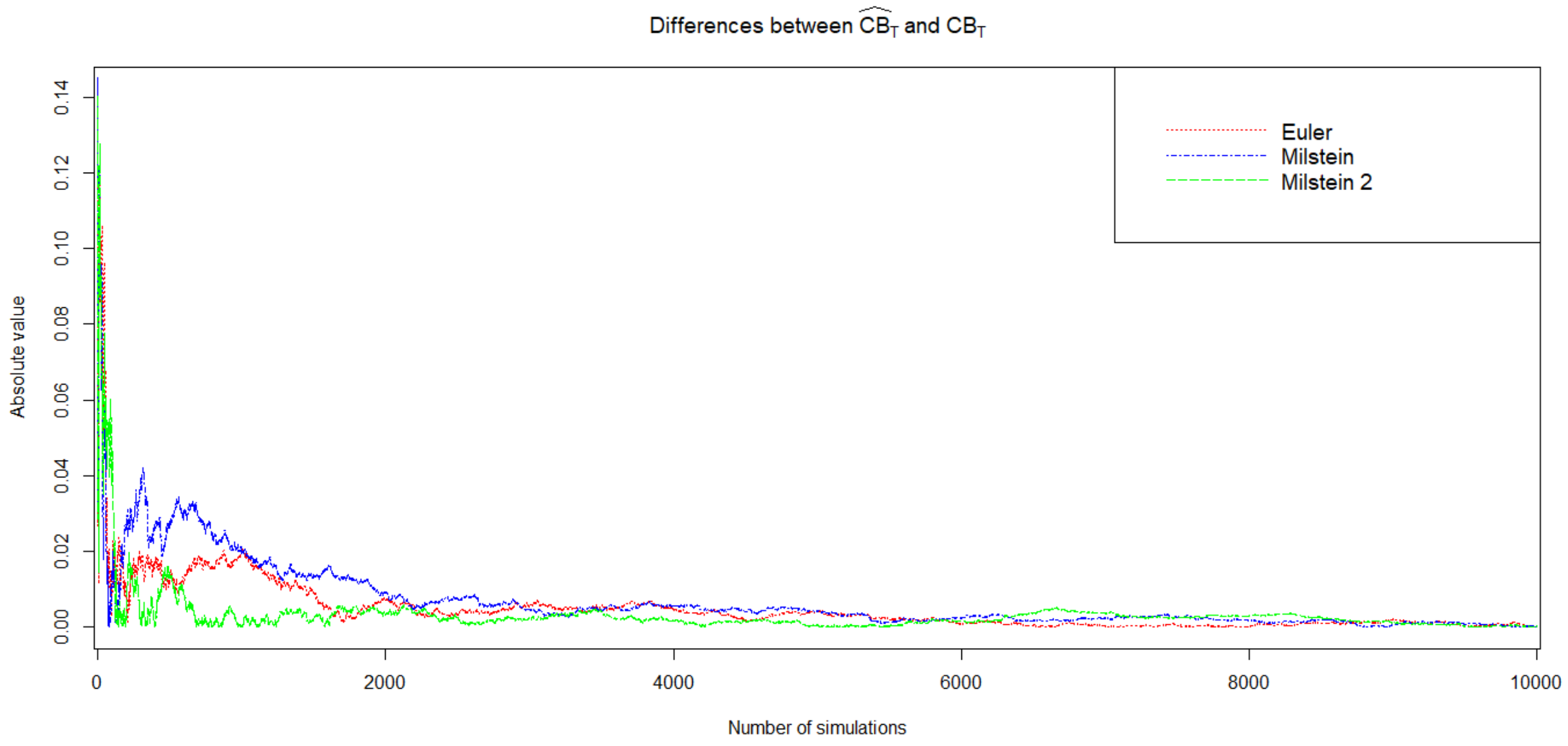

Fig. 14 - **Deflator multiplies stock over long time periods**

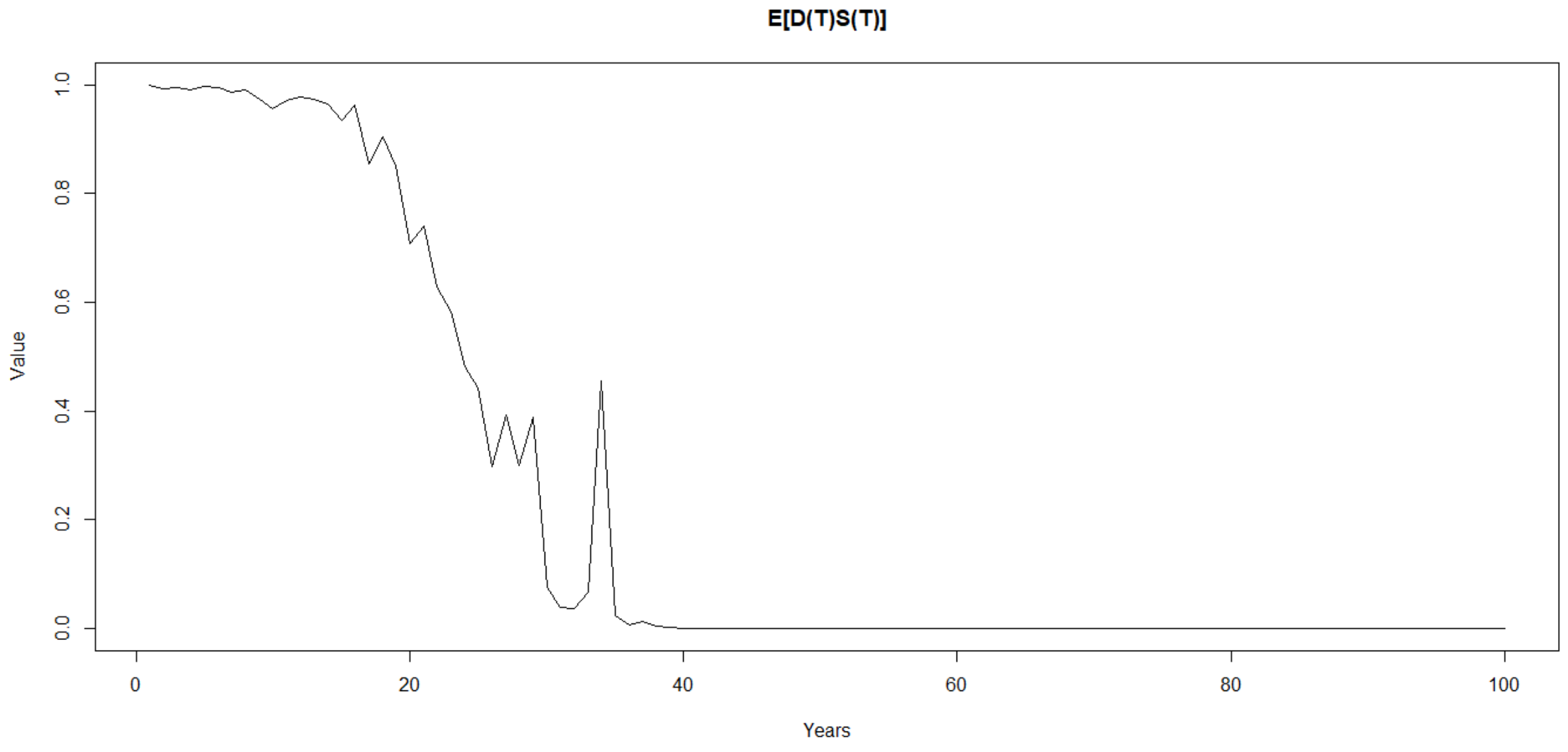

Fig. 15 - **Deflator multiplies stock over long time periods, 16 years**

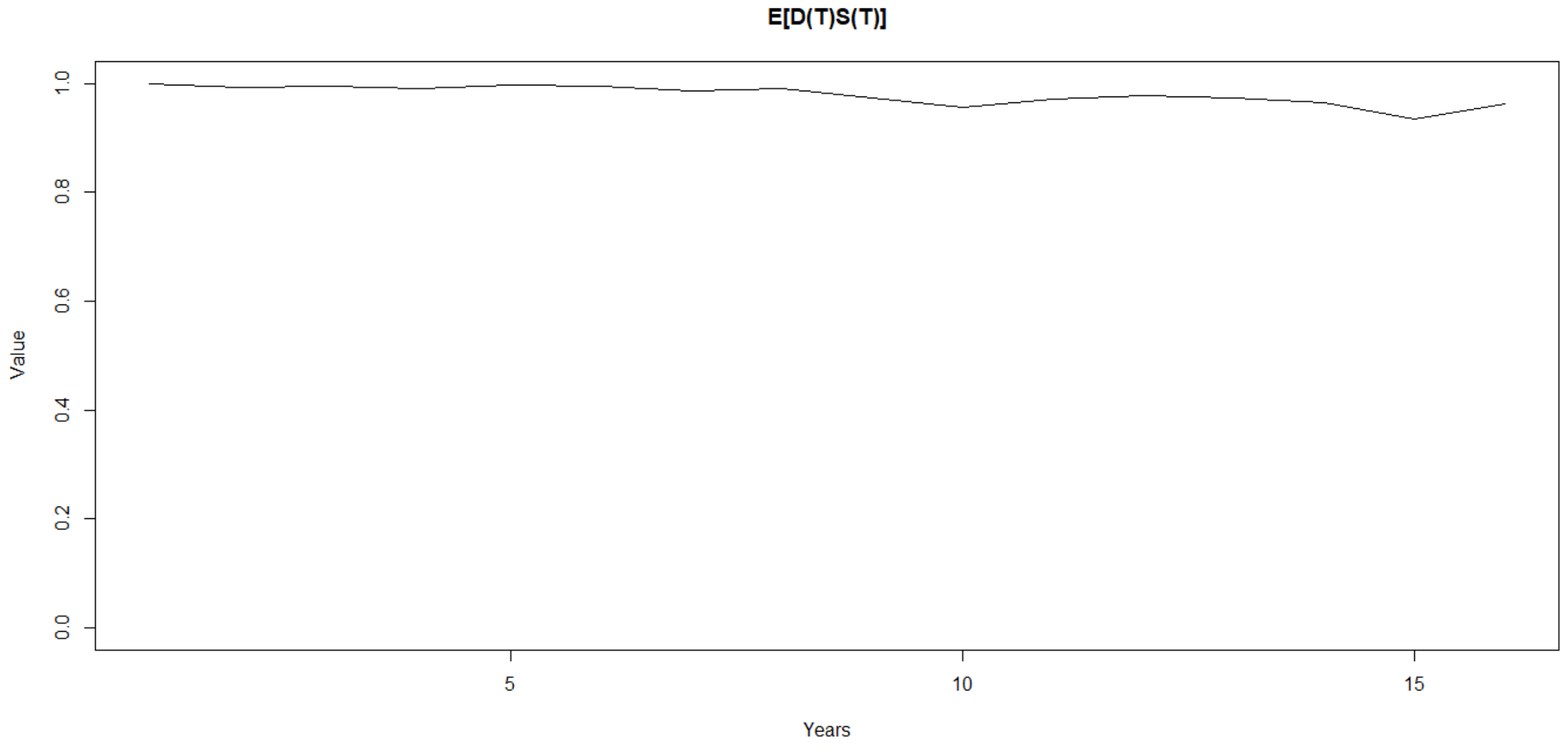

Fig. 16 - **Deflator multiplies stock over long time periods, 16 years**

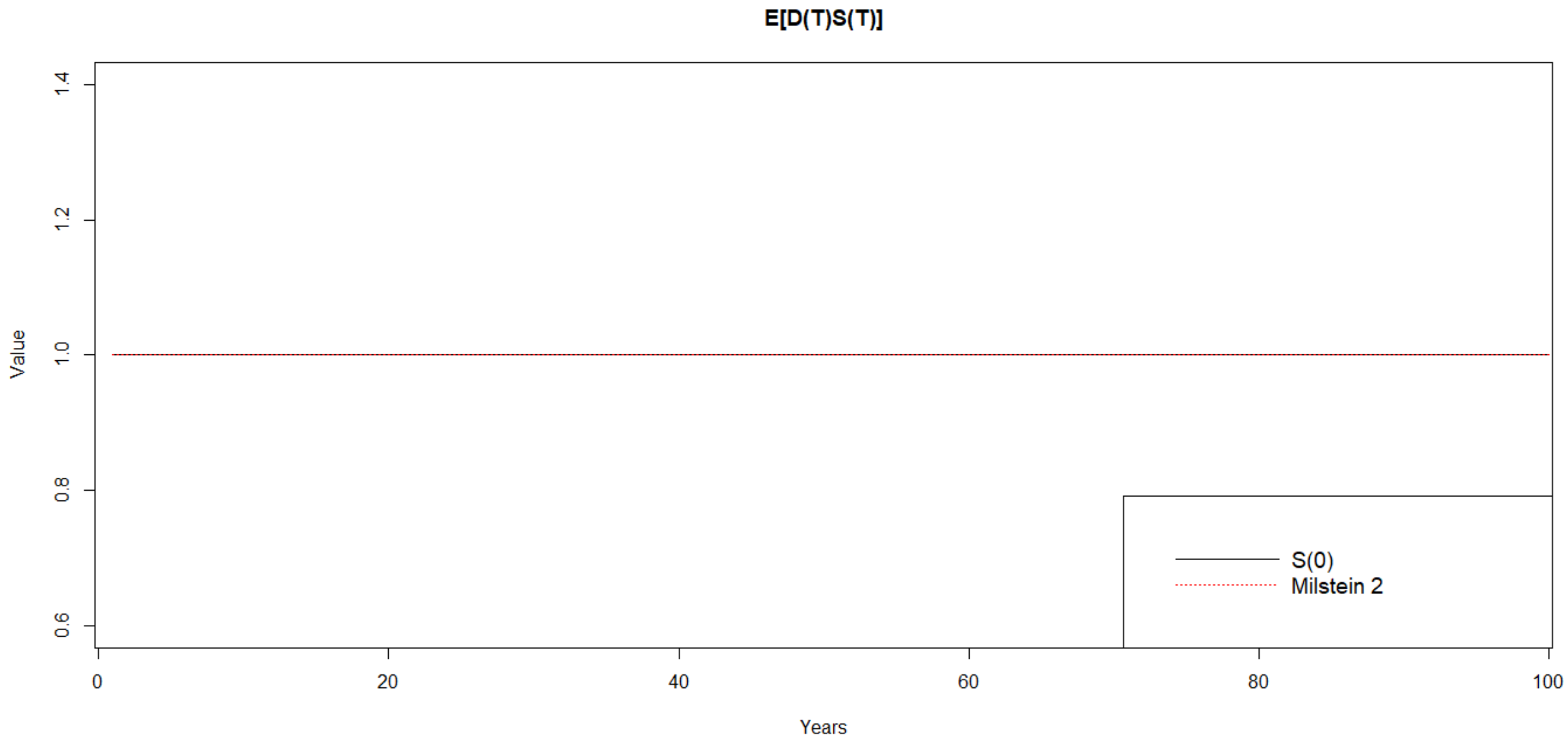

Fig. 17 - **Deflator multiplies stock over long time periods, 16 years**

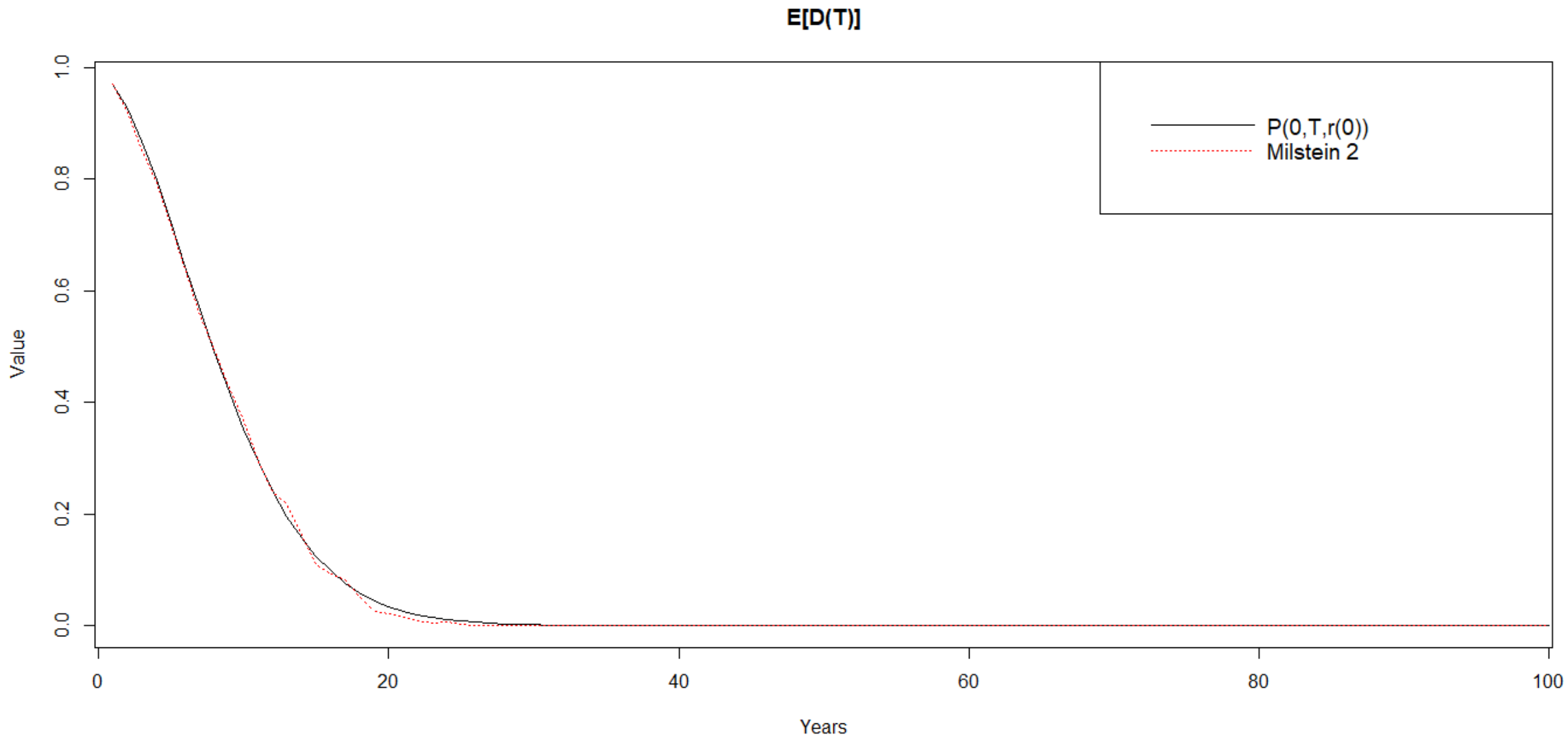

Fig. 18 - **Deflator multiplies stock over long time periods, 16 years**

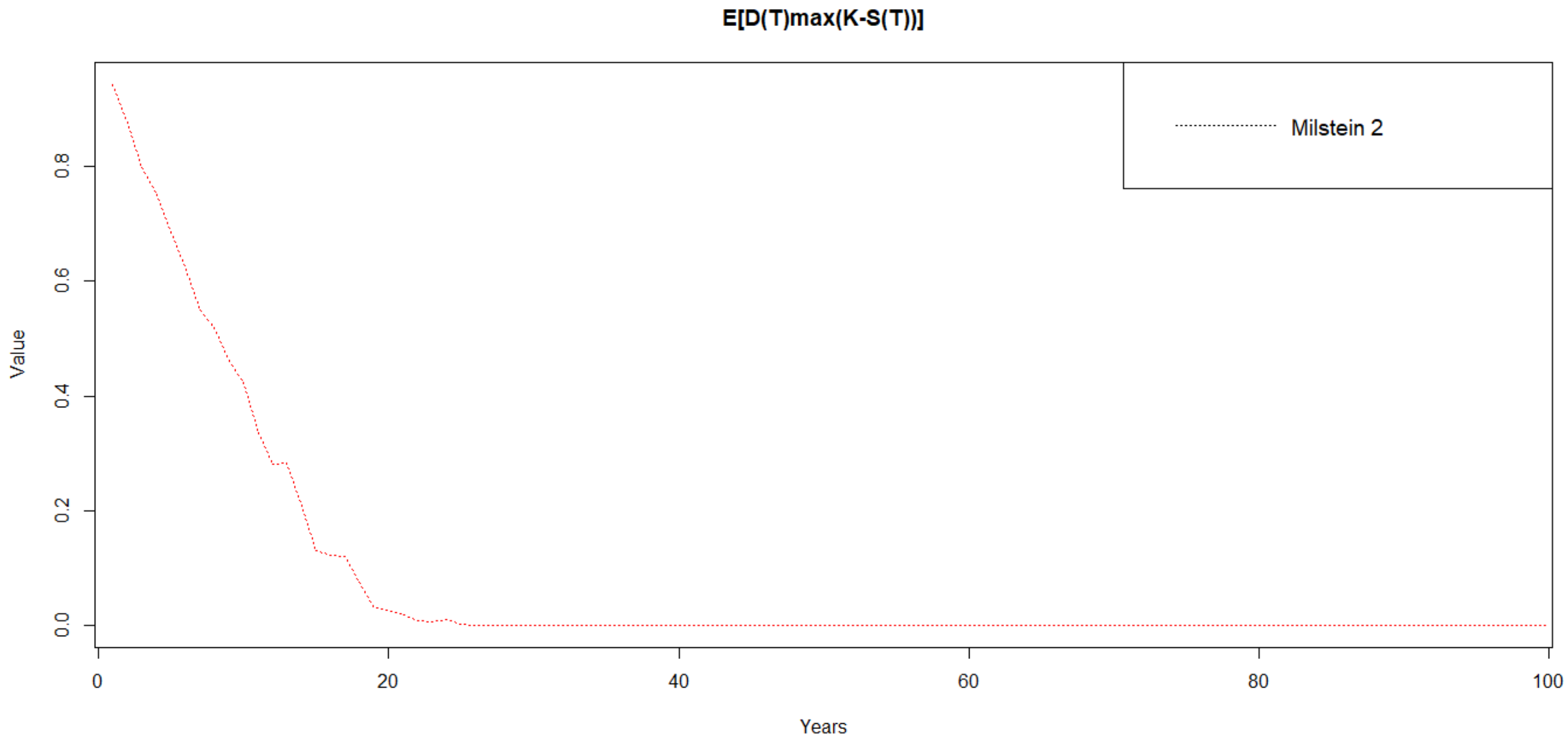

Fig. 19 - **Deflator multiplies stock over long time periods, 16 years**

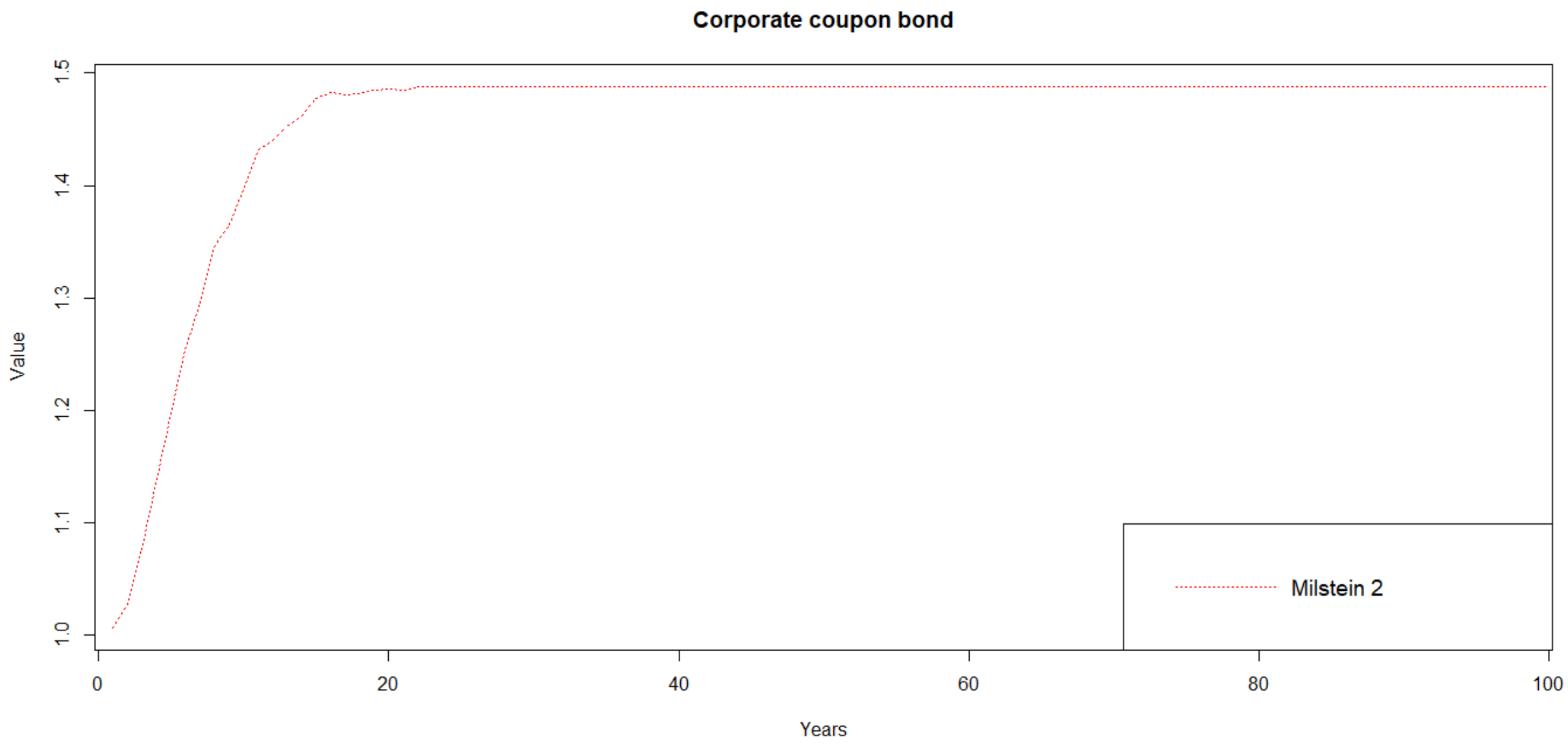

Fig. 20 - **Histogram comparison with antithetic sampling**

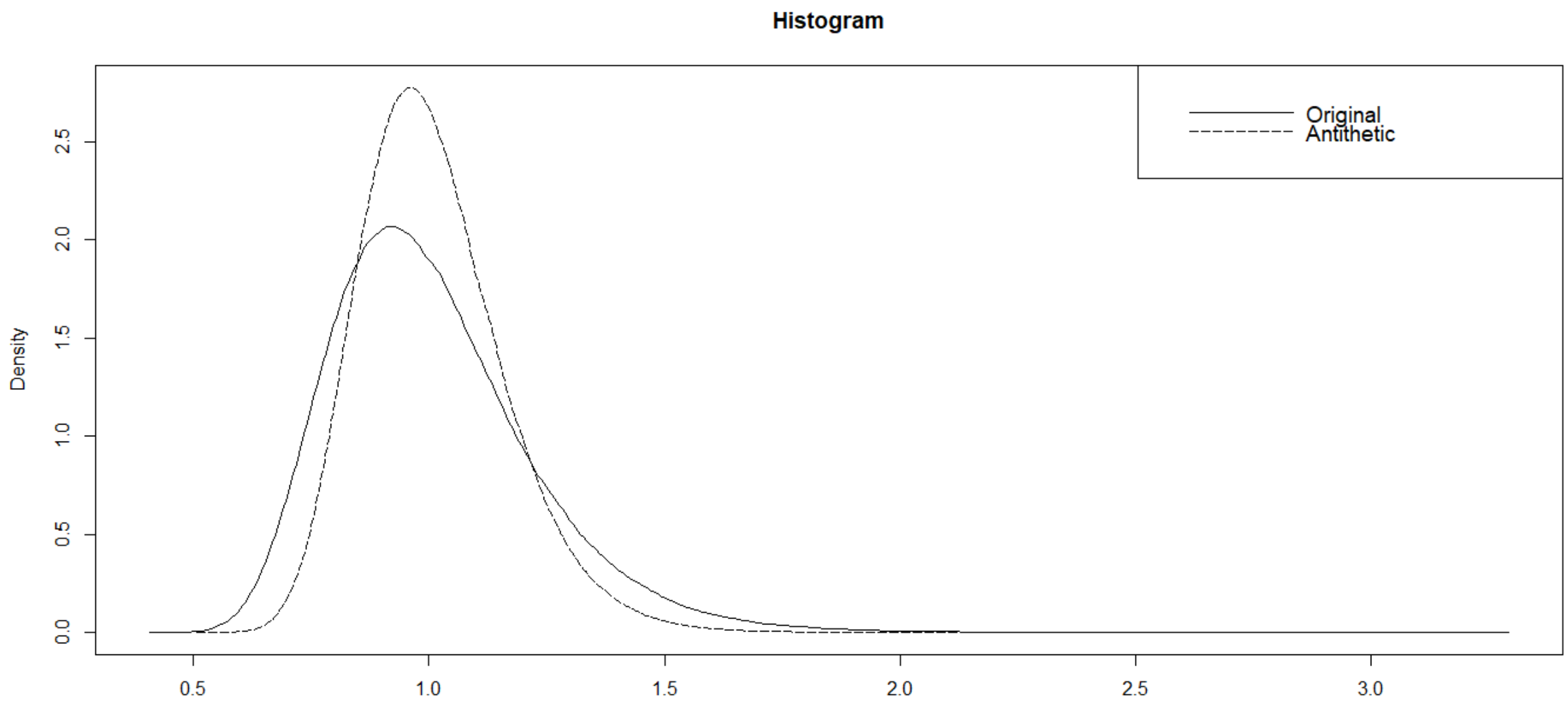

Fig. 21 - **Interest rate over long time periods**

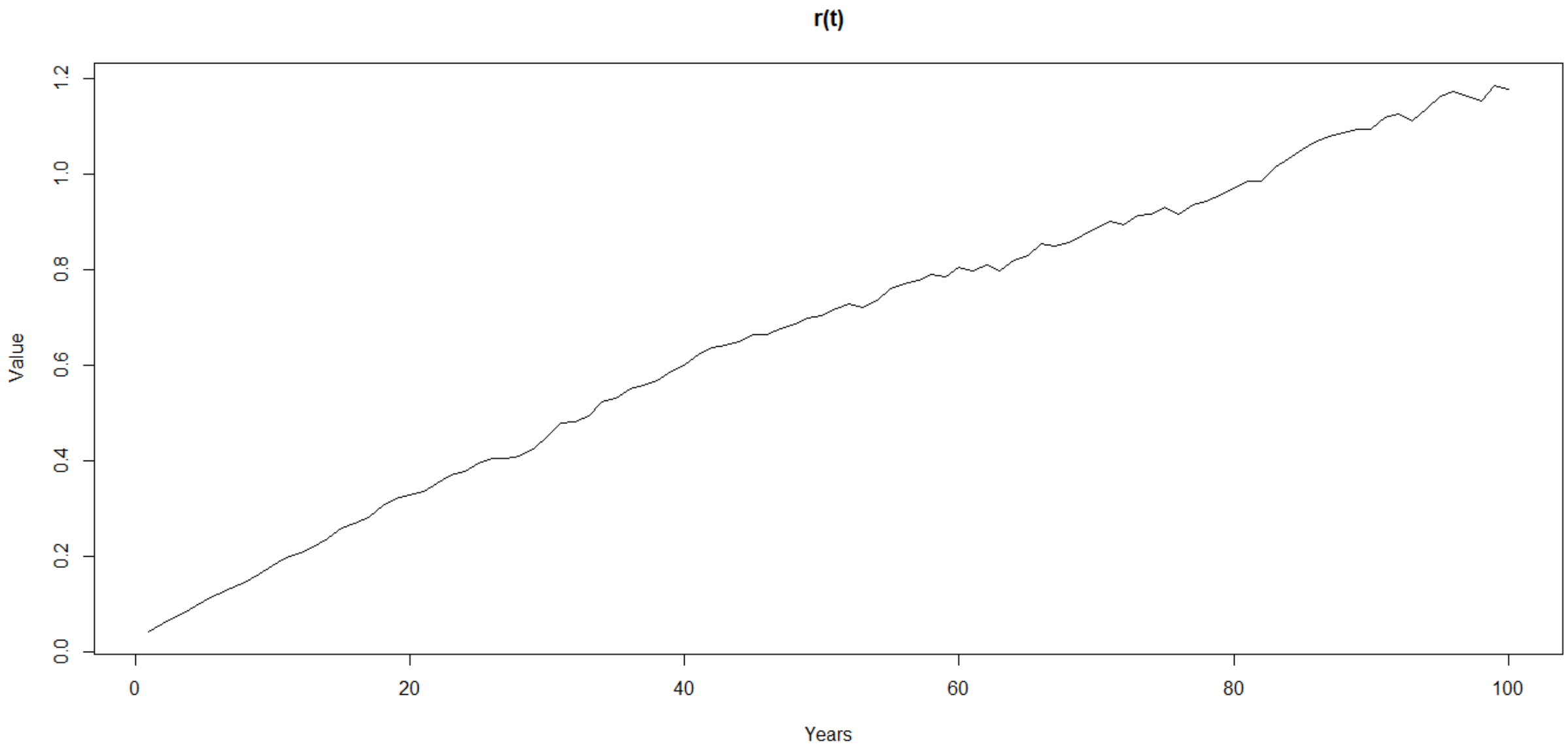

Fig. 22 - **Interest rate over long time periods, alleviated**

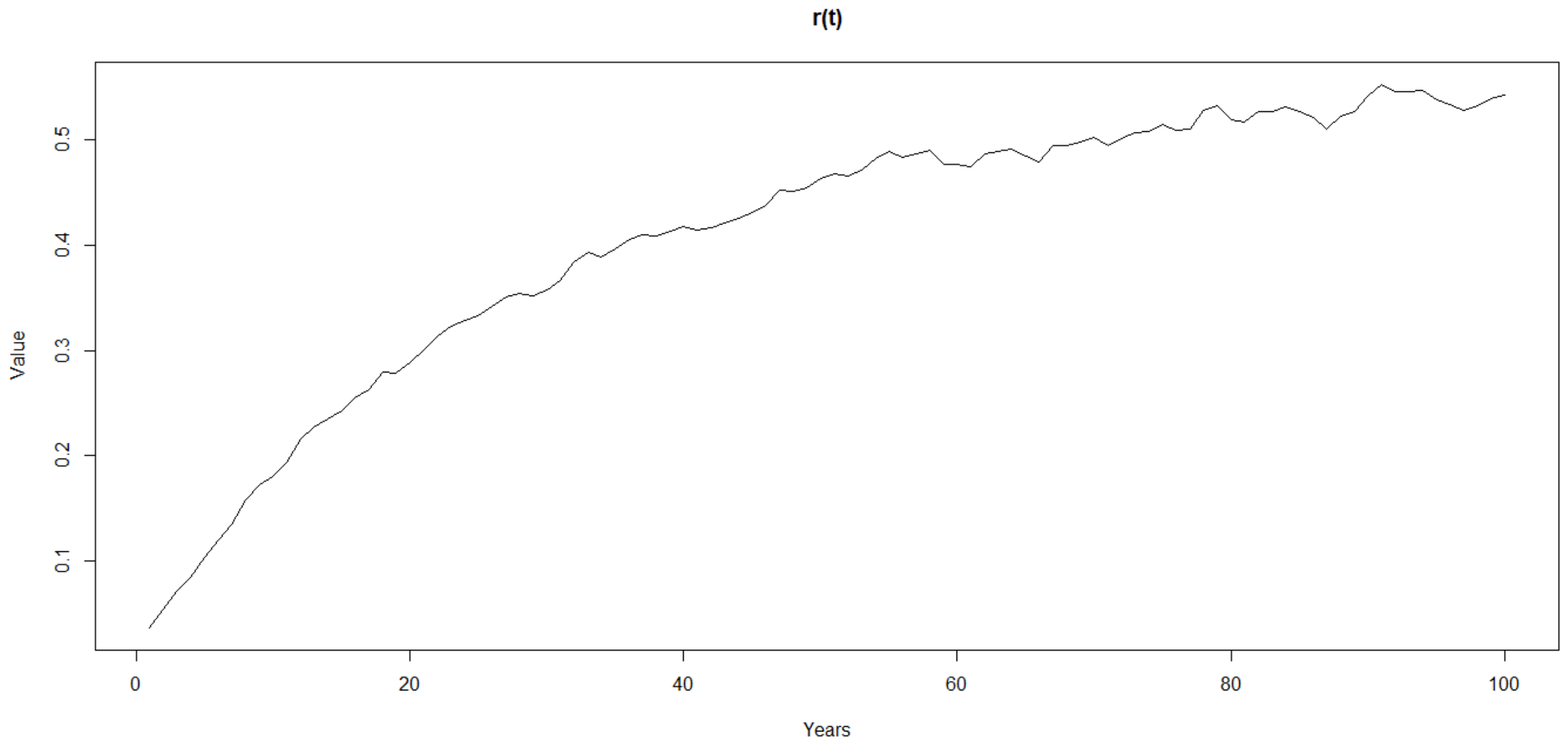

Fig. 23 - **Expected flow of benefits across time**

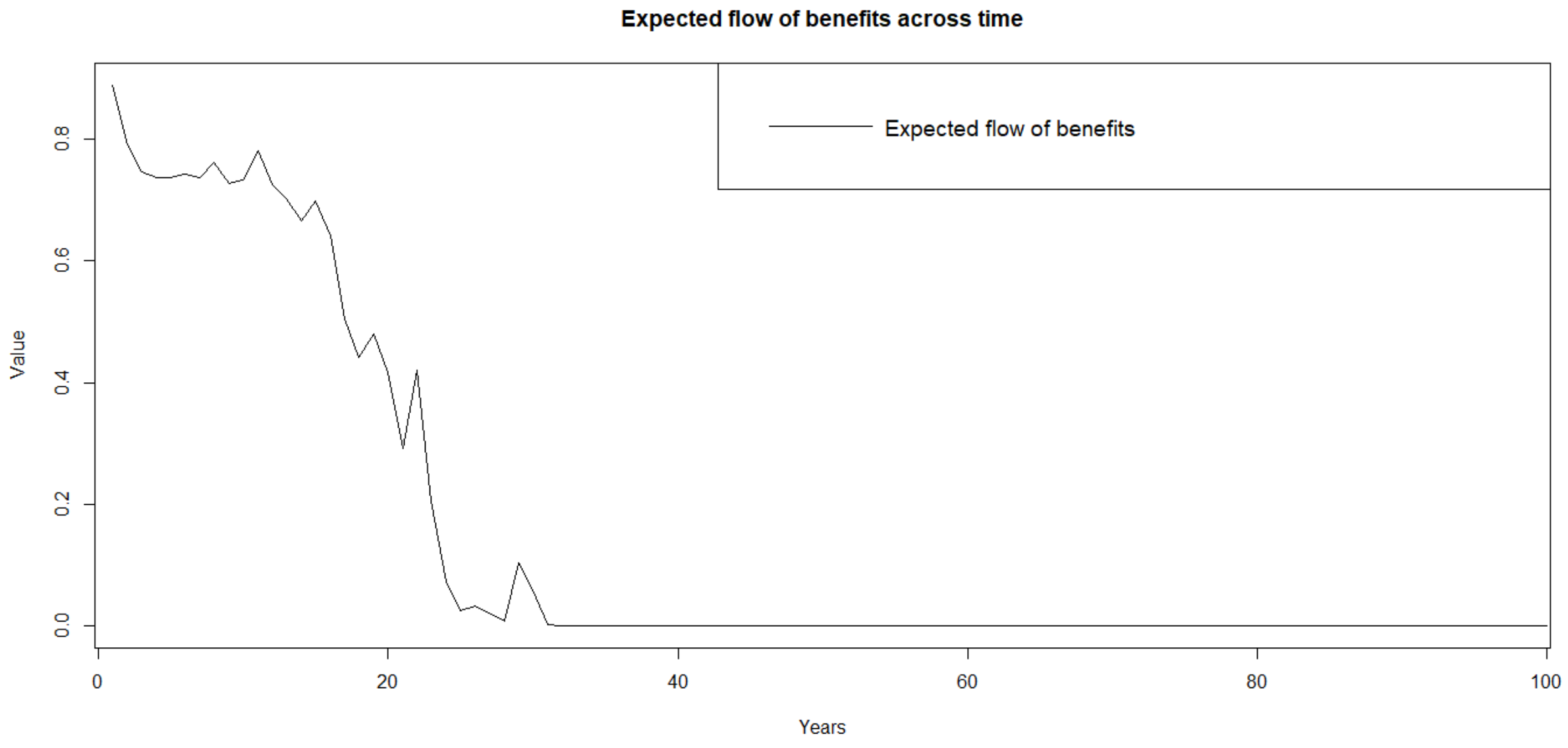

LABORATOIRE SAF SCIENCES ACTUARIELLE & FINANCIÈRE

# STOCHASTIC DEFLATOR FOR AN ECONOMIC SCENARIO GENERATOR WITH FIVE FACTORS

## SUPPLEMENTARY MATERIAL

**Po-Keng Cheng*** **Frédéric Planchet***

Univ Lyon - Université Claude Bernard Lyon 1,

ISFA, Laboratoire SAF EA2429, F-69366, Lyon, France

Prim'Act, 42 avenue de la Grande Armée, 75017 Paris, France

## Summary

* Po-Keng Cheng and Frédéric Planchet are researcher at SAF laboratory (EA n°2429). Frédéric Planchet is also consulting actuary at Prim'Act. Contact: ansd39@gmail.com / frederic@planchet.net.

## 1. APPENDIX 1: CONSTRUCT CORRELATED BROWNIAN MOTIONS

In Appendix 1, we provide details of the calculations to construct correlated Brownian motions in our model. We start with the Brownian motion parts of interest rate process and stock price process. Then we derive the Brownian motion parts of default intensity process and convenience yield process step by step. Readers are referred to Shreve (2004) Chapter 3.4 for discussions about quadratic variation.

Step 1

Let $dW_r = dW_0$, then $dW_S = \rho_{rS} dW_r + \sqrt{1-\rho_{rS}^2}\, dW_1$.

*pf*.

$$dW_r dW_S = dW_0\left(\rho_{rS} dW_0 + \sqrt{1-\rho_{rS}^2}\, dW_1\right) = \rho_{rS} dt$$

$$\begin{aligned} dW_S dW_S &= \left(\rho_{rS} dW_0 + \sqrt{1-\rho_{rS}^2}\, dW_1\right)\left(\rho_{rS} dW_0 + \sqrt{1-\rho_{rS}^2}\, dW_1\right) \\ &= \left(\rho_{rS}^2 + 1 - \rho_{rS}^2\right) dt = dt \end{aligned}$$

Step 2

$dW_\chi = \rho_{r\chi} dW_r + \rho'_{S\chi} dW_1 + \rho'_{\chi\chi} dW_2$,

where $\rho'_{S\chi} = \dfrac{\rho_{S\chi} - \rho_{rS}\rho_{r\chi}}{\sqrt{1-\rho_{rS}^2}}$, $\rho'_{\chi\chi} = \sqrt{\dfrac{1-\rho_{rS}^2 - \rho_{r\chi}^2 - \rho_{S\chi}^2 + 2\rho_{rS}\rho_{r\chi}\rho_{S\chi}}{1-\rho_{rS}^2}}$.

*pf*.

$$dW_r dW_\chi = dW_0\left(\rho_{r\chi} dW_0 + \rho'_{S\chi} dW_1 + \rho'_{\chi\chi} dW_2\right) = \rho_{r\chi} dt$$

$$\begin{aligned} dW_S dW_\chi &= \left(\rho_{rS} dW_0 + \sqrt{1-\rho_{rS}^2}\, dW_1\right)\left(\rho_{r\chi} dW_0 + \rho'_{S\chi} dW_1 + \rho'_{\chi\chi} dW_2\right) \\ &= \left(\rho_{rS}\rho_{r\chi} + \rho'_{S\chi}\sqrt{1-\rho_{rS}^2}\right) dt = \rho_{S\chi} dt \end{aligned}$$

$\rho_{rS}\rho_{r\chi} + \rho'_{S\chi}\sqrt{1-\rho_{rS}^2} = \rho_{S\chi}$, then $\rho'_{S\chi} = \dfrac{\rho_{S\chi} - \rho_{rS}\rho_{r\chi}}{\sqrt{1-\rho_{rS}^2}}$.

$$\begin{aligned} dW_\chi dW_\chi &= \left(\rho_{r\chi} dW_0 + \rho'_{S\chi} dW_1 + \rho'_{\chi\chi} dW_2\right)\left(\rho_{r\chi} dW_0 + \rho'_{S\chi} dW_1 + \rho'_{\chi\chi} dW_2\right) \\ &= \left(\rho_{r\chi}^2 + \rho'^2_{S\chi} + \rho'^2_{\chi\chi}\right) dt = dt \end{aligned}$$

$\rho_{r\chi}^2 + \rho'^2_{S\chi} + \rho'^2_{\chi\chi} = 1$, then $\rho'^2_{\chi\chi} = 1 - \rho_{r\chi}^2 - \rho'^2_{S\chi} = 1 - \rho_{r\chi}^2 - \dfrac{\left(\rho_{S\chi} - \rho_{rS}\rho_{r\chi}\right)^2}{1-\rho_{rS}^2}$.

Then, $\rho'_{\chi\chi} = \sqrt{\dfrac{1-\rho_{rS}^2 - \rho_{r\chi}^2 - \rho_{S\chi}^2 + 2\rho_{rS}\rho_{r\chi}\rho_{S\chi}}{1-\rho_{rS}^2}}$.

## Step 3

$$dW_\gamma = \rho_{r\gamma} dW_r + \rho''_{S\gamma} dW_1 + \rho''_{\chi\gamma} dW_2 + \rho''_{\gamma\gamma} dW_3,$$

where $\rho''_{S\gamma} = \dfrac{\rho_{S\gamma} - \rho_{rS}\rho_{r\gamma}}{\sqrt{1-\rho_{rS}^2}}$,

$$\rho''_{\chi\gamma} = \frac{\rho_{\chi\gamma} - \rho_{r\chi}\rho_{r\gamma} - \rho_{S\chi}\rho_{S\gamma} - \rho_{rS}^2\rho_{\chi\gamma} + \rho_{rS}\rho_{r\chi}\rho_{S\gamma} + \rho_{rS}\rho_{r\gamma}\rho_{S\chi}}{\sqrt{1+\rho_{rS}^4 - 2\rho_{rS}^3\rho_{r\chi}\rho_{S\chi} - 2\rho_{rS}^2 + \rho_{rS}^2\rho_{r\chi}^2 + \rho_{rS}^2\rho_{S\chi}^2 - \rho_{r\chi}^2 - \rho_{S\chi}^2 + 2\rho_{rS}\rho_{r\chi}\rho_{S\chi}}},$$

$$\rho''_{\gamma\gamma} = \sqrt{1-\rho_{r\gamma}^2 - \rho''^2_{S\gamma} - \rho''^2_{\chi\gamma}}.$$

*pf.*

$$dW_r dW_\gamma = dW_0\left(\rho_{r\gamma} dW_0 + \rho''_{S\gamma} dW_1 + \rho''_{\chi\gamma} dW_2 + \rho''_{\gamma\gamma} dW_3\right) = \rho_{r\gamma} dt$$

$$\begin{aligned} dW_S dW_\gamma &= \left(\rho_{rS} dW_0 + \sqrt{1-\rho_{rS}^2}\, dW_1\right)\left(\rho_{r\gamma} dW_0 + \rho''_{S\gamma} dW_1 + \rho''_{\chi\gamma} dW_2 + \rho''_{\gamma\gamma} dW_3\right) \\ &= \left(\rho_{rS}\rho_{r\gamma} + \rho''_{S\gamma}\sqrt{1-\rho_{rS}^2}\right) dt = \rho_{S\gamma} dt \end{aligned}$$

Then, $\rho''_{S\gamma} = \dfrac{\rho_{S\gamma} - \rho_{rS}\rho_{r\gamma}}{\sqrt{1-\rho_{rS}^2}}$.

$$\begin{aligned} dW_\chi dW_\gamma &= \left(\rho_{r\chi} dW_0 + \rho'_{S\chi} dW_1 + \rho'_{\chi\chi} dW_2\right)\left(\rho_{r\gamma} dW_0 + \rho''_{S\gamma} dW_1 + \rho''_{\chi\gamma} dW_2 + \rho''_{\gamma\gamma} dW_3\right) \\ &= \left(\rho_{r\chi}\rho_{r\gamma} + \rho'_{S\chi}\rho''_{S\gamma} + \rho'_{\chi\chi}\rho''_{\chi\gamma}\right) dt = \rho_{\chi\gamma} dt \end{aligned}$$

$\rho_{r\chi}\rho_{r\gamma} + \rho'_{S\chi}\rho''_{S\gamma} + \rho'_{\chi\chi}\rho''_{\chi\gamma} = \rho_{\chi\gamma}$, then $\rho''_{\chi\gamma} = \dfrac{\rho_{\chi\gamma} - \rho_{r\chi}\rho_{r\gamma} - \rho'_{S\chi}\rho''_{S\gamma}}{\rho'_{\chi\chi}}$.

$$\rho''_{\chi\gamma} = \frac{\rho_{\chi\gamma} - \rho_{r\chi}\rho_{r\gamma} - \rho_{S\chi}\rho_{S\gamma} - \rho_{rS}^2\rho_{\chi\gamma} + \rho_{rS}\rho_{r\chi}\rho_{S\gamma} + \rho_{rS}\rho_{r\gamma}\rho_{S\chi}}{\sqrt{1+\rho_{rS}^4 - 2\rho_{rS}^3\rho_{r\chi}\rho_{S\chi} - 2\rho_{rS}^2 + \rho_{rS}^2\rho_{r\chi}^2 + \rho_{rS}^2\rho_{S\chi}^2 - \rho_{r\chi}^2 - \rho_{S\chi}^2 + 2\rho_{rS}\rho_{r\chi}\rho_{S\chi}}}$$

$$\begin{aligned} dW_\gamma dW_\gamma &= \left(\rho_{r\gamma} dW_0 + \rho''_{S\gamma} dW_1 + \rho''_{\chi\gamma} dW_2 + \rho''_{\gamma\gamma} dW_3\right)\left(\rho_{r\gamma} dW_0 + \rho''_{S\gamma} dW_1 + \rho''_{\chi\gamma} dW_2 + \rho''_{\gamma\gamma} dW_3\right) \\ &= \left(\rho_{r\gamma}^2 + \rho''^2_{S\gamma} + \rho''^2_{\chi\gamma} + \rho''^2_{\gamma\gamma}\right) dt = dt \end{aligned}$$

$\rho_{r\gamma}^2 + \rho''^2_{S\gamma} + \rho''^2_{\chi\gamma} + \rho''^2_{\gamma\gamma} = 1$, then $\rho''_{\gamma\gamma} = \sqrt{1-\rho_{r\gamma}^2 - \rho''^2_{S\gamma} - \rho''^2_{\chi\gamma}}$.

$$\begin{aligned} \rho''_{\gamma\gamma} &= \sqrt{\frac{1-\rho_{rS}^2 - \rho_{r\gamma}^2 - \rho_{S\gamma}^2 + 2\rho_{rS}\rho_{r\gamma}\rho_{S\gamma}}{1-\rho_{rS}^2} - \frac{\left(\rho_{\chi\gamma} - \rho_{r\chi}\rho_{r\gamma} - \rho_{S\chi}\rho_{S\gamma} - \rho_{rS}^2\rho_{\chi\gamma} + \rho_{rS}\rho_{r\chi}\rho_{S\gamma} + \rho_{rS}\rho_{r\gamma}\rho_{S\chi}\right)^2}{1+\rho_{rS}^4 - 2\rho_{rS}^3\rho_{r\chi}\rho_{S\chi} - 2\rho_{rS}^2 + \rho_{rS}^2\rho_{r\chi}^2 + \rho_{rS}^2\rho_{S\chi}^2 - \rho_{r\chi}^2 - \rho_{S\chi}^2 + 2\rho_{rS}\rho_{r\chi}\rho_{S\chi}}} \\ &= \sqrt{\frac{D}{C}}, \text{ where } C = \left(1-\rho_{rS}^2\right)\left(1+\rho_{rS}^4 - 2\rho_{rS}^3\rho_{r\chi}\rho_{S\chi} - 2\rho_{rS}^2 + \rho_{rS}^2\rho_{r\chi}^2 + \rho_{rS}^2\rho_{S\chi}^2 - \rho_{r\chi}^2 - \rho_{S\chi}^2 + 2\rho_{rS}\rho_{r\chi}\rho_{S\chi}\right) \end{aligned}$$

$$\begin{aligned} &= 1-\rho_{rS}^6 + 2\rho_{rS}^5\rho_{r\chi}\rho_{S\chi} - \rho_{rS}^4\rho_{r\chi}^2 - \rho_{rS}^4\rho_{S\chi}^2 + 3\rho_{rS}^4 - 4\rho_{rS}^3\rho_{r\chi}\rho_{S\chi} + 2\rho_{rS}^2\rho_{r\chi}^2 + 2\rho_{rS}^2\rho_{S\chi}^2 \\ &\quad - 3\rho_{rS}^2 - \rho_{r\chi}^2 - \rho_{S\chi}^2 + 2\rho_{rS}\rho_{r\chi}\rho_{S\chi}, \end{aligned}$$

$$\begin{aligned} D = 1 &- \rho_{rS}^6 + \rho_{rS}^6\rho_{\chi\gamma}^2 - 2\rho_{rS}^5\rho_{r\chi}\rho_{S\gamma}\rho_{\chi\gamma} - 2\rho_{rS}^5\rho_{r\gamma}\rho_{S\chi}\rho_{\chi\gamma} + 2\rho_{rS}^5\rho_{r\chi}\rho_{S\chi} + 2\rho_{rS}^5\rho_{r\gamma}\rho_{S\gamma} \\ &- 2\rho_{rS}^4\rho_{r\chi}\rho_{r\gamma}\rho_{S\chi}\rho_{S\gamma} + 2\rho_{rS}^4\rho_{r\chi}\rho_{r\gamma}\rho_{\chi\gamma} + 2\rho_{rS}^4\rho_{S\chi}\rho_{S\gamma}\rho_{\chi\gamma} + \rho_{rS}^4\rho_{r\chi}^2\rho_{S\gamma}^2 + \rho_{rS}^4\rho_{r\gamma}^2\rho_{S\chi}^2 \\ &- \rho_{rS}^4\rho_{r\chi}^2 - \rho_{rS}^4\rho_{r\gamma}^2 - 3\rho_{rS}^4\rho_{\chi\gamma}^2 - \rho_{rS}^4\rho_{S\chi}^2 - \rho_{rS}^4\rho_{S\gamma}^2 + 3\rho_{rS}^4 + 4\rho_{rS}^3\rho_{r\chi}\rho_{S\gamma}\rho_{\chi\gamma} + 4\rho_{rS}^3\rho_{r\gamma}\rho_{S\chi}\rho_{\chi\gamma} \\ &- 4\rho_{rS}^3\rho_{r\chi}\rho_{S\chi} - 4\rho_{rS}^3\rho_{r\gamma}\rho_{S\gamma} + 4\rho_{rS}^2\rho_{r\chi}\rho_{r\gamma}\rho_{S\gamma}\rho_{S\chi} - 4\rho_{rS}^2\rho_{r\chi}\rho_{r\gamma}\rho_{\chi\gamma} - 4\rho_{rS}^2\rho_{S\chi}\rho_{S\gamma}\rho_{\chi\gamma} \\ &- 2\rho_{rS}^2\rho_{r\chi}^2\rho_{S\gamma}^2 - 2\rho_{rS}^2\rho_{r\gamma}^2\rho_{S\chi}^2 + 2\rho_{rS}^2\rho_{r\chi}^2 + 2\rho_{rS}^2\rho_{r\gamma}^2 + 2\rho_{rS}^2\rho_{S\chi}^2 + 2\rho_{rS}^2\rho_{S\gamma}^2 + 3\rho_{rS}^2\rho_{\chi\gamma}^2 + \rho_{r\chi}^2\rho_{S\gamma}^2 \\ &+ \rho_{r\gamma}^2\rho_{S\chi}^2 - 3\rho_{rS}^2 - \rho_{r\chi}^2 - \rho_{r\gamma}^2 - \rho_{S\chi}^2 - \rho_{S\gamma}^2 - \rho_{\chi\gamma}^2 - 2\rho_{rS}\rho_{r\chi}\rho_{S\gamma}\rho_{\chi\gamma} - 2\rho_{rS}\rho_{r\gamma}\rho_{S\chi}\rho_{\chi\gamma} \\ &- 2\rho_{r\chi}\rho_{r\gamma}\rho_{S\chi}\rho_{S\gamma} + 2\rho_{rS}\rho_{r\chi}\rho_{S\chi} + 2\rho_{rS}\rho_{r\gamma}\rho_{S\gamma} + 2\rho_{r\chi}\rho_{r\gamma}\rho_{\chi\gamma} + 2\rho_{S\chi}\rho_{S\gamma}\rho_{\chi\gamma} \end{aligned}$$

Following this recursive method, we could construct *n* correlated Brownian motions. However, the coefficients for each independent Brownian motions would become more and more tedious. In addition, if two variables are negative linearly correlated, then add a negative sign in front of one of the two variables would make them positive linearly correlated (such that the formulas here work).

Alternatively, if the correlation matrix of $n$ correlated Brownian motions $\mathbf{C}_{n\times n}$ is symmetric and positive definite, we could decompose $\mathbf{C}$ into $\mathbf{L}\mathbf{L}^T$ by Cholesky decomposition, where $\mathbf{L}_{n\times n}$ is a lower triangular matrix and $\mathbf{L}^T$ is the transpose of $\mathbf{L}$. Then we could generate $n$ correlated Brownian motions $\mathbf{B}_{n\times 1}$ by $\mathbf{LW}$, here $\mathbf{W}_{n\times 1}$ is a vector of $n$ independent Brownian motions and $\mathbf{B}\mathbf{B}^T = \mathbf{L}\mathbf{W}\mathbf{W}^T\mathbf{L}^T = \mathbf{C}$.

## 2. Appendix 2: Derive drift term and diffusion term of $P(t,T,r(t))$

In Appendix 2, we derive the drift term $\tilde{\mu}(t,r(t))$ and the diffusion term $\tilde{\sigma}(t,r(t))$ for $P(t,T,r(t))$. Let the process of zero coupon bond of no risk with maturity $T$ be $\frac{dP(t,T,r(t))}{P(t,T,r(t))} = \tilde{\mu}(t,r(t))dt + \tilde{\sigma}(t,r(t))dW_r(t)$. Since $P(t,T,r(t))$ is a function of $t$ and $r(t)$, by Itô formula, we have

$$dP(t,T,r(t)) = \frac{\partial P}{\partial t}dt + \frac{\partial P}{\partial r}dr(t) + \frac{1}{2}\frac{\partial^2 P}{\partial r^2}dr(t)dr(t)$$

where $\frac{\partial P}{\partial t}$ is the first partial derivative of $P(t,T,r(t))$ with respect to $t$, $\frac{\partial P}{\partial r}$ and $\frac{\partial^2 P}{\partial r^2}$ are the first and second partial derivative of $P(t,T,r(t))$ with respect to $r(t)$. Thus, the dynamics of $P(t,T,r(t))$ could also be written as

$$dP(t,T,r(t)) = \left(P_t + \alpha(t,r(t))P_r + \frac{1}{2}P_{rr}\beta^2(t,r(t))\right)dt + P_r\beta(t,r(t))dW_r(t),$$

$P_t = \frac{\partial P}{\partial t}$, $P_r = \frac{\partial P}{\partial r}$, $P_{rr} = \frac{\partial^2 P}{\partial r^2}$. Comparing these two representations of $dP(t,T,r(t))$, we have $P(t,T,r(t))\tilde{\sigma}(t,r(t)) = P_r\beta(t,r(t))$. As a result, we could calculate $\tilde{\sigma}(t,r(t))$ by $\frac{P_r\beta(t,r(t))}{P(t,T,r(t))}$.

Next step is to derive the drift term $\tilde{\mu}(t,r(t))$ in the physical world. $\delta(t)$ equals $e^{-\int_0^t r(s)ds}$, and let $I(t)$ equal $\int_0^t r(s)ds$, and $f(x) = e^{-x}$. Then, $dI(t)$ is equal to $r(t)dt$, $f'(x) = -f(x)$, and $f''(x) = f(x)$. By Itô formula again,

$d\delta(t) = df(I(t)) = f'(I(t))dI(t) + \frac{1}{2}f''(I(t))dI(t)dI(t) = -r(t)\delta(t)dt.$

$d\left[\delta(t)P(t,T,r(t))\right] = \delta(t)dP(t,T,r(t)) + P(t,T,r(t))d\delta(t) + d\delta(t)dP(t,T,r(t))$ by Itô

product rule. We introduce the process of market price of risk $\theta(t)$ here, and let $d\tilde{W}_r(t)=\theta(t)dt+dW_r(t)$. Let $\frac{d\mathbf{Q}'}{d\mathbf{P}}=Z(t)=\mathbf{exp}\left\{-\int_0^t\theta(s)dW_r-\frac{1}{2}\int_0^t\theta^2(s)ds\right\}$, $Z=Z(T)$, and assume $E\left(\int_0^T\theta^2(s)Z^2(s)ds\right)<\infty$. Then by Girsanov's Theorem, $E(Z)=1$, and $\tilde{W}_r(t)$, $0\le t\le T$, under probability measure $\mathbf{Q}'$ is a Brownian motion. Then

$$\begin{aligned}&d\left[\delta(t)P(t,T,r(t))\right]\\&=\left[\tilde{\mu}(t,r(t))-r(t)\right]\delta(t)P(t,T,r(t))dt+\tilde{\sigma}(t,r(t))\delta(t)P(t,T,r(t))dW_r(t)\\&=\tilde{\sigma}(t,r(t))\delta(t)P(t,T,r(t))d\tilde{W}_r(t)\end{aligned} \quad \text{(A2.1)}$$

and $\theta(t)$ equal $\frac{\tilde{\mu}(t,r(t))-r(t)}{\tilde{\sigma}(t,r(t))}$, mathematically speaking.

We rewrite $\frac{dP(t,T,r(t))}{P(t,T,r(t))}=\left(\tilde{\mu}(t,r(t))-\tilde{\sigma}(t,r(t))\theta(t)\right)dt+\tilde{\sigma}(t,r(t))d\tilde{W}_r(t)$.

$P(t,T,r(t))$ is the zero coupon bond of no risk with maturity $T$, so the drift term of $\frac{dP(t,T,r(t))}{P(t,T,r(t))}$ under probability measure $\mathbf{Q}'$ is equal to $r(t)$.[1] Hence, we have $\tilde{\mu}(t,r(t))$ equals $r(t)+\tilde{\sigma}(t,r(t))\theta(t)$.

## 3. APPENDIX 3: DERIVE GENERAL FORM OF DEFLATOR $D(t)$

In Appendix 3, we provide details of the calculations to derive the general form of deflator $D(t)$ step by step.

$$dD(t)=\Omega(t)dt+\Phi(t)dW_r(t)+\Psi(t)dW_1(t)+\Gamma(t)dW_2(t)+\mathrm{I}(t)dW_3(t) \quad \text{(A3.1)}$$

Recall that we would like to have $E^{\mathbf{Q}}\left(\delta(t)X\right)=E\left[D(t)X\right]$ for a nonnegative random variable $X$. We let $D(t)B(t)$, $D(t)P(t,T,r(t))$, $D(t)S(t)$, $D(t)\chi(t)$, and $D(t)\gamma(t)$ be $\mathbf{P}$-martingales, then their drift terms are all equal to zero.

[1] We could verify this by the intuition of change of measure with $\delta(t)$ and also the solution of $\delta(t)P(t,T,r(t))$ being exponential functions. Alternatively, we could plug $\tilde{\mu}(t,r(t))$ back to $d\left[\delta(t)P(t,T,r(t))\right]$ to verify the statement.

A3.1 Short-term saving, $D(t)B(t)$

$$
\begin{aligned}
d\left[D(t)B(t)\right] &= B(t)dD(t)+D(t)dB(t)+dB(t)dD(t) \\
&= \left[\Omega(t)B(t)+D(t)B(t)r(t)\right]dt+\Phi(t)B(t)dW_r(t)+\Psi(t)B(t)dW_1(t)+\Gamma(t)B(t)dW_2(t) \\
&\quad +\mathrm{I}(t)B(t)dW_3(t)
\end{aligned}
$$

Let the drift term of $d\left[D(t)B(t)\right]$ equal zero, $\Omega(t)B(t)+D(t)B(t)r(t)=0$, then $\Omega(t)$ is equal to $-D(t)r(t)$.

A3.2 Zero coupon bond of no risk with maturity $T$, $D(t)P(t,T,r(t))$

$$
\begin{aligned}
d\left[D(t)P(t,T,r(t))\right] &= P(t,T,r(t))dD(t)+D(t)dP(t,T,r(t))+dP(t,T,r(t))dD(t) \\
&= \begin{bmatrix} \Omega(t)P(t,T,r(t))+\Phi(t)P(t,T,r(t))\tilde{\sigma}(t,r(t))+D(t)P(t,T,r(t))r(t) \\ +D(t)P(t,T,r(t))\tilde{\sigma}(t,r(t))\theta(t) \end{bmatrix}dt \\
&\quad +\left[\Phi(t)P(t,T,r(t))+D(t)P(t,T,r(t))\tilde{\sigma}(t,r(t))\right]dW_r(t) \\
&\quad +\Psi(t)P(t,T,r(t))dW_1(t)+\Gamma(t)P(t,T,r(t))dW_2(t)+\mathrm{I}(t)P(t,T,r(t))dW_3(t)
\end{aligned}
$$

Let the drift term of $d\left[D(t)P(t,T,r(t))\right]$ equal zero and plug $\Omega(t)$ into the drift term, we have $\Phi(t)=-D(t)\theta(t)$.

A3.3 Stock, $D(t)S(t)$

$$
\begin{aligned}
d\left[D(t)S(t)\right] &= S(t)dD(t)+D(t)dS(t)+dD(t)dS(t) \\
&= \left[\Omega(t)S(t)+\Phi(t)S(t)\sigma_S(t)\rho_{rS}+\Psi(t)S(t)\sigma_S\sqrt{1-\rho_{rS}^2}+D(t)S(t)\mu_S(t)\right]dt \\
&\quad +\left[\Phi(t)S(t)+D(t)S(t)\sigma_S(t)\rho_{rS}\right]dW_r(t)+\left[\Psi(t)S(t)+D(t)S(t)\sigma_S(t)\sqrt{1-\rho_{rS}^2}\right]dW_1(t) \\
&\quad +\Gamma(t)S(t)dW_2(t)+\mathrm{I}(t)S(t)dW_3(t)
\end{aligned}
$$

Let the drift term of $d\left[D(t)S(t)\right]$ equal zero and plug $\Omega(t)$, $\Phi(t)$ into the drift term, then $\Psi(t)$ is equal to $\dfrac{D(t)\left[r(t)+\theta(t)\sigma_S(t)\rho_{rS}-\mu_S(t)\right]}{\sigma_S(t)\sqrt{1-\rho_{rS}^2}}$.

A3.4 Default density, $D(t)\chi(t)$

$$
\begin{aligned}
d\left[D(t)\chi(t)\right] &= \chi(t)dD(t)+D(t)d\chi(t)+d\chi(t)dD(t) \\
&= \left[\Omega(t)\chi(t)+\Phi(t)\sigma_\chi\rho_{r\chi}\sqrt{\chi(t)}+\Psi(t)\sigma_\chi\rho'_{S\chi}\sqrt{\chi(t)}+\Gamma(t)\sigma_\chi\rho'_{\chi\chi}\sqrt{\chi(t)}+D(t)e-D(t)f\chi(t)\right]dt \\
&\quad +\left[\Phi(t)\chi(t)+D(t)\sigma_\chi\rho_{r\chi}\sqrt{\chi(t)}\right]dW_r(t)+\left[\Psi(t)\chi(t)+D(t)\sigma_\chi\rho'_{S\chi}\sqrt{\chi(t)}\right]dW_1(t) \\
&\quad +\left[\Gamma(t)\chi(t)+D(t)\sigma_\chi\rho'_{\chi\chi}\sqrt{\chi(t)}\right]dW_2(t)+\mathrm{I}(t)\chi(t)dW_3(t)
\end{aligned}
$$

Let the drift term of $d\left[D(t)\chi(t)\right]$ equal zero and plug $\Omega(t)$, $\Phi(t)$, $\Psi(t)$ into the drift term, we have

$$\Gamma(t)=D(t)\left\{\frac{\theta(t)\rho_{r\chi}}{\rho'_{\chi\chi}}+\frac{r(t)\chi(t)-e+f\chi(t)}{\sigma_\chi\rho'_{\chi\chi}\sqrt{\chi(t)}}+\frac{\rho'_{S\chi}\left[\mu_S(t)-r(t)-\theta(t)\sigma_S(t)\rho_{rS}\right]}{\rho'_{\chi\chi}\sigma_S(t)\sqrt{1-\rho_{rS}^2}}\right\}.$$

A3.5 Convenience yield, $D(t)\gamma(t)$

$$\begin{aligned}d\left[D(t)\gamma(t)\right]&=\gamma(t)dD(t)+D(t)d\gamma(t)+d\gamma(t)dD(t)\\&=\left[\Omega(t)\gamma(t)+\Phi(t)\eta\rho_{r\gamma}+\Psi(t)\eta\rho''_{S\gamma}+\Gamma(t)\eta\rho''_{\chi\gamma}+\mathrm{I}(t)\eta\rho''_{\gamma\gamma}\right]dt+\left[\Phi(t)\gamma(t)+D(t)\eta\rho_{r\gamma}\right]dW_r(t)\\&\quad+\left[\Psi(t)\gamma(t)+D(t)\eta\rho''_{S\gamma}\right]dW_1(t)+\left[\Gamma(t)\gamma(t)+D(t)\eta\rho''_{\chi\gamma}\right]dW_2(t)\\&\quad+\left(\mathrm{I}(t)\gamma(t)+D(t)\eta\rho''_{\gamma\gamma}\right)dW_3(t)\end{aligned}$$

Let the drift term of $d\left[D(t)\gamma(t)\right]$ equal zero and plug $\Omega(t)$, $\Phi(t)$, $\Psi(t)$, $\Gamma(t)$ into the drift term, then

$$\mathrm{I}(t)=D(t)\left\{\begin{aligned}&\frac{\rho_{r\gamma}\theta(t)}{\rho''_{\gamma\gamma}}+\frac{r(t)\gamma(t)}{\eta\rho''_{\gamma\gamma}}-\frac{\rho''_{\chi\gamma}\rho_{r\chi}\theta(t)}{\rho''_{\gamma\gamma}\rho'_{\chi\chi}}+\frac{\rho''_{\chi\gamma}\left[e-r(t)\chi(t)-f\chi(t)\right]}{\rho''_{\gamma\gamma}\rho'_{\chi\chi}\sigma_\chi\sqrt{\chi(t)}}\\&+\frac{\left(\rho''_{S\gamma}\rho'_{\chi\chi}-\rho''_{\chi\gamma}\rho'_{S\chi}\right)\left[\mu_S(t)-r(t)-\rho_{rS}\theta(t)\sigma_S(t)\right]}{\rho''_{\gamma\gamma}\rho'_{\chi\chi}\sigma_S(t)\sqrt{\left(1-\rho_{rS}^2\right)}}\end{aligned}\right\}.$$

A3.6 Deflator, $D(t)$

We could derive the general form of deflator $D(t)$ now. Let $y(t)$ equal $\ln D(t)$, then the first partial derivative ($\frac{\partial y}{\partial D}$) and the second partial derivative ($\frac{\partial^2 y}{\partial D^2}$) of $y(t)$ with respect to $D(t)$ are $\frac{1}{D(t)}$ and $-\frac{1}{D^2(t)}$ respectively. By Itô formula, $dy(t)=\frac{\partial y}{\partial D}dD(t)+\frac{1}{2}\frac{\partial^2 y}{\partial D^2}dD(t)dD(t)$.

$$\begin{aligned}dy(t)&=\left[\frac{\Omega(t)}{D(t)}-\frac{\Phi^2(t)}{2D^2(t)}-\frac{\Psi^2(t)}{2D^2(t)}-\frac{\Gamma^2(t)}{2D^2(t)}-\frac{\mathrm{I}^2(t)}{2D^2(t)}\right]dt+\frac{\Phi(t)}{D(t)}dW_r(t)\\&\quad+\frac{\Psi(t)}{D(t)}dW_1(t)+\frac{\Gamma(t)}{D(t)}dW_2(t)+\frac{\mathrm{I}(t)}{D(t)}dW_3(t)\end{aligned}\qquad\text{(A3.2)}$$

Let
$$\begin{cases} K_\Psi(t)=\dfrac{\Psi(t)}{D(t)}=\dfrac{r(t)+\theta(t)\sigma_S(t)\rho_{rS}-\mu_S(t)}{\sigma_S(t)\sqrt{1-\rho_{rS}^2}} \\ K_\Gamma(t)=\dfrac{\Gamma(t)}{D(t)}=\dfrac{\theta(t)\rho_{r\chi}}{\rho'_{\chi\chi}}+\dfrac{r(t)\chi(t)-e+f\chi(t)}{\sigma_\chi\rho'_{\chi\chi}\sqrt{\chi(t)}}+\dfrac{\rho'_{S\chi}\left[\mu_S(t)-r(t)-\theta(t)\sigma_S(t)\rho_{rS}\right]}{\rho'_{\chi\chi}\sigma_S(t)\sqrt{1-\rho_{rS}^2}} \\ K_\mathrm{I}(t)=\dfrac{\mathrm{I}(t)}{D(t)}=\dfrac{\rho_{r\gamma}\theta(t)}{\rho''_{\gamma\gamma}}+\dfrac{r(t)\gamma(t)}{\eta\rho''_{\gamma\gamma}}-\dfrac{\rho''_{\chi\gamma}\rho_{r\chi}\theta(t)}{\rho''_{\gamma\gamma}\rho'_{\chi\chi}}+\dfrac{\rho''_{\chi\gamma}\left[e-r(t)\chi(t)-f\chi(t)\right]}{\rho''_{\gamma\gamma}\rho'_{\chi\chi}\sigma_\chi\sqrt{\chi(t)}} \\ \qquad+\dfrac{\left(\rho''_{S\gamma}\rho'_{\chi\chi}-\rho''_{\chi\gamma}\rho'_{S\chi}\right)\left[\mu_S(t)-r(t)-\rho_{rS}\theta(t)\sigma_S(t)\right]}{\rho''_{\gamma\gamma}\rho'_{\chi\chi}\sigma_S(t)\sqrt{1-\rho_{rS}^2}} \end{cases}$$
, and

integrate both sides of $dy(t)$, plugging $\frac{\Omega(t)}{D(t)}=-r(t)$ and $\frac{\Phi(t)}{D(t)}=-\theta(t)$ into $y(t)$. We have $y(t)$ as follows.

$$\begin{aligned} y(t)=y(0)&+\int_0^t\left[-r(s)-\frac{1}{2}\theta^2(s)-\frac{1}{2}K_\Psi^2(s)-\frac{1}{2}K_\Gamma^2(s)-\frac{1}{2}K_\mathrm{I}^2(s)\right]ds \\ &-\int_0^t\theta(s)dW_r(t)+\int_0^t K_\Psi(s)dW_1(s)+\int_0^t K_\Gamma(s)dW_2(s)+\int_0^t K_\mathrm{I}(s)dW_3(s) \end{aligned} \tag{A3.3}$$

Take exponential both sides of $y(t)$, we have the general form of deflator.

$$\begin{aligned} D(t)=D(0)\mathbf{exp}&\left\{-\int_0^t r(s)ds-\int_0^t\frac{1}{2}\left[\theta^2(s)+K_\Psi^2(s)+K_\Gamma^2(s)+K_\mathrm{I}^2(s)\right]ds\right\} \\ &\times\mathbf{exp}\left[-\int_0^t\theta(s)dW_r(s)+\int_0^t K_\Psi(s)dW_1(s)+\int_0^t K_\Gamma(s)dW_2(s)+\int_0^t K_\mathrm{I}(s)dW_3(s)\right] \end{aligned} \tag{A3.4}$$

## 4. Appendix 4: Regularity conditions for deflator $D(t)$

In Appendix 4, we provide an analysis for regularity conditions regarding the drift coefficient $\mu_S(t)$ of stock prices, $e$ in the drift term of default density $\chi(t)$, and the diffusion coefficient $\eta$ of convenience yield $\gamma(t)$. In Appendix 3, we derive the general form of deflator $D(t)$ by letting $D(t)B(t)$, $D(t)P(t,T,r(t))$, $D(t)S(t)$, $D(t)\chi(t)$, and $D(t)\gamma(t)$ be $\mathbf{P}$ -martingales. However, recall that we'd like to have $E^{\mathbf{Q}}(\delta(t)X)=E[D(t)X]$ for a nonnegative random variable $X$, in which $\delta(t)$ is a discount process equalling $e^{-\int_0^t r(s)ds}$. In our model, $B(t)$, $P(t,T,r(t))$, and $S(t)$ are the nonnegative random variables $X$ in $E^{\mathbf{Q}}(\delta(t)X)=E[D(t)X]$. In addition, we also have to consider $\chi(t)$ and $\gamma(t)$ under probability measure $\mathbf{Q}$ when deriving the deflator $D(t)$. We discuss $\delta(t)B(t)$, $\delta(t)P(t,T,r(t))$, $\delta(t)S(t)$, $\delta(t)\chi(t)$, and $\delta(t)\gamma(t)$ under probability measure $\mathbf{Q}$ as follows.

A4.1 Short-term saving, $\delta(t)B(t)$

$$d\left[\delta(t)B(t)\right]=\delta(t)dB(t)+B(t)d\delta(t)+d\delta(t)dB(t)=0$$

Thus, $\delta(t)B(t)$ is a $\mathbf{Q}$-martingale.

A4.2 Zero coupon bond of no risk with maturity $T$, $\delta(t)P(t,T,r(t))$

From (A2.1), $d\left[\delta(t)P(t,T,r(t))\right]=\tilde{\sigma}(t,r(t))\delta(t)P(t,T,r(t))d\tilde{W}_r(t)$.

Thus, $\delta(t)P(t,T,r(t))$ is a $\mathbf{Q}$-martingale.

A4.3 Stock, $\delta(t)S(t)$

$$\begin{aligned}d\left[\delta(t)S(t)\right]&=\delta(t)dS(t)+S(t)d\delta(t)+d\delta(t)dS(t)\\&=\delta(t)S(t)\left[\mu_S(t)-r(t)\right]dt+\delta(t)S(t)\sigma_S(t)\rho_{rS}dW_r(t)\\&\quad+\delta(t)S(t)\sigma_S(t)\sqrt{1-\rho_{rS}^2}dW_1(t)\\&=\delta(t)S(t)\left[\mu_S(t)-r(t)-\theta(t)\sigma_S(t)\rho_{rS}\right]dt+\delta(t)S(t)\sigma_S(t)\rho_{rS}d\tilde{W}_r(t)\\&\quad+\delta(t)S(t)\sigma_S(t)\sqrt{1-\rho_{rS}^2}dW_1(t),\ \text{under } \mathbf{Q}\end{aligned}$$

We require $\mu_S(t)=r(t)+\theta(t)\sigma_S(t)\rho_{rS}$ as regularity condition, such that $\delta(t)S(t)$ is a $\mathbf{Q}$-martingale. As a result, $K_\Psi(t)=0$.

Alternatively, let $\delta_S(t)$ be a process equal to $e^{-\int_0^t\left[\mu_S(s)-r(s)-\theta(s)\sigma_S(s)\rho_{rS}\right]ds}$. We could see that $\delta(t)\left[\delta_S(t)S(t)\right]$ is a $\mathbf{Q}$-martingale as follows.

Let $\delta_S(t)=e^{-\int_0^t\left[\mu_S(s)-r(s)-\theta(s)\sigma_S(s)\rho_{rS}\right]ds}$, $I_S(t)=\int_0^t\left[\mu_S(s)-r(s)-\theta(s)\sigma_S(s)\rho_{rS}\right]ds$, and $f(x)=e^{-x}$,
then $dI_S(t)=\left[\mu_S(t)-r(t)-\theta(t)\sigma_S(t)\rho_{rS}\right]dt$, $f'(x)=-f(x)$, and $f''(x)=f(x)$.

$$d\delta_S(t)=df\left(I_S(t)\right)=f'\left(I_S(t)\right)dI_S(t)+\frac{1}{2}f''\left(I_S(t)\right)dI_S(t)dI_S(t)=-\left[\mu_S(t)-r(t)-\theta(t)\sigma_S(t)\rho_{rS}\right]\delta_S(t)dt$$

$$\begin{aligned}d\left[\delta_S(t)S(t)\right]&=\delta_S(t)dS(t)+S(t)d\delta_S(t)+d\delta_S(t)dS(t)\\&=\delta_S(t)S(t)\left[r(t)+\theta(t)\sigma_S(t)\rho_{rS}\right]dt+\delta_S(t)S(t)\sigma_S(t)\rho_{rS}dW_r(t)+\delta_S(t)S(t)\sigma_S(t)\sqrt{1-\rho_{rS}^2}dW_1(t)\end{aligned}$$

$$\begin{aligned}d\left\{\delta(t)\left[\delta_S(t)S(t)\right]\right\}&=\delta(t)d\left[\delta_S(t)S(t)\right]+\delta_S(t)S(t)d\delta(t)+d\delta(t)d\left[\delta_S(t)S(t)\right]\\&=\delta(t)\delta_S(t)S(t)\theta(t)\sigma_S(t)\rho_{rS}dt+\delta(t)\delta_S(t)S(t)\sigma_S(t)\rho_{rS}dW_r(t)+\delta(t)\delta_S(t)S(t)\sigma_S(t)\sqrt{1-\rho_{rS}^2}dW_1(t)\\&=\delta(t)\delta_S(t)S(t)\sigma_S(t)\rho_{rS}d\tilde{W}_r(t)+\delta(t)\delta_S(t)S(t)\sigma_S(t)\sqrt{1-\rho_{rS}^2}dW_1(t),\ \text{under } \mathbf{Q}\end{aligned}$$

Let $D(t)\left[\delta_S(t)S(t)\right]$ be a $\mathbf{P}$-martingale, then $\Psi(t)=0$ (so that $K_\Psi(t)=0$).

$$\begin{aligned}d\left\{D(t)\left[\delta_S(t)S(t)\right]\right\}&=\delta_S(t)S(t)dD(t)+D(t)d\left[\delta_S(t)S(t)\right]+dD(t)d\left[\delta_S(t)S(t)\right]\\&=\delta_S(t)S(t)\left\{D(t)\left[r(t)+\theta(t)\sigma_S(t)\rho_{rS}\right]+\Omega(t)+\Phi(t)\sigma_S(t)\rho_{rS}+\Psi(t)\sigma_S(t)\sqrt{1-\rho_{rS}^2}\right\}dt\\&\quad+\delta_S(t)S(t)\left[\Phi(t)+D(t)\sigma_S(t)\rho_{rS}\right]dW_r(t)+\delta_S(t)S(t)\left[\Psi(t)+D(t)\sigma_S(t)\sqrt{1-\rho_{rS}^2}\right]dW_1(t)\\&\quad+\delta_S(t)S(t)\Gamma(t)dW_2(t)+\delta_S(t)S(t)\mathrm{I}(t)dW_3(t)\end{aligned}$$

Plug $\Omega(t),\Phi(t)$ into $D(t)\left[r(t)+\theta(t)\sigma_S(t)\rho_{rS}\right]+\Omega(t)+\Phi(t)\sigma_S(t)\rho_{rS}+\Psi(t)\sigma_S(t)\sqrt{1-\rho_{rS}^2}=0$, then $\Psi(t)=0$.

A4.4 Default density, $\delta(t)\chi(t)$

$$\begin{aligned} d\left[\delta(t)\chi(t)\right] &= \delta(t)d\chi(t)+\chi(t)d\delta(t)+d\delta(t)d\chi(t) \\ &= \delta(t)\left[e-r(t)\chi(t)-f\chi(t)\right]dt+\sigma_\chi\rho_{r\chi}\delta(t)\sqrt{\chi(t)}dW_r(t) \\ &\quad +\sigma_\chi\rho'_{S\chi}\delta(t)\sqrt{\chi(t)}dW_1(t)+\sigma_\chi\rho'_{\chi\chi}\delta(t)\sqrt{\chi(t)}dW_2(t) \\ &= \delta(t)\left[e-r(t)\chi(t)-f\chi(t)-\sigma_\chi\rho_{r\chi}\theta(t)\sqrt{\chi(t)}\right]dt+\sigma_\chi\rho_{r\chi}\delta(t)\sqrt{\chi(t)}d\tilde{W}_r(t) \\ &\quad +\sigma_\chi\rho'_{S\chi}\delta(t)\sqrt{\chi(t)}dW_1(t)+\sigma_\chi\rho'_{\chi\chi}\delta(t)\sqrt{\chi(t)}dW_2(t), \text{ under } \mathbf{Q} \end{aligned}$$

We require $e=r(t)\chi(t)+f\chi(t)+\sigma_\chi\rho_{r\chi}\theta(t)\sqrt{\chi(t)}$ as regularity condition, such that $\delta(t)\chi(t)$ is a $\mathbf{Q}$-martingale. As a result, $K_\Gamma(t)=0$. In addition, we also have to consider if the Feller condition holds, i.e. $r(t)\chi(t)+f\chi(t)+\sigma_\chi\rho_{r\chi}\theta(t)\sqrt{\chi(t)}>\frac{1}{2}\sigma_\chi^2$.

A4.5 Convenience yield, $\delta(t)\gamma(t)$

$$\begin{aligned} d\left[\delta(t)\gamma(t)\right] &= \delta(t)d\gamma(t)+\gamma(t)d\delta(t)+d\delta(t)d\gamma(t) \\ &= -\delta(t)\gamma(t)r(t)dt+\eta\rho_{r\gamma}\delta(t)dW_r(t)+\eta\rho''_{S\gamma}\delta(t)dW_1(t) \\ &\quad +\eta\rho''_{\chi\gamma}\delta(t)dW_2(t)+\eta\rho''_{\gamma\gamma}\delta(t)dW_3(t) \\ &= \delta(t)\left[-\gamma(t)r(t)-\eta\rho_{r\gamma}\theta(t)\right]dt+\eta\rho_{r\gamma}\delta(t)d\tilde{W}_r(t)+\eta\rho''_{S\gamma}\delta(t)dW_1(t) \\ &\quad +\eta\rho''_{\chi\gamma}\delta(t)dW_2(t)+\eta\rho''_{\gamma\gamma}\delta(t)dW_3(t), \text{ under } \mathbf{Q} \end{aligned}$$

We require $\eta=-\frac{\gamma(t)r(t)}{\rho_{r\gamma}\theta(t)}$ as regularity condition, such that $\delta(t)\gamma(t)$ is a $\mathbf{Q}$-martingale. As a result, $K_I(t)=0$.

From (A3.4), we rewrite the general form of deflator as

$$D(t)=D(0)\mathbf{exp}\left[-\int_0^t r(s)ds-\frac{1}{2}\int_0^t\theta^2(s)ds-\int_0^t\theta(s)dW_r(s)\right].$$

## 5. APPENDIX 5: EUROPEAN PUT OPTION PRICING UNDER CIR INTEREST RATE IN KIM (2002)

In Kim (2002), the process of CIR interest rate $dr(t)$ under probability measure $\mathbf{Q}'$ is as follows.

$$dr(t)=\left[\kappa_{Kim}\theta_{Kim}-\left(\kappa_{Kim}+\delta_{Kim}\lambda_{Kim}\right)r(t)\right]dt+\delta_{Kim}\sqrt{r(t)}d\tilde{W}_r(t) \qquad \text{(A5.1)}$$

Let $\lambda_{Kim}=1$, then $\delta_{Kim}=\sigma_r$, $\kappa_{Kim}=b_r-\sigma_r$, $\theta_{Kim}=a_r/(b_r-\sigma_r)$. Then, we could calculate the price of a European call option of stock S with strike $K$ and maturity $T$ at time zero, $Call_{Kim}(0,S(0),T,K)$.

$$
\begin{aligned}
Call_{Kim}\left(0,S(0),T,K\right) &= \left[S(0)\Phi(d_1) - K\exp\left(-\int_0^T r_t^* dt\right)\Phi(d_2)\right] \\
&+ \delta_{Kim} C_0 \left[S(0)\phi(d_1) - K\exp\left(-\int_0^T r_t^* dt\right)\left(\phi(d_2) - \sigma_S\sqrt{T}\Phi(d_2)\right)\right] \qquad (17) \\
&+ \delta_{Kim} C_1 \left[d_2 S(0)\phi(d_1) - d_1 K\exp\left(-\int_0^T r_t^* dt\right)\phi(d_2)\right] + o(\delta_{Kim}),
\end{aligned}
$$

where $\Phi(\cdot)$ and $\phi(\cdot)$ are the cumulative density function and probability density function of the standard normal distribution respectively;

$$
r_t^* = r_0 e^{-\kappa_{Kim}t} + \theta_{Kim}\left(1-e^{-\kappa_{Kim}t}\right),\ \exp\left(-\int_0^T r_t^* dt\right) = \exp\left[-\frac{(r_0-\theta_{Kim})}{\kappa_{Kim}}\left(1-e^{-\kappa_{Kim}T}\right) - \theta_{Kim}T\right];
$$

$$
d_1 = \frac{1}{\sigma_S\sqrt{T}}\left[\ln\frac{S(0)}{K} + \frac{(r_0-\theta_{Kim})}{\kappa_{Kim}}\left(1-e^{-\kappa_{Kim}T}\right) + \theta_{Kim}T + \frac{\sigma_S^2}{2}T\right],\ d_2 = d_1 - \sigma_S\sqrt{T};
$$

$$
C_0 = \frac{1}{\kappa_{Kim}\sigma_S\sqrt{T}}\left[(r_0-\theta_{Kim})\left(\frac{1-e^{-\kappa_{Kim}T}}{\kappa_{Kim}} - Te^{-\kappa_{Kim}T}\right) + \theta_{Kim}T\left(1-\frac{1-e^{-\kappa_{Kim}T}}{\kappa_{Kim}}\right)\right];
$$

$$
C_1 = -\frac{\rho_{rS}}{\sigma_S T}C_{11},\ C_{11} = \frac{2\sqrt{\theta_{Kim}}\left[\left(1+2e^{\kappa_{Kim}T}\right)\sqrt{r_0} - 3e^{\frac{\kappa_{Kim}T}{2}}\sqrt{r_0-\theta_{Kim}\left(1-e^{-\kappa_{Kim}T}\right)}\right] + \psi_{Kim}\left[\theta_{Kim}\left(1+2e^{\kappa_{Kim}T}\right) - r_0\right]}{2e^{\kappa_{Kim}T}\kappa_{Kim}^2\sqrt{\theta_{Kim}}},
$$

$$
\psi_{Kim} = \ln\left[\frac{\theta_{Kim}\left(2e^{\kappa_{Kim}T}-1\right) + r_0 + 2e^{\frac{\kappa_{Kim}T}{2}}\sqrt{\theta_{Kim}^2\left(e^{\kappa_{Kim}T}-1\right) + \theta_{Kim}r_0}}{\left(\sqrt{r_0}+\sqrt{\theta_{Kim}}\right)^2}\right].
$$

By Put-Call parity, $Call\left(0,S(0),T,K\right) + Ke^{-\int_0^T r_u du} = Put\left(0,S(0),T,K\right) + S(0)$. We could then calculate the price of the European put option at time zero $Put_{Kim}\left(0,S(0),T,K\right)$ as $Call_{Kim}\left(0,S(0),T,K\right) + KP\left(0,T,r(0)\right) - S(0)$.

## 6. APPENDIX 6: IMPLEMENTATION OF SIMPLIFIED SECOND MILSTEIN METHOD

In Appendix 6, we present the implementation of simplified Second Milstein method in our example. $dX_t = a(t,X_t)dt + b(t,X_t)dW_t$ where $X_t$, $W_t$, $a(t,X_t)$, and $b(t,X_t)$ are as follows.

$$
X_t = \begin{bmatrix} r(t) \\ \theta(t) \\ B(t) \\ P\left(t,T,r(t)\right) \\ S(t) \\ \chi(t) \\ \gamma(t) \\ D(t) \end{bmatrix}, W_t = \begin{bmatrix} W_r(t) \\ W_1(t) \\ W_2(t) \\ W_3(t) \\ W_\theta(t) \end{bmatrix} \qquad (A6.1)
$$

$$a(t,X_t)=\begin{bmatrix} a_r-b_r r(t)+\theta(t)\sigma_r\sqrt{r(t)} \\ a_\theta-b_\theta\theta(t) \\ B(t)r(t) \\ P(t,T,r(t))r(t)+P_r\sigma_r\sqrt{r(t)}\theta(t) \\ S(t)\left[r(t)+\theta(t)\sigma_S(t)\rho_{rS}\right] \\ r(t)\chi(t)+\sigma_\chi\rho_{r\chi}\theta(t)\sqrt{\chi(t)} \\ 0 \\ -r(t)D(t) \end{bmatrix} \quad \text{(A6.2)}$$

$$b(t,X_t)=\begin{bmatrix} \sigma_r\sqrt{r(t)} & 0 & 0 & 0 & 0 \\ 0 & 0 & 0 & 0 & \sigma_\theta\sqrt{\theta(t)} \\ 0 & 0 & 0 & 0 & 0 \\ P_r\sigma_r\sqrt{r(t)} & 0 & 0 & 0 & 0 \\ S(t)\sigma_S(t)\rho_{rS} & S(t)\sigma_S(t)\sqrt{1-\rho_{rS}^2} & 0 & 0 & 0 \\ \sigma_\chi\rho_{r\chi}\sqrt{\chi(t)} & \sigma_\chi\rho'_{S\chi}\sqrt{\chi(t)} & \sigma_\chi\rho'_{\chi\chi}\sqrt{\chi(t)} & 0 & 0 \\ -\dfrac{\gamma(t)r(t)}{\theta(t)} & -\dfrac{\rho''_{S\gamma}\gamma(t)r(t)}{\rho_{r\gamma}\theta(t)} & -\dfrac{\rho''_{\chi\gamma}\gamma(t)r(t)}{\rho_{r\gamma}\theta(t)} & -\dfrac{\rho''_{\gamma\gamma}\gamma(t)r(t)}{\rho_{r\gamma}\theta(t)} & 0 \\ -\theta(t)D(t) & 0 & 0 & 0 & 0 \end{bmatrix} \quad \text{(A6.3)}$$

For each $i=1,\ldots,d$ ,

$$\begin{aligned} Y_{n+1,i} &= Y_{n,i}+a_i(n,Y_n)\Delta t+\sum_{k=1}^{m}b_{ik}(n,Y_n)\Delta W_{n,k}+\frac{1}{2}L^0a_i(n,Y_n)(\Delta t)^2 \\ &+\frac{1}{2}\sum_{k=1}^{m}\left[L^k a_i(n,Y_n)+L^0b_{ik}(n,Y_n)\right]\Delta W_{n,k}\Delta t+\frac{1}{2}\sum_{k=1}^{m}\sum_{j=1}^{m}L^j b_{ik}(n,Y_n)\left(\Delta W_{n,j}\Delta W_{n,k}-V_{jk}\right). \end{aligned}$$

(A6.4)

In order to calculate $Y_{n+1}$ at each time step $n+1$, we have to calculate operators $L^0$ and $L^k$ for each $a_i(n,X_n)$ and $b_{ik}(n,X_n)$. We present two illustrative examples below as follows.

A6.1 $a_1(t, X_t)$

$$a_1(t, X_t) = a_r - b_r r(t) + \theta(t)\sigma_r\sqrt{r(t)},\quad \frac{\partial a_1(t, X_t)}{\partial x_1} = -b_r + \frac{1}{2}\theta(t)\sigma_r\frac{1}{\sqrt{r(t)}},\quad \frac{\partial a_1(t, X_t)}{\partial x_2} = \sigma_r\sqrt{r(t)}$$

$$\frac{\partial^2 a_1(t, X_t)}{\partial x_1 \partial x_1} = -\frac{1}{4}\theta(t)\sigma_r\frac{1}{\sqrt{r^3(t)}},\quad \frac{\partial^2 a_1(t, X_t)}{\partial x_1 \partial x_2} = \frac{\partial^2 a_1(t, X_t)}{\partial x_2 \partial x_1} = \frac{1}{2}\sigma_r\frac{1}{\sqrt{r(t)}}$$

$$\begin{aligned} L^0 a_1(t, X_t) &= \frac{\partial a_1(t, X_t)}{\partial t} + \sum_{i=1}^{8} a_i(t, X_t)\frac{\partial a_1(t, X_t)}{\partial x_i} + \frac{1}{2}\sum_{i,j=1}^{8}\Sigma_{t,ij}\frac{\partial^2 a_1(t, X_t)}{\partial x_i \partial x_j} \\ &= a_1(t, X_t)\left[-b_r + \frac{1}{2}\theta(t)\sigma_r\frac{1}{\sqrt{r(t)}}\right] + a_2(t, X_t)\sigma_r\sqrt{r(t)} \\ &\quad -\frac{1}{8}\Sigma_{t,11}\sigma_r\theta(t)\frac{1}{\sqrt{r^3(t)}} + \frac{1}{2}\Sigma_{t,12}\sigma_r\frac{1}{\sqrt{r(t)}} \end{aligned}$$

$$\begin{aligned} L^k a_1(t, X_t) &= \sum_{i=1}^{d} b_{ik}(t, X_t)\frac{\partial a_1(t, X_t)}{\partial x_i} \\ &= b_{1k}(t, X_t)\left[-b_r + \frac{1}{2}\theta(t)\sigma_r\frac{1}{\sqrt{r(t)}}\right] + b_{2k}(t, X_t)\sigma_r\sqrt{r(t)} \end{aligned}$$

A6.2 $b_{11}(t, X_t)$

$$b_{11}(t, X_t) = \sigma_r\sqrt{r(t)},\quad \frac{\partial b_{11}(t, X_t)}{\partial x_1} = \frac{1}{2}\sigma_r\frac{1}{\sqrt{r(t)}}$$

$$\frac{\partial^2 b_{11}(t, X_t)}{\partial x_1 \partial x_1} = -\frac{1}{4}\sigma_r\frac{1}{\sqrt{r^3(t)}}$$

$$\begin{aligned} L^0 b_{11}(t, X_t) &= \frac{\partial b_{11}(t, X_t)}{\partial t} + \sum_{i=1}^{8} a_i(t, X_t)\frac{\partial b_{11}(t, X_t)}{\partial x_i} + \frac{1}{2}\sum_{i,j=1}^{8}\Sigma_{t,ij}\frac{\partial^2 b_{11}(t, X_t)}{\partial x_i \partial x_j} \\ &= \frac{1}{2}a_1(t, X_t)\sigma_r\frac{1}{\sqrt{r(t)}} - \frac{1}{8}\Sigma_{t,11}\sigma_r\frac{1}{\sqrt{r^3(t)}} \end{aligned}$$

$$\begin{aligned} L^k b_{11}(t, X_t) &= \sum_{i=1}^{d} b_{ik}(t, X_t)\frac{\partial b_{11}(t, X_t)}{\partial x_i} \\ &= \frac{1}{2}b_{1k}(t, X_t)\sigma_r\frac{1}{\sqrt{r(t)}} \end{aligned}$$

## 7. APPENDIX 7: CORPORATE COUPON BOND PRICING IN LONGSTAFF ET AL. (2005)

In Longstaff et al. (2005), the formula to calculate price of a corporate coupon bond is as follows.

$$CB(c,\omega,T)=c\int_0^T A_{CB}(t)\exp\left(B_{CB}(t)\chi_0\right)C_{CB}(t)P(0,t,r(0))e^{-\gamma_0 t}dt$$
$$+A_{CB}(T)\exp\left(B_{CB}(T)\chi_0\right)C_{CB}(T)P(0,T,r(0))e^{-\gamma_0 T}$$
$$+(1-\omega)\int_0^T \exp\left(B_{CB}(t)\chi_0\right)C_{CB}(t)P(0,t,r(0))\left[G_{CB}(t)+H_{CB}(t)\chi_0\right]e^{-\gamma_0 t}dt$$

$$A_{CB}(t)=\exp\left[\frac{e_\chi\left(f_\chi+\phi\right)}{\sigma_\chi^2}t\right]\left(\frac{1-\kappa}{1-\kappa e^{\phi t}}\right)^{\frac{2e_\chi}{\sigma_\chi^2}},\ B_{CB}(t)=\frac{f_\chi-\phi}{\sigma_\chi^2}+\frac{2\phi}{\sigma_\chi^2\left(1-\kappa e^{\phi t}\right)},$$

$$C_{CB}(t)=\exp\left(\frac{\eta^2 t^3}{6}\right),\ G_{CB}(t)=\frac{e_\chi}{\phi}\left(e^{\phi t}-1\right)\exp\left[\frac{e_\chi\left(f_\chi+\phi\right)}{\sigma_\chi^2}t\right]\left(\frac{1-\kappa}{1-\kappa e^{\phi t}}\right)^{\frac{2e_\chi}{\sigma_\chi^2}+1},$$

$$H_{CB}(t)=\exp\left[\frac{e_\chi\left(f_\chi+\phi\right)+\phi\sigma_\chi^2}{\sigma_\chi^2}t\right]\left(\frac{1-\kappa}{1-\kappa e^{\phi t}}\right)^{\frac{2e_\chi}{\sigma_\chi^2}+2},\ \kappa=\left(f_\chi+\phi\right)/\left(f_\chi-\phi\right),$$

$$\phi=\sqrt{2\sigma_\chi^2+f_\chi^2}$$

In our numerical example, $e_\chi$ is equal to $r(0)\chi(0)+f\chi(0)+\sigma_\chi\rho_{r\chi}\theta(0)\sqrt{\chi(0)}$ and $f_\chi$ equals $f$, in which $f$ is equal to 0.1.

## 8. Appendix 8: Example with CIR model and corporate coupon bond in Section 4.1.2

For the example with CIR model and corporate coupon bond in Section 4.2, the discount process $\delta(t)$ is equal to $e^{-\int_0^t[r(s)+\chi(s)+\gamma(s)]ds}$ and $d\delta(t)=-\left[r(t)+\chi(t)+\gamma(t)\right]\delta(t)dt$. To accommodate the three risk factors (interest rate, default intensity, and convenience yield) with deflator, we let $dB(t)=B(t)\left[r(t)+\chi(t)+\gamma(t)\right]dt$. We repeat the calculation in Appendix 3.1, then we have $\Omega(t)$ equalling $-D(t)\left[r(t)+\chi(t)+\gamma(t)\right]$. Similarly, we have $\Phi(t)$ equalling $-D(t)\theta(t)$ and $\Psi(t)$, $\Gamma(t)$, $\mathrm{I}(t)$ equalling zero. We show the detailed calculations as follows.

A8.1 Short-term saving, $D(t)B(t)$

$$d\left[D(t)B(t)\right]=B(t)dD(t)+D(t)dB(t)+dB(t)dD(t)$$
$$=\left\{\Omega(t)B(t)+D(t)B(t)\left[r(t)+\chi(t)+\gamma(t)\right]\right\}dt+\Phi(t)B(t)dW_r(t)$$
$$+\Psi(t)B(t)dW_1(t)+\Gamma(t)B(t)dW_2(t)+\mathrm{I}(t)B(t)dW_3(t)$$

Let the drift term of $d\left[D(t)B(t)\right]$ equal zero, $\Omega(t)B(t)+D(t)B(t)\left[r(t)+\chi(t)+\gamma(t)\right]=0$, then $\Omega(t)$ is equal to $-D(t)\left[r(t)+\chi(t)+\gamma(t)\right]$.

Also, $d\left[\delta(t)B(t)\right]=\delta(t)dB(t)+B(t)d\delta(t)+d\delta(t)dB(t)=0$.

Thus, $\delta(t)B(t)$ is a $\mathbf{Q}$-martingale.

A8.2 Zero coupon bond of no risk with maturity $T$, $D(t)P(t,T,r(t))$

Similar to Appendix 2 (A2.1), $d\left[\delta(t)P(t,T,r(t))\right]=\tilde{\sigma}(t,r(t))\delta(t)P(t,T,r(t))d\tilde{W}_r(t)$ with $\theta(t)$ equalling $\dfrac{\tilde{\mu}(t,r(t))-\left[r(t)+\chi(t)+\gamma(t)\right]}{\tilde{\sigma}(t,r(t))}$ here. $\delta(t)P(t,T,r(t))$ is a $\mathbf{Q}$-martingale.

We rewrite $\dfrac{dP(t,T,r(t))}{P(t,T,r(t))}=\left(\tilde{\mu}(t,r(t))-\tilde{\sigma}(t,r(t))\theta(t)\right)dt+\tilde{\sigma}(t,r(t))d\tilde{W}_r(t)$.

$P(t,T,r(t))$ is the zero coupon bond of no risk with maturity $T$, so the drift term of $\dfrac{dP(t,T,r(t))}{P(t,T,r(t))}$ under probability measure $\mathbf{Q}'$ is equal to $r(t)+\chi(t)+\gamma(t)$. Hence, we have $\tilde{\mu}(t,r(t))$ equal $r(t)+\chi(t)+\gamma(t)+\tilde{\sigma}(t,r(t))\theta(t)$.

$$\frac{dP(t,T,r(t))}{P(t,T,r(t))}=\left[r(t)+\chi(t)+\gamma(t)+\tilde{\sigma}(t,r(t))\theta(t)\right]dt+\tilde{\sigma}(t,r(t))dW_r(t),\ \tilde{\sigma}(t,r(t))=\frac{P_r\beta(t,r(t))}{P(t,r(t))} \quad \text{(A8.1)}$$

$$dP(t,T,r(t))=P(t,T,r(t))\left[r(t)+\chi(t)+\gamma(t)+\tilde{\sigma}(t,r(t))\theta(t)\right]dt+P(t,T,r(t))\tilde{\sigma}(t,r(t))dW_r(t)$$

$$\begin{aligned}d\left[D(t)P(t,T,r(t))\right]&=P(t,T,r(t))dD(t)+D(t)dP(t,T,r(t))+dP(t,T,r(t))dD(t)\\&=\begin{Bmatrix}\Omega(t)P(t,T,r(t))+\Phi(t)P(t,T,r(t))\tilde{\sigma}(t,r(t))\\+D(t)P(t,T,r(t))\left[r(t)+\chi(t)+\gamma(t)+\tilde{\sigma}(t,r(t))\theta(t)\right]\end{Bmatrix}dt\\&\quad+\left[\Phi(t)P(t,T,r(t))+D(t)P(t,T,r(t))\tilde{\sigma}(t,r(t))\right]dW_r(t)\\&\quad+\Psi(t)P(t,T,r(t))dW_1(t)+\Gamma(t)P(t,T,r(t))dW_2(t)+\mathrm{I}(t)P(t,T,r(t))dW_3(t)\end{aligned}$$

Let the drift term of $d\left[D(t)P(t,T,r(t))\right]$ equal zero and plug $\Omega(t)$ into the drift term, we have $\Phi(t)=-D(t)\theta(t)$.

A8.3 Stock, $D(t)S(t)$

$$\begin{aligned}d\left[D(t)S(t)\right]&=S(t)dD(t)+D(t)dS(t)+dD(t)dS(t)\\&=\left[\Omega(t)S(t)+\Phi(t)S(t)\sigma_S(t)\rho_{rS}+\Psi(t)S(t)\sigma_S\sqrt{1-\rho_{rS}^2}+D(t)S(t)\mu_S(t)\right]dt\\&\quad+\left[\Phi(t)S(t)+D(t)S(t)\sigma_S(t)\rho_{rS}\right]dW_r(t)+\left[\Psi(t)S(t)+D(t)S(t)\sigma_S(t)\sqrt{1-\rho_{rS}^2}\right]dW_1(t)\\&\quad+\Gamma(t)S(t)dW_2(t)+\mathrm{I}(t)S(t)dW_3(t)\end{aligned}$$

Let the drift term of $d\left[D(t)S(t)\right]$ equal zero and plug $\Omega(t)$, $\Phi(t)$ into the drift term, then $\Psi(t)$ is equal to $\dfrac{D(t)\left[r(t)+\chi(t)+\gamma(t)+\theta(t)\sigma_S(t)\rho_{rS}-\mu_S(t)\right]}{\sigma_S(t)\sqrt{1-\rho_{rS}^2}}$.

$$\begin{aligned} d\left[\delta(t)S(t)\right] &= \delta(t)dS(t)+S(t)d\delta(t)+d\delta(t)dS(t) \\ &= \delta(t)S(t)\left[\mu_S(t)-r(t)-\chi(t)-\gamma(t)\right]dt+\delta(t)S(t)\sigma_S(t)\rho_{rS}dW_r(t) \\ &\quad +\delta(t)S(t)\sigma_S(t)\sqrt{1-\rho_{rS}^2}dW_1(t) \\ &= \delta(t)S(t)\left[\mu_S(t)-r(t)-\chi(t)-\gamma(t)-\theta(t)\sigma_S(t)\rho_{rS}\right]dt+\delta(t)S(t)\sigma_S(t)\rho_{rS}d\tilde{W}_r(t) \\ &\quad +\delta(t)S(t)\sigma_S(t)\sqrt{1-\rho_{rS}^2}dW_1(t), \text{ under } \mathbf{Q} \end{aligned}$$

We require $\mu_S(t)=r(t)+\chi(t)+\gamma(t)+\theta(t)\sigma_S(t)\rho_{rS}$ as regularity condition, such that $\delta(t)S(t)$ is a $\mathbf{Q}$-martingale. As a result, $\Psi(t)=0$.

A8.4 Default density, $D(t)\chi(t)$

$$\begin{aligned} d\left[D(t)\chi(t)\right] &= \chi(t)dD(t)+D(t)d\chi(t)+d\chi(t)dD(t) \\ &= \left[\Omega(t)\chi(t)+\Phi(t)\sigma_\chi\rho_{r\chi}\sqrt{\chi(t)}+\Psi(t)\sigma_\chi\rho'_{S\chi}\sqrt{\chi(t)}+\Gamma(t)\sigma_\chi\rho'_{\chi\chi}\sqrt{\chi(t)}+D(t)e-D(t)f\chi(t)\right]dt \\ &\quad +\left[\Phi(t)\chi(t)+D(t)\sigma_\chi\rho_{r\chi}\sqrt{\chi(t)}\right]dW_r(t)+\left[\Psi(t)\chi(t)+D(t)\sigma_\chi\rho'_{S\chi}\sqrt{\chi(t)}\right]dW_1(t) \\ &\quad +\left[\Gamma(t)\chi(t)+D(t)\sigma_\chi\rho'_{\chi\chi}\sqrt{\chi(t)}\right]dW_2(t)+\mathrm{I}(t)\chi(t)dW_3(t) \end{aligned}$$

Let the drift term of $d\left[D(t)\chi(t)\right]$ equal zero and plug $\Omega(t)$, $\Phi(t)$, $\Psi(t)$ into the drift term, we have $\Gamma(t)=D(t)\left\{\frac{\theta(t)\rho_{r\chi}}{\rho'_{\chi\chi}}+\frac{\left[r(t)+\chi(t)+\gamma(t)\right]\chi(t)-e+f\chi(t)}{\sigma_\chi\rho'_{\chi\chi}\sqrt{\chi(t)}}\right\}$.

$$\begin{aligned} d\left[\delta(t)\chi(t)\right] &= \delta(t)d\chi(t)+\chi(t)d\delta(t)+d\delta(t)d\chi(t) \\ &= \delta(t)\left\{e-\left[r(t)+\chi(t)+\gamma(t)\right]\chi(t)-f\chi(t)\right\}dt+\sigma_\chi\rho_{r\chi}\delta(t)\sqrt{\chi(t)}dW_r(t) \\ &\quad +\sigma_\chi\rho'_{S\chi}\delta(t)\sqrt{\chi(t)}dW_1(t)+\sigma_\chi\rho'_{\chi\chi}\delta(t)\sqrt{\chi(t)}dW_2(t) \\ &= \delta(t)\left\{e-\left[r(t)+\chi(t)+\gamma(t)\right]\chi(t)-f\chi(t)-\sigma_\chi\rho_{r\chi}\theta(t)\sqrt{\chi(t)}\right\}dt+\sigma_\chi\rho_{r\chi}\delta(t)\sqrt{\chi(t)}d\tilde{W}_r(t) \\ &\quad +\sigma_\chi\rho'_{S\chi}\delta(t)\sqrt{\chi(t)}dW_1(t)+\sigma_\chi\rho'_{\chi\chi}\delta(t)\sqrt{\chi(t)}dW_2(t), \text{ under } \mathbf{Q} \end{aligned}$$

We require $e=\left[r(t)+\chi(t)+\gamma(t)\right]\chi(t)+f\chi(t)+\sigma_\chi\rho_{r\chi}\theta(t)\sqrt{\chi(t)}$ as regularity condition, such that $\delta(t)\chi(t)$ is a $\mathbf{Q}$-martingale. As a result, $\Gamma(t)=0$. In addition, we also have to consider if the Feller condition holds (though this does not guarantee $\chi(t)$ to be positive since $e$ involves several factors and is not a constant here), i.e.

$$\left[r(t)+\chi(t)+\gamma(t)\right]\chi(t)+f\chi(t)+\sigma_\chi\rho_{r\chi}\theta(t)\sqrt{\chi(t)}>\frac{1}{2}\sigma_\chi^2.$$

A8.5 Convenience yield, $D(t)\gamma(t)$

$$\begin{aligned} d\left[D(t)\gamma(t)\right] &= \gamma(t)dD(t)+D(t)d\gamma(t)+d\gamma(t)dD(t) \\ &= \left[\Omega(t)\gamma(t)+\Phi(t)\eta\rho_{r\gamma}+\Psi(t)\eta\rho''_{S\gamma}+\Gamma(t)\eta\rho''_{\chi\gamma}+\mathrm{I}(t)\eta\rho''_{\gamma\gamma}\right]dt+\left[\Phi(t)\gamma(t)+D(t)\eta\rho_{r\gamma}\right]dW_r(t) \\ &\quad +\left[\Psi(t)\gamma(t)+D(t)\eta\rho''_{S\gamma}\right]dW_1(t)+\left[\Gamma(t)\gamma(t)+D(t)\eta\rho''_{\chi\gamma}\right]dW_2(t) \\ &\quad +\left(\mathrm{I}(t)\gamma(t)+D(t)\eta\rho''_{\gamma\gamma}\right)dW_3(t) \end{aligned}$$

Let the drift term of $d\left[D(t)\gamma(t)\right]$ equal zero and plug $\Omega(t)$, $\Phi(t)$, $\Psi(t)$, $\Gamma(t)$ into the drift term, then $\mathrm{I}(t)=D(t)\left\{\frac{\rho_{r\gamma}\theta(t)}{\rho''_{\gamma\gamma}}+\frac{\left[r(t)+\chi(t)+\gamma(t)\right]\gamma(t)}{\eta\rho''_{\gamma\gamma}}\right\}$.

$$\begin{aligned} d\left[\delta(t)\gamma(t)\right] &= \delta(t)d\gamma(t)+\gamma(t)d\delta(t)+d\delta(t)d\gamma(t) \\ &= -\delta(t)\gamma(t)\left[r(t)+\chi(t)+\gamma(t)\right]dt+\eta\rho_{r\gamma}\delta(t)dW_r(t)+\eta\rho''_{S\gamma}\delta(t)dW_1(t) \\ &\quad +\eta\rho''_{\chi\gamma}\delta(t)dW_2(t)+\eta\rho''_{\gamma\gamma}\delta(t)dW_3(t) \\ &= \delta(t)\left\{-\gamma(t)\left[r(t)+\chi(t)+\gamma(t)\right]-\eta\rho_{r\gamma}\theta(t)\right\}dt+\eta\rho_{r\gamma}\delta(t)d\tilde{W}_r(t)+\eta\rho''_{S\gamma}\delta(t)dW_1(t) \\ &\quad +\eta\rho''_{\chi\gamma}\delta(t)dW_2(t)+\eta\rho''_{\gamma\gamma}\delta(t)dW_3(t), \text{ under } \mathbf{Q} \end{aligned}$$

We require $\eta=-\frac{\gamma(t)\left[r(t)+\chi(t)+\gamma(t)\right]}{\rho_{r\gamma}\theta(t)}$ as regularity condition, such that $\delta(t)\gamma(t)$ is a $\mathbf{Q}$-martingale. As a result, $\mathrm{I}(t)=0$. In addition, readers should be very careful about the initial parameter setting since $\eta$ also involves several factors and is not a constant here. In our numerical example under Longstaff et al. (2005) in Section 4.2, $\chi(t)$ becomes negative after projecting longer than 7 years and the simulation couldn't continue, in which the large $\gamma(t)$ is the reason leads to the negative value of $\chi(t)$. Further study would be to investigate the long-term behaviours of $\chi(t)$ and $\gamma(t)$ after we require the regularity conditions.

A8.6 Implementation of time discretization

In A8.6, we show the implementation of time discretization for stochastic processes in A8.1 to A8.5 (different from the example in Section 4.1).

$$\begin{bmatrix} dB(t) \\ dP(t,T,r(t)) \\ dS(t) \\ d\chi(t) \\ d\gamma(t) \\ dD(t) \end{bmatrix} = \begin{bmatrix} B(t)\left[r(t)+\chi(t)+\gamma(t)\right] & 0 & 0 & 0 & 0 \\ P(t,T,r(t))\left[r(t)+\chi(t)+\gamma(t)\right]+P_r\sigma_r\sqrt{r(t)}\theta(t) & P_r\sigma_r\sqrt{r(t)} & 0 & 0 & 0 \\ S(t)\left[r(t)+\chi(t)+\gamma(t)+\theta(t)\sigma_S(t)\rho_{rS}\right] & S(t)\sigma_S(t)\rho_{rS} & S(t)\sigma_S(t)\sqrt{1-\rho_{rS}^2} & 0 & 0 \\ \left[r(t)+\chi(t)+\gamma(t)\right]\chi(t)+\sigma_\chi\rho_{r\chi}\theta(t)\sqrt{\chi(t)} & \sigma_\chi\rho_{r\chi}\sqrt{\chi(t)} & \sigma_\chi\rho'_{S\chi}\sqrt{\chi(t)} & \sigma_\chi\rho'_{\chi\chi}\sqrt{\chi(t)} & 0 \\ 0 & -\frac{\gamma(t)\left[r(t)+\chi(t)+\gamma(t)\right]}{\theta(t)} & -\frac{\rho''_{S\gamma}\gamma(t)\left[r(t)+\chi(t)+\gamma(t)\right]}{\rho_{r\gamma}\theta(t)} & -\frac{\rho''_{\chi\gamma}\gamma(t)\left[r(t)+\chi(t)+\gamma(t)\right]}{\rho_{r\gamma}\theta(t)} & -\frac{\rho''_{\gamma\gamma}\gamma(t)\left[r(t)+\chi(t)+\gamma(t)\right]}{\rho_{r\gamma}\theta(t)} \\ -D(t)\left[r(t)+\chi(t)+\gamma(t)\right] & -D(t)\theta(t) & 0 & 0 & 0 \end{bmatrix} \begin{bmatrix} dt \\ dW_r(t) \\ dW_1(t) \\ dW_2(t) \\ dW_3(t) \end{bmatrix}$$

### A8.6.1 Euler method

$$\begin{bmatrix} B_{i+1} \\ P_{i+1} \\ S_{i+1} \\ \chi_{i+1} \\ \gamma_{i+1} \\ D_{i+1} \end{bmatrix} = \begin{bmatrix} B_i \\ P_i \\ S_i \\ \chi_i \\ \gamma_i \\ D_i \end{bmatrix} + \begin{bmatrix} B_i\left(r_i+\chi_i+\gamma_i\right) & 0 & 0 & 0 & 0 \\ P_i\left(r_i+\chi_i+\gamma_i\right)+P_{r,t_i}\sigma_r\sqrt{r_i}\theta_i & P_{r,t_i}\sigma_r\sqrt{r_i} & 0 & 0 & 0 \\ S_i\left(r_i+\chi_i+\gamma_i+\theta_i\sigma_{S,t_i}\rho_{rS}\right) & S_i\sigma_{S,t_i}\rho_{rS} & S_i\sigma_{S,t_i}\sqrt{1-\rho_{rS}^2} & 0 & 0 \\ \left(r_i+\chi_i+\gamma_i\right)\chi_i+\sigma_\chi\rho_{r\chi}\theta_i\sqrt{\chi_i} & \sigma_\chi\rho_{r\chi}\sqrt{\chi_i} & \sigma_\chi\rho'_{S\chi}\sqrt{\chi_i} & \sigma_\chi\rho'_{\chi\chi}\sqrt{\chi_i} & 0 \\ 0 & -\frac{\gamma_i\left(r_i+\chi_i+\gamma_i\right)}{\theta_i} & -\frac{\rho''_{S\gamma}\gamma_i\left(r_i+\chi_i+\gamma_i\right)}{\rho_{r\gamma}\theta_i} & -\frac{\rho''_{\chi\gamma}\gamma_i\left(r_i+\chi_i+\gamma_i\right)}{\rho_{r\gamma}\theta_i} & -\frac{\rho''_{\gamma\gamma}\gamma_i\left(r_i+\chi_i+\gamma_i\right)}{\rho_{r\gamma}\theta_i} \\ -D_i\left(r_i+\chi_i+\gamma_i\right) & -D_i\theta_i & 0 & 0 & 0 \end{bmatrix} \begin{bmatrix} \Delta t_i \\ \Delta W_{r,i} \\ \Delta W_{1,i} \\ \Delta W_{2,i} \\ \Delta W_{3,i} \end{bmatrix}$$

### A8.6.2 Milstein method

$$\begin{bmatrix} B_{i+1} \\ P_{i+1} \\ S_{i+1} \\ \chi_{i+1} \\ \gamma_{i+1} \\ D_{i+1} \end{bmatrix} = \begin{bmatrix} B_i \\ P_i \\ S_i \\ \chi_i \\ \gamma_i \\ D_i \end{bmatrix} + \begin{bmatrix} B_i\left(r_i+\chi_i+\gamma_i\right) & 0 & 0 & 0 & 0 \\ P_i\left(r_i+\chi_i+\gamma_i\right)+P_{r,t_i}\sigma_r\sqrt{r_i}\theta_i & P_{r,t_i}\sigma_r\sqrt{r_i} & 0 & 0 & 0 \\ S_i\left(r_i+\chi_i+\gamma_i+\theta_i\sigma_{S,t_i}\rho_{rS}\right) & S_i\sigma_{S,t_i}\rho_{rS} & S_i\sigma_{S,t_i}\sqrt{1-\rho_{rS}^2} & 0 & 0 \\ \left(r_i+\chi_i+\gamma_i\right)\chi_i+\sigma_\chi\rho_{r\chi}\theta_i\sqrt{\chi_i} & \sigma_\chi\rho_{r\chi}\sqrt{\chi_i} & \sigma_\chi\rho'_{S\chi}\sqrt{\chi_i} & \sigma_\chi\rho'_{\chi\chi}\sqrt{\chi_i} & 0 \\ 0 & -\frac{\gamma_i\left(r_i+\chi_i+\gamma_i\right)}{\theta_i} & -\frac{\rho''_{S\gamma}\gamma_i\left(r_i+\chi_i+\gamma_i\right)}{\rho_{r\gamma}\theta_i} & -\frac{\rho''_{\chi\gamma}\gamma_i\left(r_i+\chi_i+\gamma_i\right)}{\rho_{r\gamma}\theta_i} & -\frac{\rho''_{\gamma\gamma}\gamma_i\left(r_i+\chi_i+\gamma_i\right)}{\rho_{r\gamma}\theta_i} \\ -D_i\left(r_i+\chi_i+\gamma_i\right) & -D_i\theta_i & 0 & 0 & 0 \end{bmatrix} \begin{bmatrix} \Delta t_i \\ \Delta W_{r,i} \\ \Delta W_{1,i} \\ \Delta W_{2,i} \\ \Delta W_{3,i} \end{bmatrix}$$

$$+\frac{1}{4}\begin{bmatrix} 0 & 0 & 0 & 0 \\ 0 & 0 & 0 & 0 \\ 2\sigma_{S,t_i}^2\rho_{rS}^2 & 2\sigma_{S,t_i}^2\left(1-\rho_{rS}^2\right) & 0 & 0 \\ \sigma_\chi^2\rho_{r\chi}^2 & \sigma_\chi^2\rho_{S\chi}'^2 & \sigma_\chi^2\rho_{\chi\chi}'^2 & 0 \\ \frac{2\gamma_i}{\theta_i^2}\left(r_i+\chi_i+\gamma_i\right)\left(r_i+\chi_i+2\gamma_i\right) & 2\gamma_i\left(\frac{\rho''_{S\gamma}}{\rho_{r\gamma}\theta_i}\right)^2\left(r_i+\chi_i+\gamma_i\right)\left(r_i+\chi_i+2\gamma_i\right) & 2\gamma_i\left(\frac{\rho''_{\chi\gamma}}{\rho_{r\gamma}\theta_i}\right)^2\left(r_i+\chi_i+\gamma_i\right)\left(r_i+\chi_i+2\gamma_i\right) & 2\gamma_i\left(\frac{\rho''_{\gamma\gamma}}{\rho_{r\gamma}\theta_i}\right)^2\left(r_i+\chi_i+\gamma_i\right)\left(r_i+\chi_i+2\gamma_i\right) \\ 2D_i\theta_i^2 & 0 & 0 & 0 \end{bmatrix} \begin{bmatrix} \left(\Delta W_{r,i}\right)^2-\Delta t_i \\ \left(\Delta W_{1,i}\right)^2-\Delta t_i \\ \left(\Delta W_{2,i}\right)^2-\Delta t_i \\ \left(\Delta W_{3,i}\right)^2-\Delta t_i \end{bmatrix}$$

### A8.6.2 Second Milstein method

Here are $X_t$, $W_t$, $a\left(t,X_t\right)$, and $b\left(t,X_t\right)$ of simplified Second Milstein method in Section 4.2, $dX_t = a\left(t,X_t\right)dt + b\left(t,X_t\right)dW_t$.

$$X_t = \begin{bmatrix} r(t) \\ \theta(t) \\ B(t) \\ P\left(t,T,r(t)\right) \\ S(t) \\ \chi(t) \\ \gamma(t) \\ D(t) \end{bmatrix}, W_t = \begin{bmatrix} W_r(t) \\ W_1(t) \\ W_2(t) \\ W_3(t) \\ W_\theta(t) \end{bmatrix} \quad \text{(A8.1)}$$

$$a\left(t,X_t\right) = \begin{bmatrix} a_r - b_r r(t) + \theta(t)\sigma_r\sqrt{r(t)} \\ a_\theta - b_\theta\theta(t) \\ B(t)\left[r(t)+\chi(t)+\gamma(t)\right] \\ P\left(t,T,r(t)\right)\left[r(t)+\chi(t)+\gamma(t)\right]+P_r\sigma_r\sqrt{r(t)}\theta(t) \\ S(t)\left[r(t)+\chi(t)+\gamma(t)+\theta(t)\sigma_S(t)\rho_{rS}\right] \\ \left[r(t)+\chi(t)+\gamma(t)\right]\chi(t)+\sigma_\chi\rho_{r\chi}\theta(t)\sqrt{\chi(t)} \\ 0 \\ -\left[r(t)+\chi(t)+\gamma(t)\right]D(t) \end{bmatrix} \quad \text{(A8.2)}$$

$$b(t,X_t)=\begin{bmatrix} \sigma_r\sqrt{r(t)} & 0 & 0 & 0 & 0 \\ 0 & 0 & 0 & 0 & 0 \\ 0 & 0 & 0 & 0 & \sigma_\theta\sqrt{\theta(t)} \\ P_r\sigma_r\sqrt{r(t)} & 0 & 0 & 0 & 0 \\ S(t)\sigma_S(t)\rho_{rS} & S(t)\sigma_S(t)\sqrt{1-\rho_{rS}^2} & 0 & 0 & 0 \\ \sigma_\chi\rho_{r\chi}\sqrt{\chi(t)} & \sigma_\chi\rho'_{S\chi}\sqrt{\chi(t)} & \sigma_\chi\rho'_{\chi\chi}\sqrt{\chi(t)} & 0 & 0 \\ -\frac{\gamma(t)\left[r(t)+\chi(t)+\gamma(t)\right]}{\theta(t)} & -\frac{\rho''_{S\gamma}\gamma(t)\left[r(t)+\chi(t)+\gamma(t)\right]}{\rho_{r\gamma}\theta(t)} & -\frac{\rho''_{\chi\gamma}\gamma(t)\left[r(t)+\chi(t)+\gamma(t)\right]}{\rho_{r\gamma}\theta(t)} & -\frac{\rho''_{\gamma\gamma}\gamma(t)\left[r(t)+\chi(t)+\gamma(t)\right]}{\rho_{r\gamma}\theta(t)} & 0 \\ -\theta(t)D(t) & 0 & 0 & 0 & 0 \end{bmatrix}$$

(A8.3)

Similar to Section A6.1, We choose $a_8(t,X_t)$ as an illustrative example here.

$$a_8(t,X_t)=-\left[r(t)+\chi(t)+\gamma(t)\right]D(t),$$

$$\frac{\partial a_8(t,X_t)}{\partial x_1}=-D(t),\ \frac{\partial a_8(t,X_t)}{\partial x_6}=-D(t),\ \frac{\partial a_8(t,X_t)}{\partial x_7}=-D(t),\ \frac{\partial a_8(t,X_t)}{\partial x_8}=-\left[r(t)+\chi(t)+\gamma(t)\right]$$

$$\frac{\partial^2 a_8(t,X_t)}{\partial x_1\partial x_8}=\frac{\partial^2 a_8(t,X_t)}{\partial x_8\partial x_1}=-1,\ \frac{\partial^2 a_8(t,X_t)}{\partial x_6\partial x_8}=\frac{\partial^2 a_8(t,X_t)}{\partial x_8\partial x_6}=-1,\ \frac{\partial^2 a_8(t,X_t)}{\partial x_7\partial x_8}=\frac{\partial^2 a_8(t,X_t)}{\partial x_8\partial x_7}=-1$$

$$\begin{aligned} L^0 a_8(t,X_t) &= \frac{\partial a_8(t,X_t)}{\partial t}+\sum_{i=1}^{8} a_i(t,X_t)\frac{\partial a_8(t,X_t)}{\partial x_i}+\frac{1}{2}\sum_{i,j=1}^{8}\Sigma_{t,ij}\frac{\partial^2 a_8(t,X_t)}{\partial x_i\partial x_j} \\ &= -a_1(t,X_t)D(t)-a_6(t,X_t)D(t)-a_7(t,X_t)D(t)-a_8(t,X_t)\left[r(t)+\chi(t)+\gamma(t)\right] \\ &\quad -\Sigma_{t,18}-\Sigma_{t,68}-\Sigma_{t,78} \end{aligned}$$

$$\begin{aligned} L^k a_8(t,X_t) &= \sum_{i=1}^{d} b_{ik}(t,X_t)\frac{\partial a_8(t,X_t)}{\partial x_i} \\ &= -b_{1k}(t,X_t)D(t)-b_{6k}(t,X_t)D(t)-b_{7k}(t,X_t)D(t)-b_{8k}(t,X_t)\left[r(t)+\chi(t)+\gamma(t)\right] \end{aligned}$$

## 9. APPENDIX 9: ONE MORE REQUIRED REGULARITY CONDITION FOR THE DIFFUSION TERM IN STOCK PRICE

In this section, we explain the one more required regularity for the diffusion term in stock price in detail. Recall that the dynamics of stock price after requiring regularity condition for $\mu_S(t)$ is as follows.

$$dS(t)=S(t)\left[r(t)+\theta(t)\sigma_S(t)\rho_{rS}\right]dt+S(t)\sigma_S(t)\rho_{rS}dW_r(t)+S(t)\sigma_S(t)\sqrt{1-\rho_{rS}^2}dW_1(t) \quad \text{(A9.1)}$$

By Itô formula,

$$\begin{aligned} d\left[\ln S(t)\right] &= \frac{\partial\left[\ln S(t)\right]}{\partial S(t)}dS(t)+\frac{1}{2}\frac{\partial^2\left[\ln S(t)\right]}{\partial\left[S(t)\right]^2}dS(t)dS(t) \\ &= \left[r(t)+\theta(t)\sigma_S(t)\rho_{rS}-\frac{1}{2}\sigma_S^2(t)\right]dt+\sigma_S(t)\rho_{rS}dW_r(t)+\sigma_S(t)\sqrt{1-\rho_{rS}^2}dW_1(t) \end{aligned}.$$

Integrate both sides of $d\left[\ln S(t)\right]$, we have

$$\ln S(t)=\ln S(0)+\int_0^t\left[r(s)+\theta(s)\sigma_S(s)\rho_{rS}-\frac{1}{2}\sigma_S^2(s)\right]ds+\int_0^t\sigma_S(s)\rho_{rS}dW_r(s)+\int_0^t\sigma_S(s)\sqrt{1-\rho_{rS}^2}dW_1(s). \quad \text{(A9.2)}$$

From Appendix 4, we have $\ln D(t)=\ln D(0)+\int_0^t\left[-r(s)-\frac{1}{2}\theta^2(s)\right]ds-\int_0^t\theta(s)dW_r(s)$.

Then,

$$\begin{aligned}\ln\left[D(t)S(t)\right]&=\ln D(t)+\ln S(t)\\&=\ln D(0)+\ln S(0)+\int_0^t\left[\theta(s)\sigma_S(s)\rho_{rS}-\frac{1}{2}\sigma_S^2(s)-\frac{1}{2}\theta^2(s)\right]ds.\\&\quad+\int_0^t\left[\sigma_S(s)\rho_{rS}-\theta(s)\right]dW_r(t)+\int_0^t\sigma_S(s)\sqrt{1-\rho_{rS}^2}dW_1(s)\end{aligned}$$

The first identity comes from that $\ln\left[D(t)S(t)\right]$ and $\ln D(t)+\ln S(t)$ are two random variables with the same characteristic function.

Take exponential both side of $\ln\left[D(t)S(t)\right]$, we have

$$\begin{aligned}D(t)S(t)=D(0)S(0)&\exp\left\{\int_0^t\left[\theta(s)\sigma_S(s)\rho_{rS}-\frac{1}{2}\sigma_S^2(s)-\frac{1}{2}\theta^2(s)\right]ds\right\}\\&\times\exp\left\{\int_0^t\left[\sigma_S(s)\rho_{rS}-\theta(s)\right]dW_r(t)+\int_0^t\sigma_S(s)\sqrt{1-\rho_{rS}^2}dW_1(s)\right\}\end{aligned}. \quad \text{(A9.3)}$$

If $\theta(t)$ and $\sigma_S(t)$ are constants, then $D(t)S(t)$ is a martingale since $W_r(t)$ and $W_1(t)$ are independent with $\exp\left\{\sigma W_i(t)-\frac{1}{2}\sigma^2 t\right\}$ being a martingale for $i=r,1$.[2]

However, $\theta(t)$ and $\sigma_S(t)$ are not constants. Let $\theta(t)\sigma_S(t)\rho_{rS}-\frac{1}{2}\sigma_S^2(t)-\frac{1}{2}\theta^2(t)$ equal zero, then both $D(t)S(t)$ and $\ln\left[D(t)S(t)\right]$ are $\mathbf{P}$-martingales.[3] Then, $\sigma_S(t)=\rho_{rS}\theta(t)\pm\theta(t)\sqrt{\rho_{rS}^2-1}$ and $\sigma_S(t)$ is a complex number if $\rho_{rS}\neq 1$ (so that $|\rho_{rS}|<1$). In our numerical example in Section 4.2, we choose $\rho_{rS}$ equalling 1 then $\sigma_S(t)$ is equal to $\theta(t)$. After plugging in $\rho_{rS}=1$ and $\sigma_S(t)=\theta(t)$ into $d\left[\ln S(t)\right]$, we have

$$d\left[\ln S(t)\right]=\left[r(t)+\frac{1}{2}\theta^2(t)\right]dt+\theta(t)dW_r(t).$$

In addition, $d\left[\ln D(t)\right]=\left[-r(t)-\frac{1}{2}\theta^2(t)\right]dt-\theta(t)dW_r(t)$. Comparing $d\left[\ln S(t)\right]$ with $d\left[\ln D(t)\right]$, we could see that the drift and the diffusion terms of $d\left[\ln S(t)\right]$ and $d\left[\ln D(t)\right]$ offset each other.

---

[2] See for example, Shreve (2004) Chapter 3.6 Theorem 3.6.1.

[3] Observe that after this required condition with the dynamics of $\ln\left[D(t)S(t)\right]$, we could derive that $D(t)S(t)$ is still a $\mathbf{P}$-martingale by Itô formula.

## 10. APPENDIX 10: INTEREST RATE IN PHYSICAL WORLD

In this section, we provide some analyses for the process of interest rate in physical world. In Section 4, the process of interest rate in $\mathbf{P}$-measure is

$$dr(t)=\left[a_r-b_r r(t)+\theta(t)\sigma_r\sqrt{r(t)}\right]dt+\sigma_r\sqrt{r(t)}dW_r(t).$$

We show that the mean and variance of the interest rate process $r(t)$ behave like the mean and variance of a CIR process asymptotically.

First, we could see that the mean and variance of the market price of risk $\theta(t)$ are $e^{-b_\theta t}\theta(0)+\frac{a_\theta}{b_\theta}\left(1-e^{-b_\theta t}\right)$ and $\frac{\sigma_\theta^2}{b_\theta}\theta(0)\left(e^{-b_\theta t}-e^{-2b_\theta t}\right)+\frac{a_\theta\sigma_\theta^2}{2b_\theta^2}\left(1-2e^{-b_\theta t}+e^{-2b_\theta t}\right)$ respectively.[4] When $t$ goes larger, the mean and variance of $\theta(t)$ asymptotically become $\frac{a_\theta}{b_\theta}$ and $\frac{a_\theta\sigma_\theta^2}{2b_\theta^2}$ respectively. We show the calculation details as follows.

i. $\theta(t)$

$$d\theta(t)=\left[a_\theta-b_\theta\theta(t)\right]dt+\sigma_\theta\sqrt{\theta(t)}dW_\theta(t)$$

Let $g(t,x)=e^{b_\theta t}x$, then $g_t(t,x)=b_\theta e^{b_\theta t}x$, $g_x(t,x)=e^{b_\theta t}$, and $g_{xx}(t,x)=0$.

$$\begin{aligned}d\left[e^{b_\theta t}\theta(t)\right]&=g_t(t,\theta(t))dt+g_x(t,\theta(t))d\theta(t)+\frac{1}{2}g_{xx}(t,\theta(t))d\theta(t)d\theta(t)\\&=b_\theta e^{b_\theta t}\theta(t)dt+e^{b_\theta t}\left[a_\theta-b_\theta\theta(t)\right]dt+e^{b_\theta t}\sigma_\theta\sqrt{\theta(t)}dW_\theta(t)\\&=a_\theta e^{b_\theta t}dt+\sigma_\theta e^{b_\theta t}\sqrt{\theta(t)}dW_\theta(t)\end{aligned}$$

Integrate both sides of $d\left[e^{b_\theta t}\theta(t)\right]$,

$$\int_0^t d\left[e^{b_\theta s}\theta(s)\right]=\int_0^t a_\theta e^{b_\theta s}ds+\int_0^t\sigma_\theta e^{b_\theta s}\sqrt{\theta(s)}dW_\theta(s)$$

$$e^{b_\theta t}\theta(t)-\theta(0)=a_\theta\int_0^t e^{b_\theta s}ds+\sigma_\theta\int_0^t e^{b_\theta s}\sqrt{\theta(s)}dW_\theta(s)$$

$$\begin{aligned}e^{b_\theta t}\theta(t)&=\theta(0)+a_\theta\int_0^t e^{b_\theta s}ds+\sigma_\theta\int_0^t e^{b_\theta s}\sqrt{\theta(s)}dW_\theta(s)\\&=\theta(0)+\frac{a_\theta}{b_\theta}\left(e^{b_\theta t}-1\right)+\sigma_\theta\int_0^t e^{b_\theta s}\sqrt{\theta(s)}dW_\theta(s)\end{aligned}$$

First, we calculate the mean of $\theta(t)$.

Take expectation both side of $e^{b_\theta t}\theta(t)$ with the expectation of an Itô intergral is zero,

then $E\left[e^{b_\theta t}\theta(t)\right]=\theta(0)+\frac{a_\theta}{b_\theta}\left(e^{b_\theta t}-1\right)$.

As a result, $E\left[\theta(t)\right]=e^{-b_\theta t}\theta(0)+\frac{a_\theta}{b_\theta}\left(1-e^{-b_\theta t}\right)$.

Particularly, $E\left[\theta(t)\right]\to\frac{a_\theta}{b_\theta}$ as $t\to\infty$.

[4] See, for example, Shreve (2004) Chapter 4.4 Example 4.4.11.

Secondly, we calculate the variance of $\theta(t)$.

Given a random variable $X$, we know that $Var(X)=E(X^2)-[E(X)]^2$.

Let $h(t,x)=x^2$, then $h_t(t,x)=0$, $h_x(t,x)=2x$, and $h_{xx}(t,x)=2$.

$$d\left\{\left[e^{b_\theta t}\theta(t)\right]^2\right\}=h_t(t,\theta(t))dt+h_x(t,\theta(t))d\left[e^{b_\theta t}\theta(t)\right]+\frac{1}{2}h_{xx}(t,\theta(t))d\left[e^{b_\theta t}\theta(t)\right]d\left[e^{b_\theta t}\theta(t)\right]$$

$$=2e^{b_\theta t}\theta(t)\left[a_\theta e^{b_\theta t}dt+\sigma_\theta e^{b_\theta t}\sqrt{\theta(t)}dW_\theta(t)\right]+\sigma_\theta^2 e^{2b_\theta t}\theta(t)dt$$

$$=\left(2a_\theta+\sigma_\theta^2\right)e^{2b_\theta t}\theta(t)dt+2\sigma_\theta e^{2b_\theta t}\theta^{\frac{3}{2}}(t)dW_\theta(t)$$

Integrate both sides of $d\left\{\left[e^{b_\theta t}\theta(t)\right]^2\right\}$,

$$\int_0^t d\left\{\left[e^{b_\theta s}\theta(s)\right]^2\right\}=\int_0^t\left(2a_\theta+\sigma_\theta^2\right)e^{2b_\theta s}\theta(s)ds+\int_0^t 2\sigma_\theta e^{2b_\theta s}\theta^{\frac{3}{2}}(s)dW_\theta(s).$$

$$\left[e^{b_\theta t}\theta(t)\right]^2-\theta^2(0)=\left(2a_\theta+\sigma_\theta^2\right)\int_0^t e^{2b_\theta s}\theta(s)ds+2\sigma_\theta\int_0^t e^{2b_\theta s}\theta^{\frac{3}{2}}(s)dW_\theta(s)$$

$$e^{2b_\theta t}\theta^2(t)=\theta^2(0)+\left(2a_\theta+\sigma_\theta^2\right)\int_0^t e^{2b_\theta s}\theta(s)ds+2\sigma_\theta\int_0^t e^{2b_\theta s}\theta^{\frac{3}{2}}(s)dW_\theta(s)$$

Suppose we could exchange integrals by the Fubini-Tonelli Theorem,
and then take expectation both sides of $e^{2b_\theta t}\theta^2(t)$ with the expectation of an Itô intergral is zero.

$$E\left[e^{2b_\theta t}\theta^2(t)\right]=\theta^2(0)+\left(2a_\theta+\sigma_\theta^2\right)\int_0^t e^{2b_\theta s}E\left[\theta(s)\right]ds$$

$$=\theta^2(0)+\left(2a_\theta+\sigma_\theta^2\right)\int_0^t\left[e^{b_\theta s}\theta(0)+\frac{a_\theta}{b_\theta}\left(e^{2b_\theta s}-e^{b_\theta s}\right)\right]ds$$

$$=\theta^2(0)+\left(2a_\theta+\sigma_\theta^2\right)\left[\frac{1}{b_\theta}e^{b_\theta s}\theta(0)+\frac{a_\theta}{b_\theta}\left(\frac{1}{2b_\theta}e^{2b_\theta s}-\frac{1}{b_\theta}e^{b_\theta s}\right)\right]\Bigg|_0^t$$

$$=\theta^2(0)+\frac{2a_\theta+\sigma_\theta^2}{b_\theta}\left[\theta(0)-\frac{a_\theta}{b_\theta}\right]\left(e^{b_\theta t}-1\right)+\frac{a_\theta\left(2a_\theta+\sigma_\theta^2\right)}{2b_\theta^2}\left(e^{2b_\theta t}-1\right)$$

As a result, $E\left[\theta^2(t)\right]=e^{-2b_\theta t}\theta^2(0)+\frac{2a_\theta+\sigma_\theta^2}{b_\theta}\left[\theta(0)-\frac{a_\theta}{b_\theta}\right]\left(e^{-b_\theta t}-e^{-2b_\theta t}\right)+\frac{a_\theta\left(2a_\theta+\sigma_\theta^2\right)}{2b_\theta^2}\left(1-e^{-2b_\theta t}\right).$

$$Var\left[\theta(t)\right]=E\left[\theta^2(t)\right]-E\left[\theta(t)\right]^2=\frac{\sigma_\theta^2}{b_\theta}\theta(0)\left(e^{-b_\theta t}-e^{-2b_\theta t}\right)+\frac{a_\theta\sigma_\theta^2}{2b_\theta^2}\left(1-2e^{-b_\theta t}+e^{-2b_\theta t}\right)$$

Particularly, $Var\left[\theta(t)\right]\to\frac{a_\theta\sigma_\theta^2}{2b_\theta^2}$ as $t\to\infty$.

Next, we calculate the mean and variance of interest rate process $r(t)$ as follows.

$ii.\ r(t)$

First, we calculate $d\left[e^{b_r t} r(t)\right]$ under risk-neutral world then tranfer into physical world.

$$\begin{cases} dr(t) = \left[a_r - b_r r(t)\right]dt + \sigma_r \sqrt{r(t)}\, d\tilde{W}_r(t) \\ d\tilde{W}_r(t) = \theta(t)dt + dW_r(t) \end{cases}$$

Similar to previous calculations in $d\left[e^{b_\theta t}\theta(t)\right]$,

$$d\left[e^{b_r t} r(t)\right] = a_r e^{b_r t}dt + \sigma_r e^{b_r t}\sqrt{r(t)}\, d\tilde{W}_r(t) = e^{b_r t}\left[a_r + \theta(t)\sigma_r\sqrt{r(t)}\right]dt + \sigma_r e^{b_r t}\sqrt{r(t)}\, dW_r(t)$$

Integrate both sides of $d\left[e^{b_r t} r(t)\right]$,

$$\int_0^t d\left[e^{b_r s} r(s)\right] = \int_0^t e^{b_r s}\left[a_r + \theta(s)\sigma_r\sqrt{r(s)}\right]ds + \int_0^t \sigma_r e^{b_r s}\sqrt{r(s)}\, dW_r(s)$$

$$e^{b_r t} r(t) - r(0) = \int_0^t e^{b_r s}\left[a_r + \theta(s)\sigma_r\sqrt{r(s)}\right]ds + \int_0^t \sigma_r e^{b_r s}\sqrt{r(s)}\, dW_r(s)$$

$$e^{b_r t} r(t) = r(0) + a_r\int_0^t e^{b_r s}ds + \sigma_r\int_0^t\left[e^{b_r s}\theta(s)\sqrt{r(s)}\right]ds + \int_0^t \sigma_r e^{b_r s}\sqrt{r(s)}\, dW_r(s)$$

$$= r(0) + \frac{a_r}{b_r}\left(e^{b_r t} - 1\right) + \sigma_r\int_0^t\left[e^{b_r s}\theta(s)\sqrt{r(s)}\right]ds + \int_0^t \sigma_r e^{b_r s}\sqrt{r(s)}\, dW_r(s)$$

$$r(t) = e^{-b_r t} r(0) + \frac{a_r}{b_r}\left(1 - e^{-b_r t}\right) + e^{-b_r t}\sigma_r\int_0^t\left[e^{b_r s}\theta(s)\sqrt{r(s)}\right]ds + e^{-b_r t}\int_0^t \sigma_r e^{b_r s}\sqrt{r(t)}\, dW_r(s)$$

Suppose we could exchange integrals by the Fubini-Tonelli Theorem,
and then take expectation both sides of $r(t)$ with the expectation of an Itô intergral is zero.

$$E\left[r(t)\right] = e^{-b_r t} r(0) + \frac{a_r}{b_r}\left(1 - e^{-b_r t}\right) + e^{-b_r t}\sigma_r\int_0^t e^{b_r s}E\left[\theta(s)\sqrt{r(s)}\right]ds$$

Then, $E\left[r(t)\right] \to \frac{a_r}{b_r}$ as $t \to \infty$.

Similar to previous calculations in $d\left\{\left[e^{b_\theta t}\theta(t)\right]^2\right\}$,

$$d\left\{\left[e^{b_r t} r(t)\right]^2\right\} = \left(2a_r + \sigma_r^2\right)e^{2b_r t} r(t)dt + 2\sigma_r e^{2b_r t} r^{\frac{3}{2}}(t)\, d\tilde{W}_r(t)$$

$$= \left(2a_r + \sigma_r^2\right)e^{2b_r t} r(t)dt + 2\sigma_r e^{2b_r t} r^{\frac{3}{2}}(t)\theta(t)dt + 2\sigma_r e^{2b_r t} r^{\frac{3}{2}}(t)\, dW_r(t).$$

Integrate both sides of $d\left\{\left[e^{b_r t} r(t)\right]^2\right\}$,

$$\int_0^t d\left\{\left[e^{b_r t} r(s)\right]^2\right\} = \int_0^t \left(2a_r + \sigma_r^2\right)e^{2b_r s} r(s)ds + \int_0^t 2\sigma_r e^{2b_r s} r^{\frac{3}{2}}(s)\theta(s)ds + \int_0^t 2\sigma_r e^{2b_r s} r^{\frac{3}{2}}(s)\, dW_r(s).$$

Then,

$$\left[e^{b_r t} r(t)\right]^2 - r^2(0) = \left(2a_r + \sigma_r^2\right)\int_0^t e^{2b_r s} r(s)\,ds + 2\sigma_r \int_0^t e^{2b_r s} r^{\frac{3}{2}}(s)\theta(s)\,ds + 2\sigma_r \int_0^t e^{2b_r s} r^{\frac{3}{2}}(s)\,dW_r(s).$$

$$\left[e^{b_r t} r(t)\right]^2 = r^2(0) + \left(2a_r + \sigma_r^2\right)\int_0^t e^{2b_r s} r(s)\,ds + 2\sigma_r \int_0^t e^{2b_r s} r^{\frac{3}{2}}(s)\theta(s)\,ds + 2\sigma_r \int_0^t e^{2b_r s} r^{\frac{3}{2}}(s)\,dW_r(s)$$

Suppose we could exchange integrals by the Fubini-Tonelli Theorem,

and then take expectation both sides of $\left[e^{b_r t} r(t)\right]^2$ with the expectation of an Itô intergral is zero.

$$E\left[e^{2b_r t} r^2(t)\right] = r^2(0) + \left(2a_r + \sigma_r^2\right)\int_0^t e^{2b_r s} E\left[r(s)\right] ds + 2\sigma_r \int_0^t e^{2b_r s} E\left[r^{\frac{3}{2}}(s)\theta(s)\right] ds$$

$$= r^2(0) + \left(2a_r + \sigma_r^2\right)\int_0^t e^{2b_r s}\left\{e^{-b_r s} r(0) + \frac{a_r}{b_r}\left(1 - e^{-b_r s}\right) + e^{-b_r s}\sigma_r \int_0^s e^{b_r u} E\left[\theta(u)\sqrt{r(u)}\right] du\right\} ds$$

$$+ 2\sigma_r \int_0^t e^{2b_r s} E\left[r^{\frac{3}{2}}(s)\theta(s)\right] ds$$

$$= r^2(0) + \frac{2a_r + \sigma_r^2}{b_r}\left[r(0) - \frac{a_r}{b_r}\right]\left(e^{b_r t} - 1\right) + \frac{a_r\left(2a_r + \sigma_r^2\right)}{2b_r^2}\left(e^{2b_r t} - 1\right)$$

$$+ \left(2a_r + \sigma_r^2\right)\sigma_r \int_0^t \left\{e^{b_r s}\int_0^s e^{b_r u} E\left[\theta(u)\sqrt{r(u)}\right] du\right\} ds + 2\sigma_r \int_0^t e^{2b_r s} E\left[r^{\frac{3}{2}}(s)\theta(s)\right] ds$$

$$e^{2b_r t} E\left[r^2(t)\right] = r^2(0) + \frac{2a_r + \sigma_r^2}{b_r}\left[r(0) - \frac{a_r}{b_r}\right]\left(e^{b_r t} - 1\right) + \frac{a_r\left(2a_r + \sigma_r^2\right)}{2b_r^2}\left(e^{2b_r t} - 1\right)$$

$$+ \left(2a_r + \sigma_r^2\right)\sigma_r \int_0^t \left\{e^{b_r s}\int_0^s e^{b_r u} E\left[\theta(u)\sqrt{r(u)}\right] du\right\} ds + 2\sigma_r \int_0^t e^{2b_r s} E\left[r^{\frac{3}{2}}(s)\theta(s)\right] ds$$

$$E\left[r^2(t)\right] = e^{-2b_r t} r^2(0) + \frac{2a_r + \sigma_r^2}{b_r}\left[r(0) - \frac{a_r}{b_r}\right]\left(e^{-b_r t} - e^{-2b_r t}\right) + \frac{a_r\left(2a_r + \sigma_r^2\right)}{2b_r^2}\left(1 - e^{-2b_r t}\right)$$

$$+ e^{-2b_r t}\left(2a_r + \sigma_r^2\right)\sigma_r \int_0^t \left\{e^{b_r s}\int_0^s e^{b_r u} E\left[\theta(u)\sqrt{r(u)}\right] du\right\} ds + e^{-2b_r t} 2\sigma_r \int_0^t e^{2b_r s} E\left[r^{\frac{3}{2}}(s)\theta(s)\right] ds$$

$$Var\left[r(t)\right] = E\left[r^2(t)\right] - E\left[r(t)\right]^2$$

$$= \frac{\sigma_r^2}{b_r} r(0)\left(e^{-b_r t} - e^{-2b_r t}\right) + \frac{a_r \sigma_r^2}{2b_r^2}\left(1 - 2e^{-b_r t} + e^{-2b_r t}\right)$$

$$+ e^{-2b_r t}\left(2a_r + \sigma_r^2\right)\sigma_r \int_0^t \left\{e^{b_r s}\int_0^s e^{b_r u} E\left[\theta(u)\sqrt{r(u)}\right] du\right\} ds + e^{-2b_r t} 2\sigma_r \int_0^t e^{2b_r s} E\left[r^{\frac{3}{2}}(s)\theta(s)\right] ds$$

$$- e^{-2b_r t}\sigma_r^2 \left\{\int_0^t e^{b_r s} E\left[\theta(s)\sqrt{r(s)}\right] ds\right\}^2 - 2e^{-2b_r t} r(0)\sigma_r \int_0^t e^{b_r s} E\left[\theta(s)\sqrt{r(s)}\right] ds$$

$$- \frac{2a_r}{b_r}\left(e^{-b_r t} - e^{-2b_r t}\right)\sigma_r \int_0^t e^{b_r s} E\left[\theta(s)\sqrt{r(s)}\right] ds$$

Then, $Var\left[r(t)\right] \to \dfrac{a_r \sigma_r^2}{2b_r^2}$ as $t \to \infty$.